# Generative Artificial Intelligence in Medical Imaging: Foundations, Progress, and Clinical Translation

Xuanru Zhou[1,2], Cheng Li[1], Shuqiang Wang[3], Ye Li[1], Tao Tan[4], Hairong Zheng[1], and Shanshan Wang*[1]

[1]Paul C. Lauterbur Research Center for Biomedical Imaging, Shenzhen Institutes of Advanced Technology, Chinese Academy of Sciences, Shenzhen, China.

[2]University of Chinese Academy of Sciences, Beijing, China.

[3]Research Center for Biomedical Information Technology, Shenzhen Institutes of Advanced Technology, Chinese Academy of Sciences, Shenzhen, China

[4]Faculty of Applied Sciences, Macao Polytechnic University, Macao, China.

* Correspondence: Shanshan Wang (ss.wang@siat.ac.cn)

Abstract: Generative artificial intelligence (AI) is rapidly transforming medical imaging by enabling capabilities such as data synthesis, image enhancement, modality translation, and spatiotemporal modeling. This review presents a comprehensive and forward-looking synthesis of recent advances in generative modeling—including generative adversarial networks (GANs), variational autoencoders (VAEs), diffusion models, and emerging multimodal foundation architectures—and evaluates their expanding roles across the clinical imaging continuum. We systematically examine how generative AI contributes to key stages of the imaging workflow, from acquisition and reconstruction to cross-modality synthesis, diagnostic support, and treatment planning. Emphasis is placed on both retrospective and prospective clinical scenarios, where generative models help address longstanding challenges such as data scarcity, standardization, and integration across modalities. To promote rigorous benchmarking and translational readiness, we propose a three-tiered evaluation framework encompassing pixel-level fidelity, feature-level realism, and task-level clinical relevance. We also identify critical obstacles to real-world deployment, including generalization under domain shift, hallucination risk, data privacy concerns, and regulatory hurdles. Finally, we explore the convergence of generative AI with large-scale foundation models, highlighting how this synergy may enable the next generation of scalable, reliable, and clinically integrated imaging systems. By charting technical progress and translational pathways, this review aims to guide future research and foster interdisciplinary collaboration at the intersection of AI, medicine, and biomedical engineering.

# 1. Introduction

## 1.1. Motivation and Clinical drivers for Generative AI

Medical imaging represents a cornerstone of modern clinical medicine, significantly contributing to all stages of healthcare, encompassing diagnostic assessment, therapeutic planning, and prognostic evaluation. In diagnosis, it enables early disease detection, classification, and quantitative assessment, supporting precision medicine[1]. During treatment, imaging guides surgical procedures, radiation therapy, and minimally invasive interventions, allowing real-time decision-making and improved outcomes[2]. For prognosis, imaging supports longitudinal disease tracking, risk assessment, and treatment response evaluation[3]. Despite remarkable technological progress, several fundamental challenges continue to hinder the full potential of medical imaging in clinical practice, as illustrated in Fig. 1.

A major challenge in medical imaging is the scarcity and heterogeneity of high-quality data. Many modalities are limited by high costs, restricted access, and technical constraints such as slow acquisition, low resolution, and motion artifacts[4,5]. To mitigate these issues, low-dose CT/PET and under-sampled strategies (e.g., compressed sensing) are used to shorten scans and reduce radiation, but they inevitably introduce noise, artifacts, and resolution loss, driving the need for advanced enhancement techniques like denoising, artifact removal, super-resolution, and reconstruction[6]. In the treatment phase, imaging underpins precision interventions and intraoperative navigation, yet challenges remain in accurate dose calculation, cross-modality synthesis, and real-time tracking. For example, MRI-to-CT translation for radiotherapy can suffer from geometric distortion and loss of detail[2], while intraoperative registration is sensitive to motion and latency, limiting guidance accuracy[7]. From a prognostic perspective, longitudinal imaging is essential for monitoring disease progression, evaluating therapeutic response, and informing risk stratification[8]. However, long-term data collection is often incomplete due to high costs, patient dropout, and inconsistent acquisition protocols across institutions[9].

These limitations underscore the need for generative models that can synthesize missing data, harmonize heterogeneous inputs, and augment incomplete datasets. The clinical demand for such capabilities constitutes a primary motivation for the integration of generative AI into medical imaging workflows.

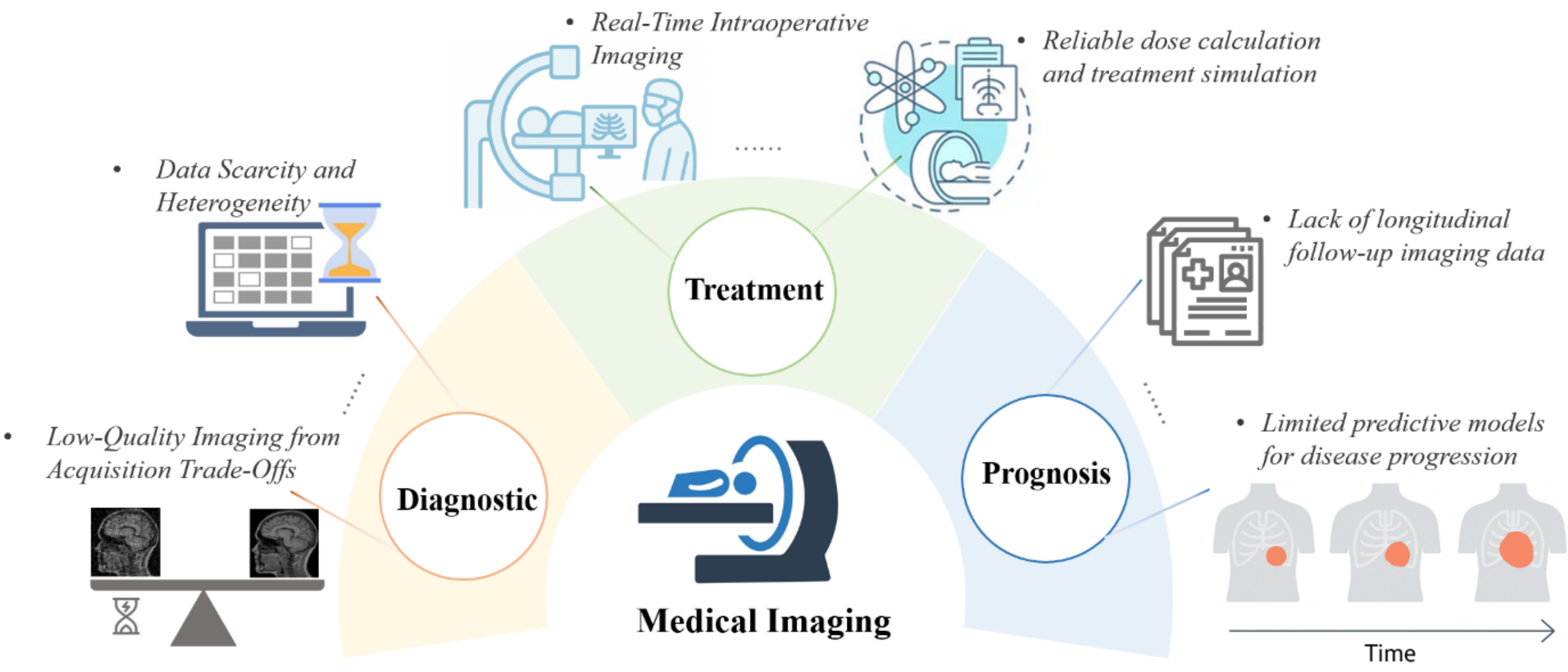

Fig. 1 | The challenges of medical imaging in clinical workflow.

## 1.2. Evolution of Generative Models in Medical Imaging

In recent years, generative artificial intelligence has emerged as a transformative force in medical imaging, revolutionizing how imaging data is generated, processed, and analyzed. Since the introduction of Generative Adversarial Networks (GANs) in 2014[10], followed by Variational Autoencoders (VAEs)[11], Diffusion Probabilistic Models (DPMs)[12], and sequence modeling architectures such as Transformers[13], Mamba[14], Autoregressive (AR) models[15,16] and foundation models[17–21], generative AI has demonstrated an unprecedented ability to model complex data distributions and generate high-quality synthetic medical images[22,23].

The integration of generative AI into medical imaging drives major advances across data augmentation, image restoration, modality translation, real-time synthesis, and prognostic modeling. Generative models synthesize realistic, high-fidelity images to address data scarcity, improving model generalization in disease detection[24,25]. AI-driven restoration enables denoising, artifact removal, super-resolution, and reconstruction, enhancing low-dose and accelerated imaging[5]. GANs and diffusion models support MRI-to-CT, PET-to-MRI, and other translations, aiding multimodal diagnosis and treatment planning[26]. Intraoperatively, real-time generative synthesis refines images for surgical navigation and radiotherapy adaptation[27–29]. For prognosis, longitudinal modeling simulates tumor growth, neurodegenerative progression, and recovery, assisting personalized treatment[30].

Overall, the integration of generative AI across the entire healthcare workflow not only revitalizes traditional medical practices but also establishes a robust foundation for the advancement of precision medicine. However, the full realization of its potential in healthcare remains constrained by several challenges. Chief among these are concerns regarding the reliability and interpretability of generative AI models. The phenomena such as hallucinations[31] can result in inaccurate outputs, while the black-box nature of many models also limits clinical trust. Generalization remains problematic, as performance often declines on unseen data or under varying imaging conditions[9,32]. Furthermore, high computational demands for training and deployment further constrain scalability in clinical settings[9,33]. Overcoming these limitations is essential to ensure the safe, effective, and ethical implementation of generative AI in medical applications.

### 1.3. Review Outline and Contributions

Generative AI is increasingly applied in medical imaging to address longstanding challenges such as limited data availability, suboptimal image quality, and insufficient temporal information. Overcoming current limitations in model generalizability, interpretability, and clinical validation is essential to advance its real-world deployment. This review aims to provide a comprehensive analysis of recent advancements in medical image generation, with a focus on their clinical applications, evaluation methodologies, and future research directions. The key contributions of this work are as follows:

- ***Comprehensive Survey of Key Generative AI Models****:* We systematically explore the theoretical foundations and practical applications of GANs, VAEs, DPMs, and sequence modeling architectures (Transformers, Mamba, AR models), as well as foundation models
- ***Integration with the Clinical Workflow:*** We analyze how generative models are applied across diagnostic, therapeutic, and prognostic stages, enabling static image synthesis, restoration, dynamic image generation, treatment planning, and disease progression modeling within clinical workflows.
- ***Proposal of a Multi-Level Evaluation Framework:*** We propose a structured evaluation framework that assesses generative models at three levels: pixel-level fidelity, feature- and distribution-level consistency, and clinical-level applicability. This framework aims to bridge technical performance with clinical utility and supports standardized benchmarking across tasks.
- ***Discussion of Challenges, Limitations, and Future Directions:*** We examine prevailing challenges that hinder the clinical translation of generative AI, including limited generalizability, high computational demands, insufficient interpretability, and regulatory uncertainty, and discuss their implications for future research and model deployment.

## 2. Key Generative AI Models in Medical Imaging

The core technologies of generative AI in medical imaging primarily include Generative Adversarial Networks (GANs)[10], Variational Autoencoders (VAEs)[11], Diffusion Probabilistic Models (DPMs)[12], and sequence modeling architectures such as Transformers[13], Mamba[14], and autoregressive models[15,16], as well as foundation models[17–21] that unify and transfer knowledge across tasks and modalities. These generative techniques have demonstrated remarkable versatility across a wide range of medical imaging tasks, including image synthesis, quality enhancement, modality translation, image reconstruction, super-resolution generation, and dynamic imaging modeling[6].

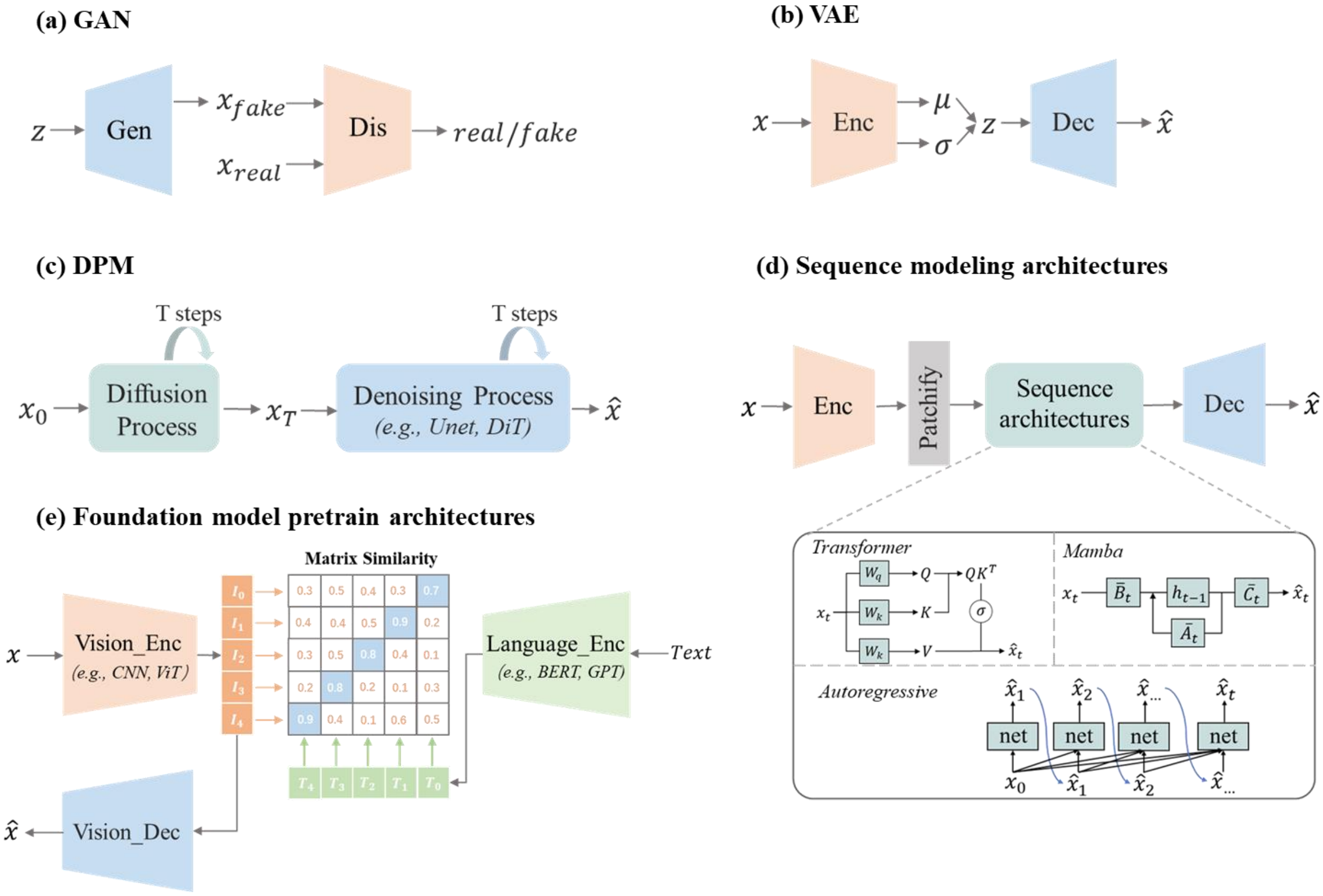

Fig. 2 | Architectures of generative AI models in medical imaging.

***Generative Adversarial Network (GANs)***, proposed by Ian Goodfellow et al. in 2014[10], represent a significant breakthrough in generative modeling by enabling the creation of realistic data distributions through an adversarial training framework. In the GAN, a generator (G) learns to produce synthetic data that closely mimics real samples, while a discriminator (D) distinguishes between real and generated data, as shown in Fig. 2(a). These networks engage in a minimax game, refining their outputs iteratively to generate high-quality, realistic data. The training process follows the objective:

$$\min_G \max_D \mathbb{E}_{x\sim p_{\text{data}}(x)}[\log D(x)] + \mathbb{E}_{z\sim p_z(z)}\left[\log\left(1 - D\left(G(z)\right)\right)\right], \quad (1)$$

where $p_{\text{data}}(x)$ is the real data distribution, $p_z(z)$ is the prior distribution on the latent vector $z$, $D(x)$ is the discriminator's probability that $x$ is real, $G(z)$ is the generated image from the latent vector $z$.

***Variational Autoencoders (VAEs)***[11], revolutionized generative modeling by combining variational inference with neural networks. A VAE consists of two components: an encoder, which maps input data to a latent space, and a decoder, which reconstructs the data from this latent representation. Fig. 2(b) shows how this structure allows VAEs to capture complex data distributions and generate new samples via latent space sampling. The training objective involves balancing reconstruction loss and Kullback-Leibler (KL) divergence[34], ensuring both accurate reconstruction and smoothness in the latent space. The objective function is expressed as:

$$\mathcal{L} = \mathbb{E}_{q_\phi(z|x)}[\log p_\theta(x \mid z)] - D_{KL}\big(q_\phi(z \mid x) \| p(z)\big), \tag{2}$$

where $q_\phi(z \mid x)$ is the encoder's approximation of the posterior distribution, $p_\theta(x \mid z)$ is the decoder's likelihood of the data given the latent variables, and $p(z)$ is the prior distribution over the latent space.

***Diffusion Probabilistic Models (DPMs)***[12], known as denoising diffusion probabilistic models, are a class of generative models inspired by non-equilibrium thermodynamics. They model data generation through a Markov chain that progressively adds Gaussian noise to the data, transforming it into a simple prior distribution, such as a standard normal distribution. The model then learns to reverse this diffusion process by progressively denoising the data, reconstructing the original data from the noisy samples. Mathematically, the forward process is expressed as:

$$q(x_t \mid x_{t-1}) = \mathcal{N}\big(x_t; \sqrt{1-\beta_t}x_{t-1}, \beta_t I\big), \tag{3}$$

where $x_0$ denotes the original data distribution, $x_t$ represents data with t step noise added, $\beta_t$ denotes the variance schedule controlling the amount of noise added at each step $t$. And the reverse process is defined as:

$$p_\theta(x_{t-1} \mid x_t) = \mathcal{N}\big(x_{t-1}; \mu_\theta(x_t, t), \Sigma_\theta(x_t, t)\big), \tag{4}$$

where $\mu_\theta$ and $\Sigma_\theta$ are the mean and covariance parameters predicted by the neural network with parameters $\theta$. The model is trained to minimize the variational bound on the negative log-likelihood, which can be expressed as:

$$\mathcal{L} = \mathbb{E}_q\big[D_{KL}\big(q(x_T \mid x_0) \| p(x_T)\big) + \sum_{t=1}^{T} D_{KL}\big(q(x_{t-1} \mid x_t, x_0) \| p_\theta(x_{t-1} \mid x_t)\big) - \log p_\theta(x_0 \mid x_1)\big], \tag{5}$$

here, $D_{KL}$ denotes the Kullback-Leibler divergence, and $p(x_T)$ is typically chosen as a standard normal distribution.

***Transformers***[13] have revolutionized deep learning by capturing long-range dependencies via self-attention, allowing them to model global relationships within data, unlike traditional CNNs that focus on localized receptive fields. The self-attention mechanism computes a sequence's representation by relating different positions within it, using query (Q), key (K), and value (V) matrices. The attention scores are calculated by the dot product of Q and K, scaled by the square root of the dimension and passed through a softmax function:

$$\text{Attention}(Q, K, V) = \text{softmax}\left(\frac{QK^T}{\sqrt{d_k}}\right)V, \tag{6}$$

***Mamba.*** The Mamba architecture, built upon state space models (SSMs)[14], has emerged as a transformative framework for medical image synthesis, addressing critical limitations of conventional models like Transformers (quadratic complexity) and CNNs (local-receptive constraints)[35]. At its core,

Mamba employs discretized state space equations to model sequential dependencies with linear computational scaling:

$$h_t = \bar{A}_t h_{t-1} + \bar{B}_t x_t, \tag{7}$$

$$y_t = \bar{C}_t h_t, \tag{8}$$

where $h_t$ denotes the hidden state, $x_t$ is the input, and $\bar{A}_t, \bar{B}_t, \bar{C}_t$ are discretized parameters derived via zero-order hold (ZOH). This formulation enables efficient integration of long-range features while maintaining the fidelity of local details, which is essential for medical imaging applications.

***Autoregressive Models***[15,16] generate images sequentially, predicting each pixel (or voxel) based on the previously generated ones. This sequential dependency modeling has proven highly effective in medical image synthesis, particularly for tasks requiring fine-grained pixel-level detail. By factorizing the joint distribution p(x) of an image into a product of conditional probabilities, AR models ensure that each generated element maintains consistency with prior context.

$$p(\mathbf{x}) = \prod_{t=1}^{T} p(x_t \mid x_{<t}), \tag{9}$$

where $x_t$ represents the $t$-th element (e.g., pixel, patch, or token) in a predefined generation order, and $x_{<t}$ denotes all previously generated elements. For high-dimensional medical images, this sequential dependency is often modeled using neural networks, such as Transformers or CNNs, to parameterize $p(x_t \mid x_{<t})$.

***Foundation Models***[17–21] are typically pretrained on large-scale datasets and designed to generalize across tasks and modalities, often requiring minimal task-specific supervision. The core idea is to bring matching image–text pairs closer together while pushing non-matching pairs apart. This framework forms the basis of many large-scale pretrained architectures as illustrated in Fig. 2 (e), enabling models to generalize across tasks with limited supervision and to support applications such as zero-shot classification, report retrieval, and text-guided image synthesis. These models are trained with a variant of the InfoNCE loss:

$$L_{\text{contrast}} = -\frac{1}{N}\sum_{i=1}^{N} \log \frac{\exp\left(\text{sim}\left(f(I_i), g(T_i)\right)/\tau\right)}{\sum_{j=1}^{N} \exp\left(\text{sim}\left(f(I_i), g(T_j)\right)/\tau\right)}, \tag{10}$$

where $f(I)$ and $g(T)$ are image and text encoders, sim(.) is a similarity (e.g., cosine), and $\tau$ is a temperature. This contrastive training ensures paired images and captions have high similarity, enabling zero-shot image classification and retrieval.

Each generative model contributes distinct advantages to medical imaging tasks. GANs and diffusion models offer high image fidelity but differ in diversity and inference efficiency. VAEs provide interpretable latent spaces with faster inference but often suffer from low visual quality. Transformers and Mamba architectures improve long-range representation, with Mamba offering better computational efficiency. Autoregressive models enable fine-grained control but are limited by sequential inference. Foundation models, pretrained on large-scale data, support cross-task generalization and multimodal integration, yet remain constrained by high computational demands and limited interpretability. A comparative overview of these methods is summarized in Table 2.

Table 2 | Summary of key strengths and limitations of representative generative models, including GANs, VAEs, diffusion models (DPMs), Transformers, Mamba, and Autoregressive models.

| Model | Strengths | Weaknesses |
|---|---|---|
| **GANs** | High fidelity;<br>High controllability;<br>Efficient inference | Mode collapse;<br>Limited diversity;<br>Training instability |
| **VAEs** | Latent space interpretability;<br>Efficient inference | Low fidelity |
| **DPMs** | High fidelity;<br>High diversity;<br>High controllability | Limited inference |
| **Transformers** | Long-range dependency;<br>Global information;<br>Multimodal adaptability | Limited local information;<br>Limited computational efficiency |
| **Mamba** | Long-range dependency;<br>High computational efficiency;<br>Efficient inference | Memory dilution;<br>Limited pre-training |
| **Autoregressive Models** | High fidelity;<br>Local information | Error accumulation;<br>Memory dilution;<br>Limited inference |
| **Foundation Models** | Cross-task generalization;<br>Multimodal understanding;<br>Zero-shot task | High training cost;<br>Data dependency |

# 3. Key Applications of Generative AI in Medical Imaging

Medical imaging face interconnected challenges: (1) the scarcity of expert-annotated data; (2) heterogeneity in image quality across different imaging devices and institutions; (3) the trade-offs involved in optimizing image quality; (4) the poor generalizability of models to rare or atypical cases. Generative AI provides promising solutions by synthesizing high-quality, realistic medical images and enhancing existing datasets with anatomically consistent variations. Now generative AI is increasingly applied across various stages of clinical workflows, including diagnosis, treatment, and prognosis. In Fig. 3, this section provides an overview of the current applications of generative AI models in medical images, with the goal of assisting researchers in analyzing the distribution of these applications and identifying potential future research directions.

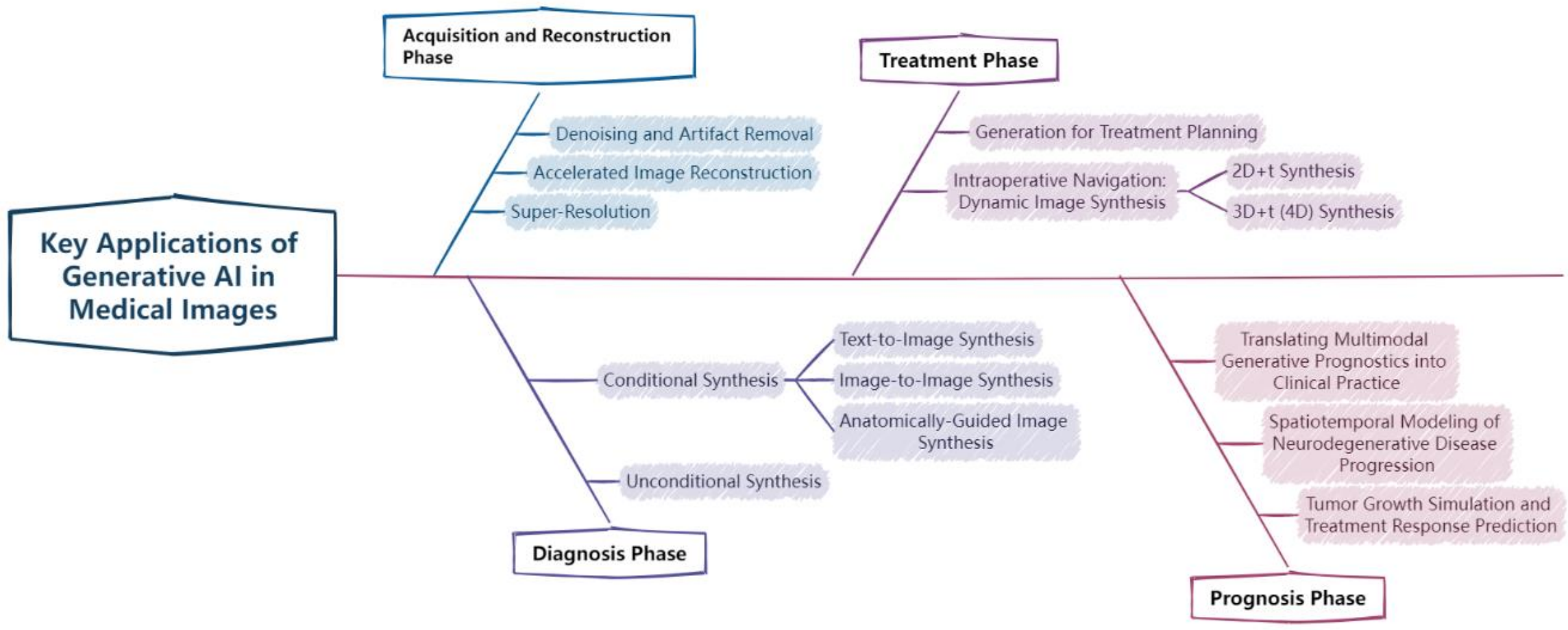

Fig. 3 | Structural taxonomy of clinical applications of generative AI in medical imaging.

## 3.1. Acquisition and Reconstruction Phase: Enhancing Data Quality and Availability

High-quality medical images are essential for accurate diagnosis, treatment planning, and disease monitoring. However, acquisition constraints, low-dose protocols, patient motion, and hardware limitations have often introduced noise, artifacts, low resolution, or incomplete data, thereby compromising clinical interpretation. To address these challenges, generative AI models have become increasingly pivotal in restoring and enhancing image quality across modalities like CT, MRI, and PET. By leveraging adversarial learning, diffusion-based modeling, and transformer architectures, these models support a wide range of restoration tasks, including denoising, artifact removal, super-resolution, and image reconstruction.

***Denoising and artifact removal.*** In low-dose CT, quantum noise and metal artifacts have obscured fine details, limiting lesion detection. Traditional filters reduced noise but blur structures. Recent generative methods performed better: the Poisson flow model[36] suppressed stochastic noise in photon-counting CT, and Wasserstein GANs removed metal-induced artifacts in dental CT, improving implant planning[37,38]. In ultra-low-dose protocols, CoreDiff[39] has been employed to reconstruct lung images, directly supporting early nodule screening. In PET, parameter-transferred GANs and diffusion models reduced noise while preserving standardized uptake values, essential for therapy monitoring[40,41]. For MRI, which is susceptible to Rician noise and motion artifacts, residual GAN has been shown to improve inter-slice consistency[42], while a reverse diffusion model[43] enhanced both resolution and noise suppression for clearer anatomical detail.

***Accelerated image reconstruction.*** Reducing acquisition time or dose often leads to sparse or incomplete data, risking loss of critical diagnostic features. In CT, GAN-based sinogram inpainting restored missing projections, enabling accurate lung screening under limited angles[44], while diffusion priors outperformed iterative methods in detecting subtle hemorrhages[45]. In PET, a CycleGAN[46] improved metabolic feature alignment across modalities, and a VAE-based method[47] reduced PET–MRI registration errors, supporting precise multimodal assessment. Dynamic PET reconstruction with deep generative models restored temporal fidelity essential for therapy monitoring[48]. For MRI, transformer-based and diffusion-informed architectures accelerated cine MRI acquisition while preserving lesion visibility[49,50], and the state-space framework such as Mamba integrated uncertainty quantification for safer clinical decision-making[51].

***Super-resolution.*** Limited resolution constrains lesion detection and functional assessment, especially in dynamic organs. Temporal super-resolution has been vital for cardiac or respiratory imaging, where diffusion-based deformation models captured complex motion and suppress irregular artifacts[52]. Spatial SR addressed structural clarity: GAN-CIRCLE improved CT texture fidelity[53], and diffusion-based dual-stream models enhanced MRI resolution while preserving anatomical consistency[54]. These advances provide higher diagnostic confidence in early disease detection and treatment planning.

Generative models have mitigated noise, artifacts, sparsity, and resolution limits while preserving diagnostic integrity. CT, PET, and MRI all benefit through more reliable reconstructions, faster acquisition, and improved lesion visibility. Super-resolution further enhances anatomical detail and temporal dynamics, reducing the need for higher dose or longer scans. These advances secure image fidelity at the acquisition stage and set the stage for the next focus: “Diagnosis Phase: Enriching Diagnostic Imaging”, where generative AI shifts from restoration to synthesis to address data scarcity and enhance diagnostic utility.

## 3.2. Diagnosis Phase: Enriching Diagnostic Imaging

Static image synthesis techniques are instrumental in addressing the challenges of data scarcity and domain adaptation. These techniques generate medical images either unconditionally (without explicit constraints) or conditionally (guided by clinical parameters, textual descriptions, etc.), providing solutions to enhance training datasets and improve model generalizability. Below, we categorize these methods based on their underlying approach: unconditional synthesis and conditional synthesis.

***Unconditional synthesis*** generates medical images directly from noise distributions, enabling the creation of diverse datasets without requiring annotations. Early GANs produced structures such as vascular surfaces and brain MRI volumes[55,56] , but suffered from mode collapse and low resolution. Advances such as StyleGAN[57,58] introduced structured latent spaces, improving fidelity and controllability. More recently, diffusion probabilistic models (DDPMs) have become the leading approach, offering stable training and higher diversity. For example, Medical diffusion models[59,60] generated high-resolution CT and MRI data with improved anatomical detail, supporting tumor detection and segmentation. These advances demonstrate unconditional synthesis as a critical tool for addressing data scarcity, though lack of explicit control limits its direct clinical use.

***Conditional synthesis.*** In contrast to unconditional synthesis, which learns image distributions independently of external inputs, conditional synthesis incorporates domain-specific priors such as clinical text, imaging data, anatomical structures, or physiological parameters into the generative process. This improves the relevance, controllability, and diagnostic value of the synthesized outputs.

- *Text-to-Image Synthesis.* Radiology reports and clinical metadata often contain valuable diagnostic cues but lack paired imaging for direct use. To bridge this gap, latent diffusion models have enabled text-to-image generation, aligning textual findings with synthetic images. For example, Chest-diffusion[61] generated chest X-rays from reports, enriching datasets for rare pathologies and improving interpretability. Extending this idea, MediSyn[62] generalized across modalities, creating diverse synthetic scans guided by textual or clinical prompts. These approaches improve the alignment between clinical documentation and imaging, expanding data availability for diagnostic model training.
- *Image-to-Image Synthesis.* In clinical workflows, missing or degraded modalities (e.g., unavailable CT in PET/MRI workflows) compromise diagnosis and treatment planning. Image-to-image synthesis has addressed this by translating between modalities while preserving structural fidelity[63]. CycleGAN[64–66] demonstrated the feasibility of bidirectional mappings between CT, PET, and MRI without paired data, while transformer-based models such as ResViT[67] further improved spatial consistency and cross-modality alignment. More recently, diffusion-based methods[39,68–70] further enhanced anatomical preservation, enabling robust modality completion and zero-shot translation. These methods have directly reduced the impact of incomplete or inconsistent imaging in clinical pipelines.
- *Anatomically-Guided Synthesis.* A persistent limitation of generative synthesis is the risk of anatomically implausible outputs. To overcome this, anatomical priors such as segmentation masks or vascular maps have been embedded into the generation process. For instance, the vascular-guided GAN[71] preserved fine vessel structures in retinal fundus images, while the segmentation-guided diffusion model[72] allowed controllable synthesis across multiple organs and modalities. By integrating structural constraints, these methods enhanced both interpretability and clinical reliability, making synthetic data more suitable for lesion augmentation and rare disease modeling.

Unconditional synthesis has expanded datasets without annotations, using GANs and diffusion models to generate diverse, anatomy-preserving images that strengthen model robustness under data scarcity. Conditional synthesis adds clinical control: text-driven methods align reports and demographics with synthetic images; image-to-image translation and completion recover missing or degraded modalities; anatomically guided generation enforces structural plausibility for lesion-level augmentation and rare-disease scenarios. Together, these approaches move beyond restoration to enrich training distributions, improve domain generalization, and tighten the link between clinical context and image content.

## 3.3. Treatment Phase: Enabling Precision Interventions

In the treatment phase of clinical care, the integration of generative AI into radiotherapy and intraoperative navigation offers transformative potential for precision medicine. By modeling complex anatomical variations, capturing physiological motion, and supporting real-time clinical decision-making, generative models are increasingly bridging the gap between static preoperative imaging and dynamic, adaptive interventions. This section explores two key areas: dose prediction and planning in radiotherapy, and dynamic image synthesis for intraoperative navigation.

***Generation for treatment planning.*** In radiotherapy, inter-patient anatomical variability and tumor motion have complicated precise dose delivery, often risking damage to adjacent organs[73,74]. Generative models have emerged as powerful tools for predicting individualized dose maps and simulating treatment anatomy. Early frameworks such as DoseNet[75] applied fully convolutional networks to rapidly generate 3D dose distributions, while TransDose[76] introduced transformers to capture long-range spatial dependencies and improve conformity around critical organs. More recently, diffusion-based approaches such as DiffDP[77] have enabled the generation of multiple plausible dose distributions from CT and segmentation inputs, supporting flexible planning in anatomically complex cases. Similarly, MD-dose[78] enhanced both sampling speed and accuracy through its Mamba-based architecture, supporting real-time adaptive planning. Foundation and generative models have extended beyond dose prediction, contributing to imaging tasks like synthetic image generation via a self-improving model[24] and CBCT-based tumor tracking[79], which collectively enhanced adaptive radiotherapy workflows. Collectively, these approaches reduce trial-and-error costs, enhanced personalization, and lay the foundation for real-time adaptive radiotherapy.

***Intraoperative navigation: dynamic image synthesis.*** Real-time intraoperative imaging must capture both anatomy and motion, but conventional acquisitions are constrained by slow speed, radiation dose, and motion artifacts. Generative models have been explored to synthesize dynamic sequences from limited inputs. In cardiac MRI, the GAN-based framework[80] accelerated cine reconstruction while preserving morphology. DragNet[81], a registration-driven method, recovered full cardiac cycles from static frames, reducing motion blur. A cascaded video diffusion model[82] refined motion and texture using semantic cues, producing smoother and more realistic echocardiograms. Multimodal conditioning, for example combining ECG with imaging, has enabled personalized cardiac motion synthesis in the HeartBeat[83]. At the volumetric level, a temporally aware GAN[84] integrated respiratory compensation into dynamic 3D cardiac MRI, effectively reducing motion-induced artifacts. Cross-modal strategies further advanced adaptability in radiotherapy: one study synthesized 4D CT from sparse CBCT[85], while another translated CBCT into 4D MRI[86]. Despite these advances, current approaches still struggle with nonlinear motion and real-time deployment. A recent text-driven method[87] that incorporated disease descriptions into cardiac cine MRI illustrates a promising path toward controllable, pathology-specific motion generation, bridging dynamic imaging with intelligent intervention.

## 3.4. Prognosis Phase: Longitudinal & Personalized Medicine

Generative medical imaging techniques have demonstrated significant clinical potential in longitudinal prognostic analysis and personalized medicine. By leveraging deep modeling of patients' multi-temporal imaging data, these approaches can simulate dynamic disease progression, predict tissue degenerative changes, and quantify prognostic risk, thereby providing data-driven support for clinical decision-making.

***Tumor growth simulation and treatment response prediction.*** Precise modeling of tumor evolution is vital for planning adaptive therapies, yet variability in growth patterns and treatment response limits conventional approaches. A treatment-aware diffusion probabilistic model[30] simulated glioma growth from longitudinal MRI and molecular data, improving future tumor prediction accuracy by over 16%. To address incomplete follow-up scans, SADM[88] introduced autoregressive sequence generation, enabling robust modeling despite missing data. Synthetic tumor framework further enhanced radiomics-based survival prediction in glioblastoma, supporting patient-specific radiotherapy[89]. More recently, foundation model[90] has trained across tumor types established unified prognostic platforms, aiding therapy selection and risk stratification.

***Spatiotemporal modeling of neurodegenerative disease progression.*** For disorders such as Alzheimer's disease, monitoring structural brain changes over time is essential for staging and therapy. Generative synthesis of longitudinal MRI has enabled visualization of subtle degenerative trajectories[91,92]. A hybrid DCGAN–SRGAN framework[93] generated synthetic MRI sequences across disease stages, achieving high classification accuracy and supporting progression modeling. The Temporal-Aware Diffusion Model (TADM) [94] further reduced brain volume prediction error by 24% compared with conventional baselines, improving anatomical fidelity in longitudinal imaging. These methods have offered quantitative and visual tools to track disease progression and guide optimal intervention timing.

**Translating multimodal generative prognostics into clinical practice.** Integrating imaging with clinical variables remains a challenge for prognosis. A conditional GAN[95] has been used to synthesize cardiac aging images, improving early detection of diastolic dysfunction, while a diffusion model[96] improved brain volume prediction in Alzheimer's by over 20%. In oncology, a foundation model[24] enhanced breast cancer stratification by increasing HER2 and EGFR sensitivity. For cerebrovascular disease, a synthetic CT-based deep model[97] predicted hematoma expansion with an AUC of 0.91, supporting early clinical decision-making. Radiomics features extracted from synthetic MRI also improved glioblastoma survival prediction across centers[89]. These advances highlight the clinical value of generative prognostics, though large-scale translation will depend on improving domain adaptation, interpretability, and workflow integration.

By capturing dynamic, multimodal disease trajectories, generative imaging models offer powerful tools for prognosis across tumor, neurological, and cardiovascular domains. Nonetheless, clinical translation at scale requires further work in domain adaptation, temporal modeling, and model interpretability. Future progress in these areas is expected to enhance robustness and generalizability across diverse clinical environments, reinforcing the role of generative models in precision medicine and personalized care.

# 4. Overview of Public Datasets

The rapid development of generative AI in medical imaging has been largely driven by large-scale, high-quality, multi-modal public datasets, which provide both essential training resources and standardized benchmarks for generalization and clinical applicability.

Representative repositories such as UK Biobank[98] and TCIA[99] encompass diverse modalities (MRI, CT, ultrasound, PET, fundus) and tumor types, enabling image synthesis, modality translation, and anomaly simulation. Grand Challenge and Kaggle platforms further facilitate reproducible benchmarking across a wide range of imaging tasks.

For specific anatomical regions, landmark datasets like DeepLesion[100], PreCT-160K[101], TotalSegmentator[102,103], and BraTS21[104] offer unprecedented scale or fine-grained annotations, supporting lesion synthesis, longitudinal prediction, and anatomically guided generation. In cardiovascular imaging, EchoNet-Dynamic[82] and ACDC[105] enable dynamic 2D+t/3D+t modeling, while datasets such as AutoPET[106] and HECKTOR[107] facilitate PET-CT fusion for tumor-focused tasks. Beyond radiology, large-scale resources in pathology (e.g., PatchCamelyon[108], Quilt-1M[109]), ophthalmology (e.g., OCT2017[110], ODIR-5K[110]), and multimodal image–text corpora (e.g., CheXpertPlus[111], MedICaT[112], Medtrinity-25M[113]) have become indispensable for foundation models and vision–language pretraining.

While these datasets have enabled substantial advances, challenges such as domain shift, annotation inconsistency, and limited dynamic or longitudinal data remain. Addressing these gaps through standardization, collaborative curation, and responsible synthetic data integration will be crucial for reliable deployment of generative models in clinical practice (see Supplementary Note S5 and Table S9 for the full dataset catalogue).

# 5. Evaluation Methods for Generative Models in Medical Imaging

Evaluation remains a key challenge for generative AI in medical imaging. Conventional pixel-level metrics often fail to capture anatomical plausibility or clinical utility, while inconsistent standards hinder fair comparison across tasks and modalities. Reliable evaluation is therefore critical for both methodological benchmarking and clinical translation. To address this, we adopt a three-level hierarchical evaluation framework (Fig. 4) that integrates complementary strategies at different abstraction levels: pixel fidelity, feature and distribution consistency, and clinical relevance. This structure provides a more systematic way to assess image quality, semantic realism, and diagnostic utility.

***Low-level evaluation*** focuses on pixel-wise similarity between generated and reference images. Metrics such as MSE, MAE, PSNR, and RMSE[114] are widely used in reconstruction and denoising tasks but correlate poorly with human perception. Structural metrics like SSIM[115], MS-SSIM[116], and FSIM[117] incorporate luminance, contrast, and texture, offering improved alignment with visual perception. Advanced variants such as IW-SSIM[118] and CACI[119] further emphasize diagnostically relevant regions. These metrics effectively assess structural integrity and visual fidelity in tasks like denoising, reconstruction, and compression. However, their focus on low-level features limits detection of semantic inconsistencies, anatomical errors, and clinically irrelevant content critical to evaluating diagnostic utility.

***Mid-level evaluation*** assesses feature-level similarity and distribution alignment using pretrained models. Metrics such as FID[120], KID[121], MMD[122], and Inception Score evaluate global structure and diversity but depend on the domain of the feature extractor. Perceptual similarity measures like LPIPS[123], and multimodal embedding scores such as CLIP Similarity[124] and MedCLIP-score[125], help detect hallucinations by assessing image–text coherence. Other metrics like RQI[119], AHI[119], and BmU[126] evaluate restoration quality and semantic alignment, while FVD[127] and FVMD[128] extend assessment to temporal coherence in dynamic imaging. Mid-level evaluations bridge pixel fidelity and clinical relevance, offering insights into perceptual and statistical realism. However, their effectiveness depends on pretrained model alignment and task complexity, making them more useful when combined with low- and high-level assessments for comprehensive validation.

***High-level evaluation*** represents clinically significant stage in assessing generative models for medical imaging. Unlike lower-level metrics that assess pixel accuracy or feature similarity, this stage focuses on clinical applicability in tasks such as diagnosis, treatment planning, and disease monitoring. It includes expert assessment, where radiologists evaluate realism and anatomical plausibility, and downstream validation through segmentation or classification performance. Radiologists can judge realism, anatomical correctness, and diagnostic plausibility, as demonstrated in studies like GenerateCT[129] and MINIM[24], where expert feedback guided model improvement. In downstream tasks, synthetic images have shown strong performance in segmentation, classification, and regression, indicating they preserve clinically relevant features. For example, synthetic MRIs supported accurate tumor segmentation, and synthetic breast cancer images improved diagnostic classification accuracy in data-limited scenarios.

Evaluating generative models in medical imaging requires balancing visual quality with clinical relevance. Pixel-level metrics are easy to compute but miss perceptual and diagnostic accuracy. Feature-based measures like FID and LPIPS better capture semantics but depend on pretrained model choice and dataset size. Expert reviews offer direct diagnostic insight but remain subjective. Combining complementary strategies is essential: objective metrics, expert ratings, and task-based validation together ensure technical and clinical utility. For example, MINIM[24] integrates all three, showing how multi-level evaluation supports models that are both statistically robust and clinically meaningful, highlighting the need for standardized, multi-faceted protocols for real-world deployment.

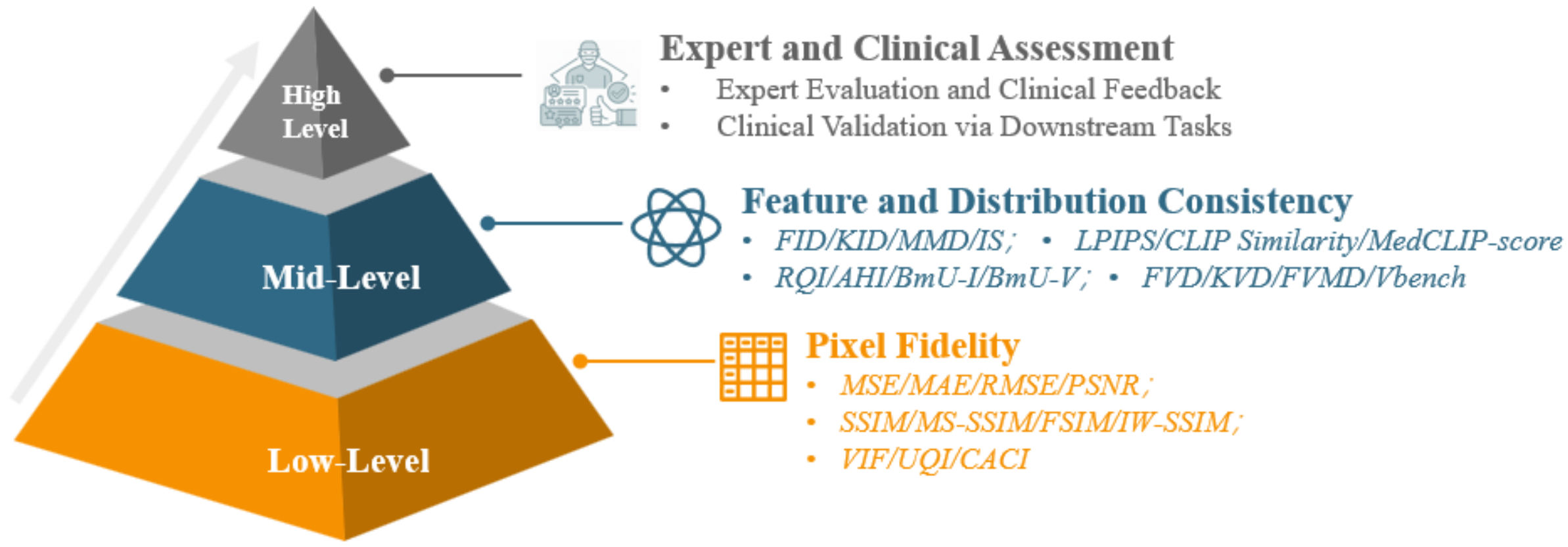

Fig. 4 | Three-Level Evaluation Pyramid for Generative Models in Medical Imaging. This figure illustrates a hierarchical evaluation framework comprising low-level, mid-level, and high-level metrics. The structure emphasizes a progression from basic image quality toward clinical applicability.

# 6. Discussion and Future Directions

Generative models in medical imaging face significant hurdles from both technical and clinical perspectives. Technically, these models grapple with challenges such as limited generalization, high computational demands, opaque decision-making processes, dependence on high-quality data, and the risk of generating misleading "hallucinations." Clinically, concerns revolve around ensuring model reliability and trustworthiness, enhancing interpretability for informed decision-making, seamlessly integrating AI into existing workflows, and addressing regulatory and ethical constraints. These challenges highlight the intricate balance between advancing AI-driven imaging technologies and meeting the stringent requirements of clinical practice.

## 6.1. Technical Challenges and Limitations

***Limited Generalization and Bias:*** Generative models often perform well on benchmark datasets but struggle when applied to different institutions, modalities, or demographics due to training data bias. For example, models trained mainly on adult CT scans may generalize poorly to pediatric or low-resource settings. Addressing this requires more diverse and representative data, including rare diseases and multi-center cohorts. Foundation models[24] have shown potential by synthesizing multi-organ or cross-modality images from text prompts and generalizing to unseen domains. However, eliminating bias remains difficult, and careful dataset curation is necessary to prevent reinforcing healthcare disparities[130].

***High Computational Demands:*** Modern generative models like GANs and diffusion models are resource-intensive, especially for high-resolution or 3D images. This limits their use in time-sensitive clinical settings such as emergency or intraoperative care. Optimizing efficiency is critical—recent works[131] on model compression, architectural improvements, and knowledge distillation aims to reduce inference time without compromising quality. At the same time, real-time performance is especially critical for clinical tasks such as intraoperative guidance and bedside diagnostics, where delays of even a few seconds can affect decision-making. So, improving computational efficiency will be essential for enabling the widespread adoption of generative models in routine clinical workflows.

***Lack of Interpretability:*** Many generative models operate as black boxes, offering little transparency into how specific outputs are produced. For instance, a model might generate a nonexistent tumor with no explanation, raising concerns in fields like radiology where trust and accuracy are vital. Scientifically, the internal logic of these models remains opaque; clinically, their lack of transparency hinders adoption. To address this, researches are exploring the use of attention maps[132,133], saliency visualization[134], and hybrid designs[135] that combine deep learning with more interpretable components. As a result, there is increasing demand for explainable AI methods that clarify which features the model has relied on and how they influenced the output.

***Data Scarcity and Privacy:*** High-quality annotated medical images are essential for training robust models, yet access is often restricted by privacy laws and institutional policies. Datasets covering rare or underrepresented conditions are especially limited. Although synthetic data may offer partial relief, initial training still requires real-world clinical input. Federated learning[136,137] has emerged as a privacy-preserving approach, allowing models to learn from distributed data sources without sharing sensitive information. Nevertheless, federated learning presents its own technical challenges, such as communication overhead, inconsistency in data quality, and difficulties in synchronizing model updates across sites.

***Hallucinations and Uncertainty:*** One major risk of generative models in medical imaging is the creation of hallucinated features—structures that appear realistic but are not present in the original image. These hallucinations can be subtle and may not be detected by standard evaluation metrics, yet they carry significant clinical risk. To manage this, researchers[138–140] are developing uncertainty estimation methods such as confidence maps, Bayesian modeling, and ensemble predictions to highlight unreliable regions. Besides, some researchers also explore statistical indicators like a hallucination index[31] to quantify the likelihood of fabricated content. Reducing these risks requires improved training strategies, including the use of diverse datasets and regularization techniques that promote anatomical fidelity. While early results are promising, the reliable detection and prevention of hallucinations in complex, real-world settings remain an open challenge.

## 6.2. Clinical Challenges and Limitations

***Reliability and Trustworthiness：*** Clinicians' primary concern is whether AI-generated images and results can be trusted for diagnosis and treatment planning. Medical decisions often rely on subtle findings, and errors such as missing a tumor or adding a false lesion can have serious consequences. Even infrequent mistakes may undermine confidence in the system. Thus, generative models must ensure not just accuracy but also consistent performance in rare or high-risk cases. Studies conducted in recent years highlight clinicians' openness to AI, while also emphasizing the importance of understanding its failure modes[141,142]. Maintaining a human-in-the-loop approach, where AI augments rather than replaces expert judgment, remains essential until reliability is firmly established.

***Explainability for Decision-Making:*** Clinicians and regulatory bodies increasingly demand that AI decisions be explainable. For generative models, this means clarifying how outputs are produced—such as why a lesion is synthesized or how an MRI is converted to a CT. Explainability is tied closely to the interpretability issues discussed above, but here the emphasis is on the end-user perspective. For instance, if a generative model highlights an area on a PET scan as malignant (by enhancing it or annotating it), the oncologist will need to understand the basis for that suggestion – was it a particular texture, intensity pattern, or a correlation with other data? Without such context, the physician cannot confidently incorporate the AI's output into their decision.

***Integration into Clinical Workflow:*** Even highly capable generative models may have limited clinical value if they cannot be integrated seamlessly into existing workflows. Hospitals and imaging centers depend on established systems like radiology information platforms and standardized diagnostic protocols. Introducing such tools raises practical concerns: Can the model deliver real-time analysis during image acquisition? Is it compatible with hospital IT infrastructure for secure access and storage? Does it create delays or add steps for clinicians? Interoperability and intuitive design are therefore essential in clinical workflows.

***Regulatory and Ethical Constraints:*** Generative AI in medicine must adhere to strict regulatory and ethical standards. Unlike fixed-function devices, they can evolve or behave unpredictably, complicating approval. Recent regulations, such as the EU AI Act, treat diagnostic AI as high risk and require transparency, human oversight, and risk controls[143]. Ethical issues include consent for generated data, privacy concerns from synthetic images, and unclear responsibility when errors occur. Fairness is also a concern, as model performance may vary between populations. Addressing these challenges requires diverse training data, consistent bias monitoring, and safeguards for equitable and responsible use.

## 6.3. The Emergence of Multimodal Foundation Models

Recently, foundation models (FMs) have been transforming medical image generation by enabling unified and transferable solutions across the clinical continuum. Pre-trained on large, diverse datasets, they exhibit strong generalization and zero-shot capabilities, making them effective across multi-modal and multi-stage imaging tasks. Unlike conventional models restricted to narrow objectives, FMs provide a flexible backbone for image synthesis, enhancement, and interpretation spanning diagnostic, therapeutic, and prognostic needs.

***Modality-specific foundation models.*** In CT, MedDiff-FM[144] leveraged diffusion-based architectures to generate high-resolution volumes under variable acquisition conditions, supporting both denoising and anatomical completion for diagnostic and radiotherapy planning. For MRI, Triad[145] jointly optimized segmentation, classification, and registration within a unified 3D framework, enforcing anatomical consistency and improving robustness across pathologies. In ophthalmology, RETFound-DE[146] showed that data-efficient pretraining on limited fundus datasets yielded strong generalization in diabetic retinopathy screening. In pathology, BEPH[147] trained on over 11 million whole-slide patches and generalized to cancer detection and survival prediction, while Prov-GigaPath[21] scaled to 1.3 billion tiles, setting new benchmarks across 26 pathology tasks. These works have highlighted how modality-specific FMs improve quality, scalability, and downstream utility.

***Vision–language foundation models.*** Vision–language FMs bridge clinical semantics with imaging, enabling interpretable, context-aware synthesis. RoentGe[19] generated chest X-rays conditioned on radiology reports, supporting standardized screening protocols. A radiograph–report FM[18] aligned images and text through masked contrastive learning, enhancing both interpretation and automated reporting. Beyond 2D tasks, MINIM[24] integrated multimodal pretraining to synthesize high-fidelity CT, MRI, and OCT from partial inputs or clinical prompts, advancing diagnosis, report generation, and cross-modality synthesis. These approaches have demonstrated the potential of vision–language FMs to unify textual and visual data, improving both clinical interpretability and workflow integration.

Foundation models mark a paradigm shift in medical imaging by offering generalizable frameworks across modalities and tasks. They enable multi-purpose synthesis and analysis, from CT denoising to multimodal report generation. However, their reliance on massive datasets introduces challenges in data availability, annotation quality, computational cost, and clinical integration. Addressing issues such as hallucinations, interpretability, and domain transfer will be key for safe deployment. Future directions include efficient pretraining, federated learning, and regulatory-aligned evaluation to ensure foundation models can support precision healthcare at scale.

# 7. Outlook

Generative models offer significant promise in medical imaging, enabling data augmentation, modality translation, and disease progression simulation. But their deployment in real-world clinical environments remains limited. This limitation arises from a combination of unresolved technical and clinical challenges. On the technical side, generative models continue to struggle with generalization across institutions and modalities, high computational requirements, limited interpretability, reliance on sensitive annotated data, and the risk of producing hallucinated features. Clinically, concerns remain regarding the reliability of model outputs, the transparency required for informed decision-making, the seamless integration of AI tools into established workflows, and adherence to evolving regulatory and ethical standards.

While generative AI offers significant opportunities to transform medical imaging, realizing its full potential will require addressing persistent challenges related to generalizability, computational efficiency, reliability, and interpretability. To advance from research prototypes to widespread clinical adoption, future work must focus on developing models that are robust to data heterogeneity, informed by anatomical and physiological priors, and capable of providing uncertainty quantification[148]. Enhancing model transparency through explainable design is essential for building clinical trust, while multi-task learning and domain adaptation strategies may improve efficiency and robustness across varied imaging tasks and institutions. Real-time inference capabilities will be critical for applications such as image-guided interventions and emergency diagnostics. Achieving these goals will rely on the development of foundation models trained on large-scale, diverse, multi-institutional datasets, and their seamless integration into clinical systems. Moving forward, close collaboration between technical, clinical, and regulatory communities will be essential to ensure that generative models meet the rigorous standards required for safe, effective, and ethical deployment in healthcare. We hope this review can serve as a valuable resource for researchers and practitioners and inspire continued innovation in this rapidly advancing field.

## Acknowledgements

This research was partly supported by the National Natural Science Foundation of China (No. 62222118, No. U22A2040), Shenzhen Medical Research Fund (No. B2402047), National Key R&D Program of China (No. 2023YFA1011400), Key Laboratory for Magnetic Resonance and Multimodality Imaging of Guangdong Province (No. 2023B1212060052), and Youth lnnovation Promotion Association CAS.

## Competing interests

The authors declare no competing interests.

## References

1. Kumar, Y., Koul, A., Singla, R. & Ijaz, M. F. Artificial intelligence in disease diagnosis: A systematic literature review, synthesizing framework and future research agenda. *J. Ambient Intell. Hum. Comput.* **14**, 8459–8486 (2023).
2. Sun, H. *et al.* Research on new treatment mode of radiotherapy based on pseudo-medical images. *Comput. Methods Programs Biomed.* **221**, 106932 (2022).
3. Kazmierski, M. *et al.* Multi-institutional prognostic modeling in head and neck cancer: evaluating impact and generalizability of deep learning and radiomics. *Cancer Res. Commun.* **3**, 1140–1151 (2023).
4. Schäfer, R. *et al.* Overcoming data scarcity in biomedical imaging with a foundational multi-task model. *Nat. Comput. Sci.* **4**, 495–509 (2024).
5. Ibrahim, M. *et al.* Generative AI for synthetic data across multiple medical modalities: A systematic review of recent developments and challenges. Preprint at https://doi.org/10.48550/arXiv.2407.00116 (2024).

6. Sun, Y., Wang, L., Li, G., Lin, W. & Wang, L. A foundation model for enhancing magnetic resonance images and downstream segmentation, registration and diagnostic tasks. *Nat. Biomed. Eng.* 1–18 (2024) doi:10.1038/s41551-024-01283-7.
7. Fan, X., Zhu, Q., Tu, P., Joskowicz, L. & Chen, X. A review of advances in image-guided orthopedic surgery. *Phys. Med. Biol.* **68**, 2TR01 (2023).
8. Yun, S. H. & Kwok, S. J. J. Light in diagnosis, therapy and surgery. *Nat. Biomed. Eng.* **1**, 1–16 (2017).
9. Bi, W. L. *et al.* Artificial intelligence in cancer imaging: Clinical challenges and applications. *Ca* **69**, 127–157 (2019).
10. Goodfellow, I. *et al.* Generative adversarial nets.
11. Kingma, D. P. & Welling, M. An introduction to variational autoencoders. *Found. Trends® Mach. Learn.* **12**, 307–392 (2019).
12. Ho, J., Jain, A. & Abbeel, P. Denoising diffusion probabilistic models. *Adv. Neural Inf. Process. Syst.* **33**, 6840–6851 (2020).
13. Dosovitskiy, A. *et al.* An image is worth 16x16 words: transformers for image recognition at scale. *Arxiv Prepr. Arxiv:2010,11929* (2020).
14. Gu, A. & Dao, T. Mamba: linear-time sequence modeling with selective state spaces. *Arxiv Prepr. Arxiv:2312,00752* (2023).
15. Van Den Oord, A., Kalchbrenner, N. & Kavukcuoglu, K. Pixel recurrent neural networks. in *International conference on machine learning* 1747–1756 (PMLR, 2016).
16. Gregor, K., Danihelka, I., Mnih, A., Blundell, C. & Wierstra, D. Deep autoregressive networks. in *International Conference on Machine Learning* 1242–1250 (PMLR, 2014).
17. Christensen, M., Vukadinovic, M., Yuan, N. & Ouyang, D. Vision–language foundation model for echocardiogram interpretation. *Nat. Med.* **30**, 1481–1488 (2024).
18. Huang, W. *et al.* Enhancing representation in radiography-reports foundation model: a granular alignment algorithm using masked contrastive learning. *Nat. Commun.* **15**, 7620 (2024).
19. Bluethgen, C. *et al.* A vision–language foundation model for the generation of realistic chest x-ray images. *Nat. Biomed. Eng.* 1–13 (2024) doi:10.1038/s41551-024-01246-y.
20. Vorontsov, E. *et al.* A foundation model for clinical-grade computational pathology and rare cancers detection. *Nat. Med.* **30**, 2924–2935 (2024).
21. Xu, H. *et al.* A whole-slide foundation model for digital pathology from real-world data. *Nature* **630**, 181–188 (2024).
22. Gupta, M. Advances in AI: employing deep generative models for the creation of synthetic healthcare datasets to improve predictive analytics. in *Int. Conf. Commun., Secur. Artif. Intell., ICCSAI* 1026–1030 (Institute of Electrical and Electronics Engineers Inc., 2023). doi:10.1109/ICCSAI59793.2023.10421464.
23. van Breugel, B., Liu, T., Oglic, D. & van der Schaar, M. Synthetic data in biomedicine via generative artificial intelligence. *Nat. Rev. Bioeng.* **2**, 991–1004 (2024).
24. Wang, J. *et al.* Self-improving generative foundation model for synthetic medical image generation and clinical applications. *Nat. Med.* **31**, 1–9 (2024).
25. Guo, P. *et al.* MAISI: Medical AI for synthetic imaging. Preprint at https://doi.org/10.48550/arXiv.2409.11169 (2024).
26. Dayarathna, S. *et al.* Deep learning based synthesis of MRI, CT and PET: review and analysis. *Med. Image Anal.* **92**, 103046 (2024).
27. Takahashi, W., Oshikawa, S. & Mori, S. Real-time markerless tumour tracking with patient-specific deep learning using a personalised data generation strategy: proof of concept by phantom study. *Br. J. Radiol.* **93**, (2020).
28. Cao, R., Wang, J. & Liu, Y.-H. SR-mamba: effective surgical phase recognition with state space model. Preprint at https://doi.org/10.48550/arXiv.2407.08333 (2024).
29. Tan, Y. *et al.* Generation and applications of synthetic computed tomography images for neurosurgical planning. *J. Neurosurg.* **141**, 742–751 (2024).
30. Liu, Q. *et al.* Treatment-aware diffusion probabilistic model for longitudinal MRI generation and diffuse glioma growth prediction. *IEEE Trans. Med. Imag.* 1–1 (2025) doi:10.1109/TMI.2025.3533038.

31. Tivnan, M. *et al.* Hallucination index: an image quality metric for generative reconstruction models. in *International Conference on Medical Image Computing and Computer-assisted Intervention* 449–458 (Springer, 2024).
32. Olin, A. B. *et al.* Robustness and generalizability of deep learning synthetic computed tomography for positron emission tomography/magnetic resonance imaging–based radiation therapy planning of patients with head and neck cancer. *Adv. Radiat. Oncol.* **6**, (2021).
33. Koohi-Moghadam, M. & Bae, K. T. Generative AI in medical imaging: Applications, challenges, and ethics. *J. Med. Syst.* **47**, 94 (2023).
34. Van Erven, T. & Harremos, P. Rényi divergence and kullback-leibler divergence. *IEEE Trans. Inf. Theory* **60**, 3797–3820 (2014).
35. Liu, J. *et al.* Swin-UMamba†: adapting mamba-based vision foundation models for medical image segmentation. *IEEE Trans. Med. Imag.* 1–1 (2024) doi:10.1109/TMI.2024.3508698.
36. Hein, D. *et al.* Noise suppression in photon-counting computed tomography using unsupervised poisson flow generative models. *Visual Comput. Ind., Biomed., Art* **7**, (2024).
37. Hu, Z. *et al.* Artifact correction in low-dose dental CT imaging using wasserstein generative adversarial networks. *Med. Phys.* **46**, 1686–1696 (2019).
38. Jiang, C. *et al.* Wasserstein generative adversarial networks for motion artifact removal in dental CT imaging. in *Progr. Biomed. Opt. Imaging Proc. SPIE* (eds. Schmidt T.G., Chen G.-H., & Bosmans H.) vol. 10948 (SPIE, 2019).
39. Gao, Q., Li, Z., Zhang, J., Zhang, Y. & Shan, H. CoreDiff: contextual error-modulated generalized diffusion model for low-dose CT denoising and generalization. *IEEE Trans. Med. Imag.* **43**, 745–759 (2024).
40. Gong, Y. *et al.* Parameter-transferred wasserstein generative adversarial network (PT-WGAN) for low-dose PET image denoising. *IEEE Trans. Radiat. Plasma Med. Sci.* **5**, 213–223 (2021).
41. Gong, K., Johnson, K., El Fakhri, G., Li, Q. & Pan, T. PET image denoising based on denoising diffusion probabilistic model. *Eur. J. Nucl. Med. Mol. Imaging* **51**, 358–368 (2024).
42. Ran, M. *et al.* Denoising of 3D magnetic resonance images using a residual encoder–decoder wasserstein generative adversarial network. *Med. Image Anal.* **55**, 165–180 (2019).
43. Chung, H., Lee, E. S. & Ye, J. C. MR image denoising and super-resolution using regularized reverse diffusion. *IEEE Trans. Med. Imag.* **42**, 922–934 (2022).
44. Li, Z. *et al.* Promising generative adversarial network based sinogram inpainting method for ultra-limited-angle computed tomography imaging. *Sens. Switz.* **19**, (2019).
45. Liu, J. *et al.* Dolce: a model-based probabilistic diffusion framework for limited-angle ct reconstruction. in *Proceedings of the IEEE/CVF International Conference on Computer Vision* 10498–10508 (2023).
46. Lei, Y. *et al.* Whole-body PET estimation from low count statistics using cycle-consistent generative adversarial networks. *Phys. Med. Biol.* **64**, (2019).
47. Gautier, V. *et al.* Bimodal PET/MRI generative reconstruction based on VAE architectures. *Phys. Med. Biol.* **69**, (2024).
48. Zhang, Q. *et al.* Deep generalized learning model for PET image reconstruction. *IEEE Trans. Med. Imag.* **43**, 122–134 (2024).
49. Huang, J. *et al.* Swin transformer for fast MRI. *Neurocomputing* **493**, 281–304 (2022).
50. Chen, L. *et al.* Joint coil sensitivity and motion correction in parallel MRI with a self-calibrating score-based diffusion model. *Med. Image Anal.* **102**, 103502 (2025).
51. Huang, J. *et al.* MambaMIR: an arbitrary-masked mamba for joint medical image reconstruction and uncertainty estimation. Preprint at https://doi.org/10.48550/arXiv.2402.18451 (2024).
52. Kim, B. & Ye, J. C. Diffusion deformable model for 4D temporal medical image generation. in *Lect. Notes Comput. Sci.* (eds. Wang L., Dou Q., Fletcher P.T., Speidel S., & Li S.) vol. 13431 LNCS 539–548 (Springer Science and Business Media Deutschland GmbH, 2022).
53. You, C. *et al.* CT Super-Resolution GAN Constrained by the Identical, Residual, and Cycle Learning Ensemble (GAN-CIRCLE). *IEEE Trans. Med. Imag.* **39**, 188–203 (2020).
54. Chu, Y., Zhou, L., Luo, G., Qiu, Z. & Gao, X. Topology-preserving computed tomography super-resolution based on dual-stream diffusion model. in *Medical Image Computing and Computer Assisted Intervention – MICCAI 2023* (eds. Greenspan, H. et al.) vol. 14229 260–270 (Springer Nature Switzerland, Cham, 2023).

55. Danu, M., Nita, C.-I., Vizitiu, A., Suciu, C. & Itu, L. M. Deep learning based generation of synthetic blood vessel surfaces. in *Int. Conf. Syst. Theory, Control Comput., ICSTCC - Proc.* (ed. Precup R.-E.) 662–667 (Institute of Electrical and Electronics Engineers Inc., 2019). doi:10.1109/ICSTCC.2019.8885576.
56. Chong, C. K. & Ho, E. T. W. Synthesis of 3D MRI brain images with shape and texture generative adversarial deep neural networks. *IEEE Access* **9**, 64747–64760 (2021).
57. Hong, S. *et al.* 3D-StyleGAN: a style-based generative adversarial network for generative modeling of three-dimensional medical images. in *Deep Generative Models, and Data Augmentation, Labelling, and Imperfections: First Workshop, DGM4MICCAI 2021, and First Workshop, DALI 2021, Held in Conjunction with MICCAI 2021, Strasbourg, France, October 1, 2021, Proceedings 1* 24–34 (Springer, 2021).
58. Zuo, L. *et al.* Synthesizing realistic brain mr images with noise control. in *Lect. Notes Comput. Sci.* (eds. Burgos N., Svoboda D., Wolterink J.M., & Zhao C.) vol. 12417 LNCS 21–31 (Springer Science and Business Media Deutschland GmbH, 2020).
59. Khader, F. *et al.* Medical diffusion: Denoising diffusion probabilistic models for 3D medical image generation. Preprint at https://doi.org/10.48550/arXiv.2211.03364 (2023).
60. Wang, H. *et al.* 3D MedDiffusion: A 3D medical diffusion model for controllable and high-quality medical image generation. Preprint at https://doi.org/10.48550/arXiv.2412.13059 (2024).
61. Huang, P. *et al.* Chest-diffusion: A light-weight text-to-image model for report-to-CXR generation. in *2024 IEEE International Symposium on Biomedical Imaging (isbi)* 1–5 (2024). doi:10.1109/ISBI56570.2024.10635417.
62. Cho, J. *et al.* MediSyn: A generalist text-guided latent diffusion model for diverse medical image synthesis. Preprint at https://doi.org/10.48550/arXiv.2405.09806 (2025).
63. Zhou, X. *et al.* Multimodality MRI synchronous construction based deep learning framework for MRI-guided radiotherapy synthetic CT generation. *Comput. Biol. Med.* **162**, 107054 (2023).
64. Dong, X. *et al.* Synthetic CT generation from non-attenuation corrected PET images for whole-body PET imaging. *Phys. Med. Biol.* **64**, (2019).
65. Yang, H. *et al.* Unsupervised MR-to-CT synthesis using structure-constrained CycleGAN. *IEEE Trans. Med. Imag.* **39**, 4249–4261 (2020).
66. Gong, K. *et al.* MR-based attenuation correction for brain PET using 3-D cycle-consistent adversarial network. *IEEE Trans. Radiat. Plasma Med. Sci.* **5**, 185–192 (2021).
67. Dalmaz, O., Yurt, M. & Cukur, T. ResViT: residual vision transformers for multimodal medical image synthesis. *IEEE Trans. Med. Imag.* **41**, 2598–2614 (2022).
68. Wang, Z. *et al.* Mutual information guided diffusion for zero-shot cross-modality medical image translation. *IEEE Trans. Med. Imag.* **43**, 2825–2838 (2024).
69. Ferreira, V. R. S. *et al.* Diffusion model for generating synthetic contrast enhanced CT from non-enhanced heart axial CT images. in *International Conference on Enterprise Information Systems, ICEIS - Proceedings* (eds. Filipe J., Smialek M., Brodsky A., & Hammoudi S.) vol. 1 857–864 (Science and Technology Publications, Lda, 2024).
70. Fu, L. *et al.* Energy-guided diffusion model for CBCT-to-CT synthesis. *Comput. Med. Imaging Graphics* **113**, (2024).
71. Jia, Y., Chen, G. & Chi, H. Retinal fundus image super-resolution based on generative adversarial network guided with vascular structure prior. *Sci. Rep.* **14**, 22786 (2024).
72. Konz, N., Chen, Y., Dong, H. & Mazurowski, M. A. Anatomically-controllable medical image generation with segmentation-guided diffusion models. Preprint at https://doi.org/10.48550/arXiv.2402.05210 (2024).
73. Lai, X. Modelling, inference and simulation of personalised breast cancer treatment. (2019).
74. Savić, M., Kurbalija, V., Balaz, I. & Ivanović, M. Heterogeneous tumour modeling using PhysiCell and its implications in precision medicine. in *Cancer, Complexity, Computation* (eds. Balaz, I. & Adamatzky, A.) vol. 46 157–189 (Springer International Publishing, Cham, 2022).
75. Kearney, V., Chan, J. W., Haaf, S., Descovich, M. & Solberg, T. D. DoseNet: a volumetric dose prediction algorithm using 3D fully-convolutional neural networks. *Phys. Med. Biol.* **63**, 235022 (2018).
76. Jiao, Z. *et al.* TransDose: transformer-based radiotherapy dose prediction from CT images guided by super-pixel-level GCN classification. *Med. Image Anal.* **89**, 102902 (2023).

77. Feng, Z. *et al.* DiffDP: radiotherapy dose prediction via a diffusion model. in *Medical Image Computing and Computer Assisted Intervention – MICCAI 2023* (eds. Greenspan, H. et al.) vol. 14225 191–201 (Springer Nature Switzerland, Cham, 2023).
78. Fu, L. *et al.* MD-dose: a diffusion model based on the mamba for radiotherapy dose prediction. *Arxiv E-prints* arXiv-2403 (2024).
79. Pan, S. *et al.* Patient-specific CBCT synthesis for real-time tumor tracking in surface-guided radiotherapy. Preprint at https://doi.org/10.48550/arXiv.2410.23582 (2024).
80. Yoon, S. *et al.* Accelerated cardiac MRI cine with use of resolution enhancement generative adversarial inline neural network. *Radiology* **307**, e222878 (2023).
81. Zakeri, A. *et al.* DragNet: learning-based deformable registration for realistic cardiac MR sequence generation from a single frame. *Med. Image Anal.* **83**, (2023).
82. Reynaud, H. *et al.* Feature-conditioned cascaded video diffusion models for precise echocardiogram synthesis. in *Medical Image Computing and Computer Assisted Intervention – MICCAI 2023* vol. 14229 142–152 (Springer Nature Switzerland, Cham, 2023).
83. Zhou, X. *et al.* HeartBeat: towards controllable echocardiography video synthesis with multimodal conditions-guided diffusion models. in *Lect. Notes Comput. Sci.* (eds. Linguraru M.G. et al.) vol. 15007 LNCS 361–371 (Springer Science and Business Media Deutschland GmbH, 2024).
84. Ghodrati, V. *et al.* Temporally aware volumetric generative adversarial network-based MR image reconstruction with simultaneous respiratory motion compensation: initial feasibility in 3D dynamic cine cardiac MRI. *Magn. Reson. Med.* **86**, 2666–2683 (2021).
85. Thummerer, A. *et al.* Deep learning–based 4D-synthetic CTs from sparse-view CBCTs for dose calculations in adaptive proton therapy. *Med. Phys.* **49**, 6824–6839 (2022).
86. Quintero, P., Wu, C., Otazo, R., Cervino, L. & Harris, W. On-board synthetic 4D MRI generation from 4D CBCT for radiotherapy of abdominal tumors: a feasibility study. *Med. Phys.* **51**, 9194–9206 (2024).
87. Liu, C., Yuan, X., Yu, Z. & Wang, Y. Texdc: text-driven disease-aware 4d cardiac cine mri images generation. in *Proceedings of the Asian Conference on Computer Vision* 3005–3021 (2024).
88. Yoon, J. S., Zhang, C., Suk, H.-I., Guo, J. & Li, X. SADM: sequence-aware diffusion model for longitudinal medical image generation. in *Information Processing in Medical Imaging* (eds. Frangi, A., de Bruijne, M., Wassermann, D. & Navab, N.) 388–400 (Springer Nature Switzerland, Cham, 2023). doi:10.1007/978-3-031-34048-2_30.
89. Moya-Sáez, E. *et al.* Synthetic MRI improves radiomics-based glioblastoma survival prediction. *NMR Biomed.* **35**, (2022).
90. Lei, W. *et al.* A data-efficient pan-tumor foundation model for oncology CT interpretation. Preprint at https://doi.org/10.48550/arXiv.2502.06171 (2025).
91. Elazab, A. *et al.* GP-GAN: brain tumor growth prediction using stacked 3D generative adversarial networks from longitudinal MR images. *Neural Networks* **132**, 321–332 (2020).
92. Ravi, D., Alexander, D. C. & Oxtoby, N. P. Degenerative adversarial NeuroImage nets: generating images that mimic disease progression. in *Lect. Notes Comput. Sci.* (eds. Shen D. et al.) vol. 11766 LNCS 164–172 (Springer Science and Business Media Deutschland GmbH, 2019).
93. SinhaRoy, R. & Sen, A. A hybrid deep learning framework to predict alzheimer's disease progression using generative adversarial networks and deep convolutional neural networks. *Arabian J. Sci. Eng.* **49**, 3267–3284 (2024).
94. Litrico, M., Guarnera, F., Giuffrida, M. V., Ravì, D. & Battiato, S. TADM: temporally-aware diffusion model for neurodegenerative progression on brain MRI. in *Medical Image Computing and Computer Assisted Intervention – MICCAI 2024* (eds. Linguraru, M. G. et al.) vol. 15002 444–453 (Springer Nature Switzerland, Cham, 2024).
95. Campello, V. M. *et al.* Cardiac aging synthesis from cross-sectional data with conditional generative adversarial networks. *Front. Cardiovasc. Med.* **9**, (2022).
96. Puglisi, L., Alexander, D. C. & Ravì, D. Enhancing spatiotemporal disease progression models via latent diffusion and prior knowledge. in *Medical Image Computing and Computer Assisted*

*Intervention – MICCAI 2024* (eds. Linguraru, M. G. et al.) vol. 15002 173–183 (Springer Nature Switzerland, Cham, 2024).
97. Yalcin, C. *et al.* Hematoma expansion prediction in intracerebral hemorrhage patients by using synthesized CT images in an end-to-end deep learning framework. *Comput. Med. Imaging Graphics* **117**, (2024).
98. Bycroft, C. *et al.* The UK biobank resource with deep phenotyping and genomic data. *Nature* **562**, 203–209 (2018).
99. Clark, K. *et al.* The cancer imaging archive (TCIA): maintaining and operating a public information repository. *J. Digital Imaging* **26**, 1045–1057 (2013).
100. Yan, K., Wang, X., Lu, L. & Summers, R. M. DeepLesion: automated mining of large-scale lesion annotations and universal lesion detection with deep learning. *J. Med. Imaging* **5**, 36501–36501 (2018).
101. Wu, L., Zhuang, J. & Chen, H. Large-scale 3D medical image pre-training with geometric context priors. Preprint at https://doi.org/10.48550/arXiv.2410.09890 (2024).
102. Wasserthal, J. *et al.* TotalSegmentator: robust segmentation of 104 anatomic structures in CT images. *Radiol.: Artif. Intell.* **5**, e230024 (2023).
103. D'Antonoli, T. A. *et al.* TotalSegmentator MRI: robust sequence-independent segmentation of multiple anatomic structures in MRI. *Radiology* **314**, e241613 (2025).
104. Baid, U. *et al.* The RSNA-ASNR-MICCAI BraTS 2021 benchmark on brain tumor segmentation and radiogenomic classification. Preprint at https://doi.org/10.48550/arXiv.2107.02314 (2021).
105. Bernard, O. *et al.* Deep learning techniques for automatic MRI cardiac multi-structures segmentation and diagnosis: is the problem solved? *IEEE Trans. Med. Imag.* **37**, 2514–2525 (2018).
106. Gatidis, S. *et al.* A whole-body FDG-PET/CT dataset with manually annotated tumor lesions. *Sci. Data* **9**, 601 (2022).
107. Andrearczyk, V. *et al.* Overview of the HECKTOR challenge at MICCAI 2022: automatic head and neck tumor segmentation and outcome prediction in PET/CT. in *Head and Neck Tumor Segmentation and Outcome Prediction* (eds. Andrearczyk, V., Oreiller, V., Hatt, M. & Depeursinge, A.) vol. 13626 1–30 (Springer Nature Switzerland, Cham, 2023).
108. Veeling, B. S., Linmans, J., Winkens, J., Cohen, T. & Welling, M. Rotation equivariant CNNs for digital pathology. in *Medical Image Computing and Computer Assisted Intervention – MICCAI 2018* (eds. Frangi, A. F., Schnabel, J. A., Davatzikos, C., Alberola-López, C. & Fichtinger, G.) vol. 11071 210–218 (Springer International Publishing, Cham, 2018).
109. Ikezogwo, W. *et al.* Quilt-1m: one million image-text pairs for histopathology. *Adv. Neural Inf. Process. Syst.* **36**, 37995–38017 (2023).
110. Kermany, D. Labeled optical coherence tomography (oct) and chest x-ray images for classification. *Mendeley Data* (2018).
111. Chambon, P. *et al.* CheXpert plus: augmenting a large chest X-ray dataset with text radiology reports, patient demographics and additional image formats. Preprint at https://doi.org/10.48550/arXiv.2405.19538 (2024).
112. Subramanian, S. *et al.* MedICaT: a dataset of medical images, captions, and textual references. Preprint at https://doi.org/10.48550/arXiv.2010.06000 (2020).
113. Xie, Y. *et al.* Medtrinity-25m: a large-scale multimodal dataset with multigranular annotations for medicine. *Arxiv Prepr. Arxiv:2408,02900* (2024).
114. Eskicioglu, A. M. & Fisher, P. S. Image quality measures and their performance. *IEEE Trans. Commun.* **43**, 2959–2965 (1995).
115. Wang, Z., Bovik, A. C., Sheikh, H. R. & Simoncelli, E. P. Image quality assessment: from error visibility to structural similarity. *IEEE Trans. Image Process.* **13**, 600–612 (2004).
116. Wang, Z., Simoncelli, E. P. & Bovik, A. C. Multiscale structural similarity for image quality assessment. in *The Thrity-seventh Asilomar Conference on Signals, Systems & Computers, 2003* vol. 2 1398–1402 (Ieee, 2003).
117. Zhang, L., Zhang, L., Mou, X. & Zhang, D. FSIM: a feature similarity index for image quality assessment. *IEEE Trans. Image Process.* **20**, 2378–2386 (2011).

118. Wang, Z. & Li, Q. Information content weighting for perceptual image quality assessment. *IEEE Trans. Image Process.* **20**, 1185–1198 (2010).
119. Bercea, C. I., Wiestler, B., Rueckert, D. & Schnabel, J. A. Evaluating normative representation learning in generative AI for robust anomaly detection in brain imaging. *Nat. Commun.* **16**, 1624 (2025).
120. Yu, Y., Zhang, W. & Deng, Y. Frechet inception distance (fid) for evaluating gans. *China Univ. Min. Technol. Beijing Grad. Sch.* **3**, (2021).
121. Wenliang, L., Moskovitz, T., Kanagawa, H. & Sahani, M. Amortised learning by wake-sleep. in *International Conference on Machine Learning* 10236–10247 (PMLR, 2020).
122. Gretton, A., Borgwardt, K. M., Rasch, M. J., Schölkopf, B. & Smola, A. A kernel two-sample test. *J. Mach. Learn. Res.* **13**, 723–773 (2012).
123. Dosovitskiy, A. & Brox, T. Generating images with perceptual similarity metrics based on deep networks. *Adv. Neural Inf. Process. Syst.* **29**, (2016).
124. Radford, A. *et al.* Learning transferable visual models from natural language supervision. in *International Conference on Machine Learning* 8748–8763 (PmLR, 2021).
125. Wang, Z., Wu, Z., Agarwal, D. & Sun, J. Medclip: contrastive learning from unpaired medical images and text. in *Proceedings of the Conference on Empirical Methods in Natural Language Processing. Conference on Empirical Methods in Natural Language Processing* vol. 2022 3876 (2022).
126. Sun, W. *et al.* Bora: Biomedical generalist video generation model. Preprint at https://doi.org/10.48550/arXiv.2407.08944 (2024).
127. Unterthiner, T. *et al.* FVD: a new metric for video generation. (2019).
128. Liu, J. *et al.* Fréchet video motion distance: a metric for evaluating motion consistency in videos. Preprint at https://doi.org/10.48550/arXiv.2407.16124 (2024).
129. Hamamci, I. E. *et al.* GenerateCT: Text-conditional generation of 3D chest CT volumes. Preprint at https://doi.org/10.48550/arXiv.2305.16037 (2024).
130. Seo, I., Bae, E., Jeon, J.-Y., Yoon, Y.-S. & Cha, J. The era of foundation models in medical imaging is approaching : a scoping review of the clinical value of large-scale generative AI applications in radiology. Preprint at https://doi.org/10.48550/arXiv.2409.12973 (2024).
131. Menghani, G. Efficient deep learning: a survey on making deep learning models smaller, faster, and better. *ACM Comput. Surv.* **55**, 1–37 (2023).
132. Chung, M., Won, J. B., Kim, G., Kim, Y. & Ozbulak, U. Evaluating visual explanations of attention maps for transformer-based medical imaging. in *Medical Image Computing and Computer Assisted Intervention – MICCAI 2024 Workshops* (eds. Celebi, M. E., Reyes, M., Chen, Z. & Li, X.) vol. 15274 110–120 (Springer Nature Switzerland, Cham, 2025).
133. Gotkowski, K., Gonzalez, C., Bucher, A. & Mukhopadhyay, A. M3d-CAM: a PyTorch library to generate 3D attention maps for MedicalDeep learning. in *Bildverarbeitung Für Die Medizin 2021* (eds. Palm, C. et al.) 217–222 (Springer Fachmedien Wiesbaden, Wiesbaden, 2021). doi:10.1007/978-3-658-33198-6_52.
134. Arun, N. *et al.* Assessing the trustworthiness of saliency maps for localizing abnormalities in medical imaging. *Radiol.: Artif. Intell.* **3**, e200267 (2021).
135. Jin, W., Li, X. & Hamarneh, G. One map does not fit all: evaluating saliency map explanation on multi-modal medical images. Preprint at https://doi.org/10.48550/arXiv.2107.05047 (2021).
136. Guan, H., Yap, P.-T., Bozoki, A. & Liu, M. Federated learning for medical image analysis: a survey. *Pattern Recognit.* 110424 (2024).
137. Krishnan, R., Rajpurkar, P. & Topol, E. J. Self-supervised learning in medicine and healthcare. *Nat. Biomed. Eng.* **6**, 1346–1352 (2022).
138. Nie, D. & Shen, D. Adversarial confidence learning for medical image segmentation and synthesis. *Int. J. Comput. Vision* **128**, 2494–2513 (2020).
139. Zhou, Q., Yu, T., Zhang, X. & Li, J. Bayesian inference and uncertainty quantification for medical image reconstruction with poisson data. *SIAM J. Imag. Sci.* **13**, 29–52 (2020).
140. Musaev, J., Anorboev, A., Seo, Y.-S., Nguyen, N. T. & Hwang, D. ICNN-ensemble: an improved convolutional neural network ensemble model for medical image classification. *IEEE Access* **11**, 86285–86296 (2023).

141. Rosenbacke, R. Cognitive challenges in human-AI collaboration: a study on trust, errors, and heuristics in clinical decision-making. (2025).
142. Contaldo, M. T. *et al.* AI in radiology: navigating medical responsibility. *Diagnostics* **14**, 1506 (2024).
143. Chen, Y. & Esmaeilzadeh, P. Generative AI in medical practice: in-depth exploration of privacy and security challenges. *J. Med. Internet Res.* **26**, e53008 (2024).
144. Yu, Y., Gu, Y., Zhang, S. & Zhang, X. MedDiff-FM: A diffusion-based foundation model for versatile medical image applications. Preprint at https://doi.org/10.48550/arXiv.2410.15432 (2024).
145. Wang, S. *et al.* Triad: vision foundation model for 3D magnetic resonance imaging. Preprint at https://doi.org/10.48550/arXiv.2502.14064 (2025).
146. Sun, Y. *et al.* A data-efficient strategy for building high-performing medical foundation models. *Nat. Biomed. Eng.* 1–13 (2025) doi:10.1038/s41551-025-01365-0.
147. Yang, Z. *et al.* A foundation model for generalizable cancer diagnosis and survival prediction from histopathological images. *Nat. Commun.* **16**, 2366 (2025).
148. Zhang, A., Xing, L., Zou, J. & Wu, J. C. Shifting machine learning for healthcare from development to deployment and from models to data. *Nat. Biomed. Eng.* **6**, 1330–1345 (2022).

# Supplementary

## S1. More Details on Review Outline and Contributions

***Search Criteria.*** This review is based on a systematic survey of recent advancements in medical image generation. We conducted a systematic literature search using PubMed, Scopus, Google Scholar, and DBLP databases to identify relevant articles published from 2019 to 2025 February. The search terms included combinations such as ("medical imag*" OR "MRI" OR "PET" OR "CT" OR "US") AND ("genera*" OR "synth*" OR "pseudo*") AND ("diagnosis" OR "treatment" OR "prognosis") AND "deep learning" in the title, abstract, or keywords. This strategy aimed to capture a broad range of studies across peer-reviewed journals, conference papers, and preprints.

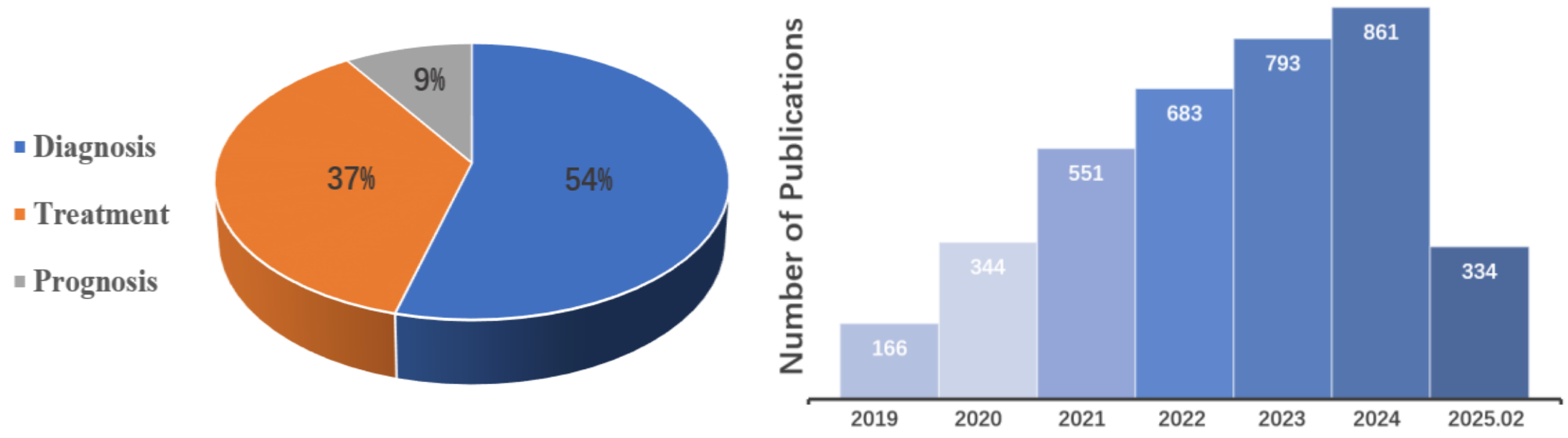

(a) (b)

**Figure S1. Statistics of the generative AI models in medical imaging.** (a) Statistics of Publications in Clinical Workflow: Diagnosis, Treatment, and Prognosis. (b) Categorization by year of publication (2019-2025, sourced from PubMed).

## S2. More Details on Overview of Related Survey

In recent years, with the burgeoning development of generative AI in the field of medical images, a multitude of surveys have been published to provide overviews. Table S1 categorizes these surveys into technology-oriented and task-oriented perspectives. The technology-oriented surveys focus on key generative models, including GANs[1–6], VAEs[7,8], diffusion models (DPMs)[9–13], and sequence modeling architectures (e.g., Transformers, Mamba, autoregressive models)[14–17],and foundation models[18–22], summarizing their core principles and applications. And the task-oriented surveys emphasize practical implementations such as data augmentation[23–26,26–28], modality translation[4,29–32], image restoration[33–37], evaluation of generated results and so on[38–41]. While these surveys offer valuable insights, they tend to either concentrate on theoretical advancements in generative models or focus on specific application scenarios. However, a comprehensive perspective that systematically maps the applications of generative AI across the entire clinical workflow, including diagnosis, treatment, and prognosis, has not been fully explored. Moreover, existing evaluations of generative models in medical imaging are often fragmented, relying on conventional image quality metrics without adequately considering clinical interpretability or downstream utility. The absence of a standardized, multi-tiered evaluation framework further impedes the clinical adoption of generative AI, raising concerns about the reliability and trustworthiness of synthesized medical images.

To address these issues, we provide a structured and comprehensive examination of generative AI in medical imaging, emphasizing its integration across different clinical phases, including diagnosis, treatment, and prognosis. We systematically explore the role of key generative models, including GANs, VAEs, diffusion models, sequence modeling architectures, and foundation models in these phases. Additionally, we propose a three-tiered evaluation framework encompassing low-level image fidelity, mid-level feature consistency, and high-level clinical relevance to ensure a more standardized assessment of generative models. Finally, we discuss the current challenges, limitations, and future directions of generative AI in medical imaging, highlighting key considerations for advancing these technologies toward real-world clinical applications.

**Table S1**. Overview of Generative AI Surveys in Medical Imaging.

| Category | Subcategory | Publication | Core Content |
|---|---|---|---|
| Technique-Oriented Surveys | **GANs** | 1–6 | -**Characteristics**: Adversarial training mechanism<br>-**Applications**: Modality synthesis, data augmentation, image denoising |
| | **VAEs** | 7,8 | -**Characteristics**: Latent space design, probabilistic generation, and feature disentanglement<br>-**Applications**: Data augmentation, image generation and analysis |
| | **DPMs** | 9–13 | -**Characteristics**: Progressive denoising process, integration of physical priors<br>-**Applications**: High-fidelity image synthesis, data augmentation, image denoising/artifact removal, reconstruction |
| | **Sequence Modeling Architectures** | 14–17 | -**Characteristics**: Transformers (global context modeling), Mamba (long-sequence processing), autoregressive generation<br>-**Applications**: Time-series prediction, reconstruction, dynamic image generation |
| Task-Oriented Surveys | **Data Synthesis & Augmentation** | 23–26,26–28 | -Multimodal data generation<br>-2D/3D data synthesis<br>- Scarce data augmentation |
| | **Modality Translation** | 4,29–32 | -Pseudo-CT generation<br>-Cross-modality translation |
| | **Image Restoration** | 33–37 | -Low-dose CT/PET reconstruction<br>-Undersampled MRI reconstruction, fast MRI imaging<br>- Super-resolution generation |
| | **Evaluation & Ethics** | 38–41 | -Generated image reliability assessment (generalizability and interpretability)<br>-Ethical risks of synthetic data |

# S3. More Details on Key Generative AI Models in Medical Imaging

Here we develop in more details on key generative AI models in medical imaging.

## S3.1. Generative Adversarial Networks

Generative Adversarial Network (GANs), proposed by Ian Goodfellow et al. in 2014[42], represent a significant breakthrough in generative modeling by enabling the creation of realistic data distributions through an adversarial training framework. In the GAN, a generator (G) learns to produce synthetic data that closely mimics real samples, while a discriminator (D) distinguishes between real and generated data, as shown in Fig. 2(a). These networks engage in a minimax game, refining their outputs iteratively to generate high-quality, realistic data. The training process follows the objective:

$$\min_G \max_D \mathbb{E}_{x \sim p_{\text{data}}(x)}[\log D(x)] + \mathbb{E}_{z \sim p_z(z)}\left[\log\left(1 - D\big(G(z)\big)\right)\right], \qquad (1)$$

where $p_{\text{data}}(x)$ is the real data distribution, $p_z(z)$ is the prior distribution on the latent vector $z$, $D(x)$ is the discriminator's probability that $x$ is real, $G(z)$ is the generated image from the latent vector $z$.

The field of GAN-based medical image generation has experienced sustained growth, with a steady increase in published studies since the introduction of GANs. As research advances, various GAN variants have been developed, each offering distinct advantages for medical imaging. Deep Convolutional GAN (DCGAN)[43] improves spatial feature learning and stabilizes training by replacing fully connected layers with convolutional operations. It performs well in high-resolution medical image synthesis and is widely used for chest X-ray generation[44], helping with lung nodule detection and segmentation. CycleGAN[45] enables modality conversion without requiring paired datasets, making it useful for MRI-to-CT synthesis in radiation therapy planning[46] and improving treatment workflows. StyleGAN[47], with its style-based architecture, allows fine-grained control over image features, making it effective for generating highly realistic and diverse medical images. Researchers frequently use it in cross-modality analysis, such as generating CT images with MRI-like textures[48], and for creating synthetic datasets that improve model generalization across different imaging systems and clinical environments[49,50].

Despite these advancements, GANs still face several challenges that hinder their clinical adoption. Training instability and mode collapse can result in inconsistent outputs with limited diversity, while the high computational demands of adversarial training add to the complexity of deployment. Furthermore, the lack of a clearly structured latent space constrains the ability to perform controlled image modifications, making it challenging to maintain anatomical accuracy and pathological consistency, both of which are critical for clinical applicability.

Furthermore, concerns regarding image diversity and reproducibility highlight the need for further refinements to enhance the robustness, interpretability, and real-world applicability of GAN-generated medical images.

## S3.2. Variational Autoencoders

Variational Autoencoders (VAEs)[51], revolutionized generative modeling by combining variational inference with neural networks. A VAE consists of two components: an encoder, which maps input data to a latent space, and a decoder, which reconstructs the data from this latent representation. Fig. 2(b) shows how this structure allows VAEs to capture complex data distributions and generate new samples via latent space sampling. The training objective involves balancing reconstruction loss and Kullback-Leibler (KL) divergence[52], ensuring both accurate reconstruction and smoothness in the latent space. The objective function is expressed as:

$$\mathcal{L} = \mathbb{E}_{q_\phi(z|x)}[\log p_\theta(x \mid z)] - D_{KL}\big(q_\phi(z \mid x) \| p(z)\big), \quad (2)$$

where $q_\phi(z \mid x)$ is the encoder's approximation of the posterior distribution, $p_\theta(x \mid z)$ is the decoder's likelihood of the data given the latent variables, and $p(z)$ is the prior distribution over the latent space.

In medical imaging, Variational Autoencoders (VAEs) have been widely used for tasks such as image synthesis and cross-modality translation. Over time, various VAE variants have been developed to address specific challenges and improve performance. β-VAE[53] , for instance, introduces a hyperparameter β to enhance latent space disentanglement, allowing each dimension to represent distinct semantic features. This has proven effective in brain MRI analysis, where it helps distinguish Alzheimer's-related atrophy from age-related changes[54]. Conditional VAE (CVAE)[55] incorporates conditional variables, such as disease labels or anatomical landmarks, enabling controlled generation of medical images. This capability is particularly valuable in cross-modality synthesis and rare disease research[56]. Vector Quantized VAE (VQ-VAE)[57] discretizes the latent space into a finite codebook, improving image quality and facilitating applications such as low-dose CT reconstruction[58] and 4D heart MRI sequences generation[59]. Additionally, the Hybrid VAE-GAN[60] combines VAE's structured latent space modeling with the adversarial training of GANs to enhance image clarity, with notable applications in glioma MRI synthesis[61].

For VAEs, the KL divergence constraint can lead to over-smoothing in the latent space, often resulting in blurry image generation. VQ-VAE alleviates this by introducing a discrete codebook, while Hybrid VAE-GAN enhances texture details. However, maintaining anatomical alignment in generated images remains difficult, often necessitating anatomical prior losses or segmentation masks to preserve structural consistency. Additionally, training 3D VAEs on large volumetric datasets requires substantial computational resources. A common strategy to improve efficiency is hierarchical training, where models are initially trained in 2D before being progressively fine-tuned for 3D applications[62].

## S3.3. Diffusion Probabilistic Models

Diffusion Probabilistic Models (DPMs)[63], known as denoising diffusion probabilistic models, are a class of generative models inspired by non-equilibrium thermodynamics. They model data generation through a Markov chain that progressively adds Gaussian noise to the data, transforming it into a simple prior distribution, such as a standard normal distribution. The model then learns to reverse this diffusion process by progressively denoising the data, reconstructing the original data from the noisy samples. Mathematically, the forward process is expressed as:

$$q(x_t \mid x_{t-1}) = \mathcal{N}\left(x_t; \sqrt{1-\beta_t}x_{t-1}, \beta_t I\right), \quad (3)$$

where $x_0$ denotes the original data distribution, $x_t$ represents data with t step noise added, $\beta_t$ denotes the variance schedule controlling the amount of noise added at each step $t$. And the reverse process is defined as:

$$p_\theta(x_{t-1} \mid x_t) = \mathcal{N}\left(x_{t-1}; \mu_\theta(x_t, t), \Sigma_\theta(x_t, t)\right), \quad (4)$$

where $\mu_\theta$ and $\Sigma_\theta$ are the mean and covariance parameters predicted by the neural network with parameters $\theta$. The model is trained to minimize the variational bound on the negative log-likelihood, which can be expressed as:

$$\mathcal{L} = \mathbb{E}_q\left[D_{KL}\left(q(x_T \mid x_0) \| p(x_T)\right) + \sum_{t=1}^{T} D_{KL}\left(q(x_{t-1} \mid x_t, x_0) \| p_\theta(x_{t-1} \mid x_t)\right) - \log p_\theta(x_0 \mid x_1)\right], \quad (5)$$

Here, $D_{KL}$ denotes the Kullback-Leibler divergence, and $p(x_T)$ is typically chosen as a standard normal distribution. DPMs often integrate with models like VAE[51] and VQ-VAE[57] for latent space compression, enhancing their capacity to model complex data distributions. The reverse diffusion process is then guided by various conditions such as text prompts[64–66], images[67,68], or other modalities[69,70], making DPMs adaptable for diverse tasks. Additionally, diffusion probabilistic models can decouple image components by separating spatial features, improving their ability to learn detailed image characteristics and enhancing their effectiveness in medical image generation

In the context of medical imaging, DPMs have demonstrated exceptional performance across a variety of tasks, including denoising[71,72], super-resolution[73,74], image synthesis[75,76], reconstruction[77–79], and so on. For instance, the Denoising Diffusion Probabilistic Model (DDPM)[71] has been widely utilized for noise reduction in X-ray imaging, improving image clarity and aiding in the detection of small fractures or nodules. The Denoising Diffusion MRI (DDM2)[72] further advances MRI denoising, effectively addressing complex and spatially varying noise patterns. For super-resolution tasks, SR3[73] has been successfully applied to optical coherence tomography, enhancing low-resolution retinal images into high-resolution ones, which is crucial for early disease detection in ophthalmology. Similarly, DiffIR[74] employs diffusion techniques to enhance ultrasound images, providing high-resolution scans that reveal subtle pathological changes, such as small tumors. In 3D imaging, 3D-DDPM[75] has been used to generate detailed brain models from limited data, facilitating neurosurgical planning and decision-making. Additionally, DPMs, through their physical heuristic modeling of diffusion processes, have proven valuable in medical image reconstruction from a physics-based perspective. They are applied in low-dose CT/PET imaging[77,78] and undersampled MRI reconstruction[79], where they help to recover high-fidelity images from incomplete data.

Collectively, these advancements underscore the growing impact of DPMs in medical imaging[80], offering state-of-the-art solutions for denoising, resolution enhancement, and high-fidelity image synthesis. However, DPMs' multi-step inference process results in slower generation speeds and higher computational costs. Moreover, the latent noise inherent in the diffusion process complicates model interpretability compared to the more structured latent features in GANs. Nonetheless, DPMs remain a powerful tool in medical imaging, producing high-quality outputs with rich detail that hold immense potential for clinical applications. As a result, DPMs have become one of the most popular and cutting-edge generative techniques in image generation today.

## S3.4. Sequence modeling architectures

In medical image generative models, sequence modeling architectures are crucial for tasks involving temporal or sequential data, such as dynamic imaging, or longitudinal studies. Key models include Transformers[81], Mamba[82], and Autoregressive (AR) models[83,84]. Fig. 2(d) presents their core structures, which will be discussed in the following sections.

### *S3.4.1. Transformer*

Transformers have revolutionized deep learning by capturing long-range dependencies via self-attention, allowing them to model global relationships within data, unlike traditional CNNs that focus on localized receptive fields. The self-attention mechanism computes a sequence's representation by relating different positions within it, using query (Q), key (K), and value (V) matrices. The attention scores are calculated by the dot product of Q and K, scaled by the square root of the dimension and passed through a softmax function:

$$\text{Attention}(Q, K, V) = \text{softmax}\left(\frac{QK^T}{\sqrt{d_k}}\right)V, \quad (6)$$

The attention mechanism in Transformers allows tokens to dynamically model relationships across entire sequences, making them particularly effective in medical imaging, where convolutional neural networks (CNNs) struggle with limited receptive fields. For instance, 3D MedDiffusion[76] integrates self-attention within diffusion steps to enhance chest CT synthesis, ensuring structural consistency across slices. However, the quadratic complexity $O(L^2)$ of standard Transformers poses scalability challenges for high-resolution 3D volumes. To overcome this, MedFormer[85] employs axial attention, significantly reducing computational costs while maintaining global inter-slice correlations. Similarly, hybrid architectures like SwinGAN[86] combine convolutional layers for local texture refinement with Swin Transformer blocks to capture global contextual information, improving MRI reconstruction performance. Beyond image synthesis, Transformers have become a cornerstone of generative AI in medical imaging due to their scalability and multi-modal processing capabilities. For example, ViT-GPT2[87] integrates a vision Transformer with a GPT-2 decoder to generate radiology reports from X-rays, while DALL-E in Medicine[88] adapts Transformer models to synthesize anatomically consistent X-rays from text descriptions, achieving high clinical relevance scores.

The success of Transformers is largely attributed to their architectural flexibility, where self-attention layers enable deep feature extraction and positional embeddings retain spatial and temporal relationships. However, their high computational demands remain a key limitation, particularly for real-time medical applications. To address this, more efficient alternatives like Mamba have been introduced, aiming to reduce computational complexity while retaining the core benefits of Transformer-based models in medical imaging.

### S3.4.2. Mamba

The Mamba architecture, built upon state space models (SSMs)[82], has emerged as a transformative framework for medical image synthesis, addressing critical limitations of conventional models like Transformers (quadratic complexity) and CNNs (local-receptive constraints)[89]. At its core, Mamba employs discretized state space equations to model sequential dependencies with linear computational scaling:

$$h_t = \bar{A}_t h_{t-1} + \bar{B}_t x_t, \tag{7}$$

$$y_t = \bar{C}_t h_t, \tag{8}$$

where $h_t$ denotes the hidden state, $x_t$ is the input, and $\bar{A}_t, \bar{B}_t, \bar{C}_t$ are discretized parameters derived via zero-order hold (ZOH). This formulation enables efficient integration of long-range features while maintaining the fidelity of local details, which is essential for medical imaging applications.

To improve Mamba's application in medical image classification, researchers introduced MedMamba[90], the first Vision Mamba model tailored for this task. Leveraging the efficiency of state-space models (SSMs), MedMamba aims to set a new benchmark in medical image classification. Meanwhile, the Vision Mamba Denoising Diffusion Probabilistic Model (VM-DDPM)[91] has shown exceptional performance in medical image synthesis, integrating CNN-based local feature extraction with SSM-driven global modeling, while maintaining linear computational complexity, making it well-suited for high-resolution imaging. Beyond classification and synthesis, Mamba-based models have proven valuable in radiotherapy planning. The MD-Dose model[92] employs Mamba encoders to efficiently propagate tumor-bed contextual information, significantly reducing errors in 3D dose maps. In video generation, VideoMamba[93] is optimized for long-video modeling, operating six times faster than TimeSformer[94] while efficiently adapting to multi-modal generative tasks. Additionally, MambaMixer[95] refines multi-dimensional data modeling, making it highly effective for multi-modal video generation.

These developments underscore Mamba's capability to serve as an alternative to Transformers or complement them in resource-constrained environments. Its selective state transitions filter out irrelevant information while preserving critical anatomical dependencies. While Mamba offers efficient long-range sequence modeling, its limited pre-training ecosystem compared to Transformers constrains its application in multi-modal medical imaging. Additionally, the state compression mechanism, although computationally advantageous, may hinder the retention of fine-grained local information, potentially leading to memory dilution or information forgetting in complex clinical tasks.

### S3.4.3. Autoregressive Models

Autoregressive (AR) models[83,84] generate images sequentially, predicting each pixel (or voxel) based on the previously generated ones. This sequential dependency modeling has proven highly effective in medical image synthesis, particularly for tasks requiring fine-grained pixel-level detail. By factorizing the joint distribution p(x) of an image into a product of conditional probabilities, AR models ensure that each generated element maintains consistency with prior context.

$$p(\mathbf{x}) = \prod_{t=1}^{T} p(x_t \mid x_{<t}), \tag{9}$$

where $x_t$ represents the $t$-th element (e.g., pixel, patch, or token) in a predefined generation order, and $x_{<t}$ denotes all previously generated elements. For high-dimensional medical images, this sequential dependency is often modeled using neural networks, such as Transformers or CNNs, to parameterize $p(x_t \mid x_{<t})$.

In longitudinal medical imaging, particularly for studying aging processes and disease progression, the Sequence-Aware Diffusion Model (SADM)[96] combines autoregressive (AR) models with diffusion processes to synthesize aging-aligned brain MRI sequences. For accelerated MRI reconstruction, the Autoregressive Image Diffusion (AID) model[97] enforces k-space consistency through retrospective sampling, suppressing aliasing artifacts compared to standard diffusion models. Additionally, recent advancements, such as MambaRoll[98] integrate AR with state-space model (SSM) latent states at the patch level, improving performance in medical image reconstruction while maintaining computational efficiency.

The "causal constraint" inherent in AR models makes them naturally suited for medical image synthesis, particularly when preserving anatomical continuity. However, these models often rely on external frameworks, such as Mamba or Transformer-based models[98], [96], to provide global priors in order to prevent error accumulation and ensure the accuracy of the synthesized data. This dependency highlights the importance of integrating both local and global contextual information for improving the performance and robustness of AR-based medical imaging applications.

Overall, Transformers effectively capture global dependencies through self-attention mechanisms but are limited by quadratic complexity, which hinders processing long sequences. Mamba addresses this limitation by utilizing a linear state-space model, trading some global awareness for computational efficiency. Autoregressive models excel at handling local dependencies but lack bidirectional context. The relationships among these approaches, as illustrated in Figure S2, highlight their complementary nature and suggest that integration can yield significant benefits. For instance, combining Transformers with state-space models[99] can reduce training time while preserving global context. Platforms like the MONAI Model Zoo[100] facilitate such integrations by providing pre-trained models, enabling rapid fine-tuning and advancing medical imaging towards more generalizable paradigms.

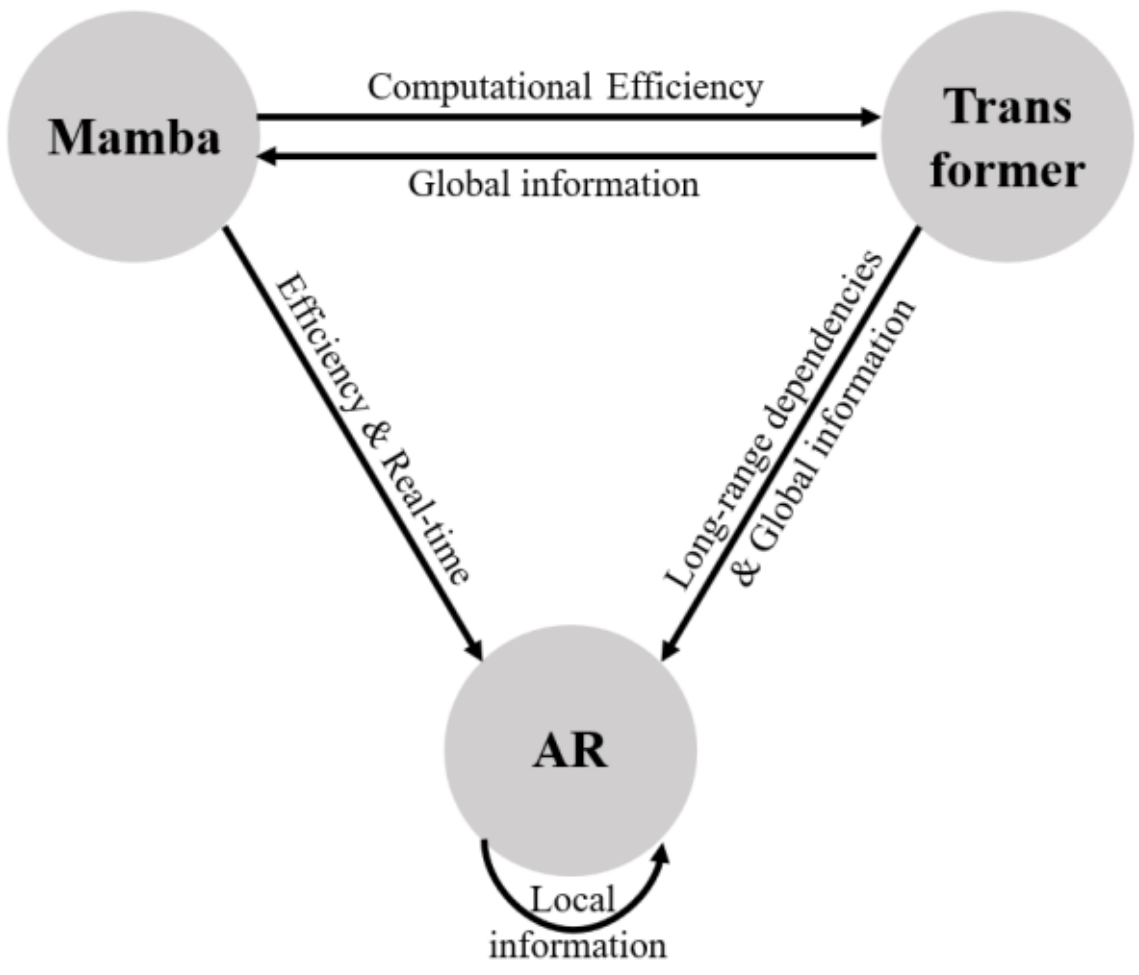

**Figure S2.** A relationship diagram illustrating the integration of Transformer, Mamba, and AR models

## S3.5. Emerging trends of Multimodal Foundation Models

With the growing demand for unified, scalable, and data-efficient solutions in medical imaging, foundation models have emerged as a promising paradigm. These models are typically pretrained on large-scale datasets and designed to generalize across tasks and modalities, often requiring minimal task-specific supervision. Building on the progress of earlier image-only generative models such as GANs and diffusion models, the field has increasingly shifted toward foundation models that integrate both visual and textual information[18–22]. These foundation models can be broadly categorized into three types: (1) vision-only models trained via self-supervised learning (e.g., masked autoencoder[101]); (2) language-only models designed for medical report understanding or generation[102–104]; and (3) vision-language models, which are of particular interest in medical imaging due to their ability to learn from paired image–text data, such as radiology images and associated reports[19,66,105].

In medical imaging, vision-language foundation models typically adopt a dual-encoder architecture that includes a visual encoder (e.g., convolutional neural networks or Vision Transformers) and a text encoder (e.g., transformer-based language models)[106–108]. These two modalities are jointly trained using contrastive learning to align their representations in a shared embedding space. The core idea is to bring matching image–text pairs closer together while pushing non-matching pairs apart. This framework forms the basis of many large-scale pretrained architectures as illustrated in Fig. 2 (e), enabling models to generalize across tasks with limited supervision and to support applications such as zero-shot classification, report retrieval, and text-guided image synthesis. These models are trained with a variant of the InfoNCE loss:

$$L_{\text{contrast}} = -\frac{1}{N}\sum_{i=1}^{N} \log \frac{\exp\left(\text{sim}\left(f(I_i), g(T_i)\right)/\tau\right)}{\sum_{j=1}^{N} \exp\left(\text{sim}\left(f(I_i), g(T_j)\right)/\tau\right)}, \tag{10}$$

where $f(I)$ and $g(T)$ are image and text encoders, sim(.) is a similarity (e.g., cosine), and $\tau$ is a temperature. This contrastive training ensures paired images and captions have high similarity, enabling zero-shot image classification and retrieval.

In the field of medical AI, foundation models are emerging as a transformative paradigm for building more generalizable and versatile systems. A representative example is CheXzero, which leverages contrastive learning on paired chest X-rays and free-text radiology reports to enable zero-shot multi-label classification of thoracic diseases. By aligning visual and textual modalities in a shared latent space, CheXzero[109] bypasses the need for explicit disease annotation and instead learns directly from natural language supervision. This approach highlights a key advantage of foundation models: the ability to scale learning through self-supervision on large, unstructured datasets. More broadly, such models pave the way toward Generalist Medical AI (GMAI)—systems trained across diverse modalities (imaging, text, clinical data)[110] to perform a wide range of tasks with minimal task-specific supervision. As a result, the development of large-scale, self-supervised, multimodal foundation models has become a central trend in medical AI research.

# S4. More Details on Key Applications of Generative AI in Medical Imaging

## S4.1. Acquisition and Reconstruction Phase: Enhancing Data Quality and Availability

### *S4.1.1. Denoising and artifact removal*

In medical imaging, noise and artifacts are major obstacles in medical imaging, often obscuring critical anatomical details and reducing diagnostic reliability. As different modalities exhibit unique noise characteristics, generative AI models are increasingly tailored to address modality-specific degradation while aligning with clinical demands.

In low-dose CT (LDCT), quantum noise and metal-induced streak artifacts remain major barriers to visualizing fine anatomical structures. Traditional denoising filters often failed to preserve structural details. To overcome this, unsupervised Poisson flow models have been introduced for photon-counting CT, effectively suppressing stochastic noise while preserving tissue contrast[111]. In dental imaging, cycle-free invertible architectures have shown superior performance in removing metal-induced beam-hardening artifacts, enhancing implant planning accuracy[112,113]. More recent frameworks, such as progressive Wasserstein GANs combined with residual encoder–decoder architectures, have demonstrated improved delineation of bronchial walls and subtle lesions[114]. For cardiac CT, TT U-Net[115] leveraged temporal transformers to reduce motion artifacts and enhance phase consistency by learning spatiotemporal features from pseudo all-phase dataset. Meanwhile, diffusion-based models like CoreDiff[116] simulated noise decay dynamics to reconstruct high-fidelity images under ultra-low-dose conditions, notably aiding in lung nodule detection. Further advancements such as DenoMamba[117] employed state-space modeling to capture short- and long-range dependencies for effective denoising while preserving diagnostic quality.

In PET imaging, statistical noise has interfered with the accurate quantification of radiotracer uptake, which is essential for tumor detection and assessment of therapeutic response. Earlier attempts to transfer CT-based denoising models to PET[118,119] suffered from poor adaptability to tracer dynamics. To address this limitation, parameter-transferred Wasserstein GANs have been developed to improve noise suppression while preserving quantitative radiotracer uptake[120]. More recently, diffusion-based models have been applied to PET denoising by treating noise as a stochastic process. For instance, denoising diffusion probabilistic models enhanced image quality and anatomical fidelity without compromising uptake accuracy[121] while ControlNet-guided 3D diffusion models enabled adaptive denoising of whole-body PET scans using low-dose inputs as conditional priors[122]. These approaches improve the reconstruction of biologically realistic signal distributions, thereby reducing noise-related bias in quantitative analysis and enhancing the reliability of PET imaging for tumor evaluation.

In MRI, images are prone to Rician noise and motion artifacts, particularly in dynamic acquisitions. Conventional denoising often blurred structural boundaries. To combat this, a residual Wasserstein GAN has been employed to model inter-slice anatomical consistency, improving both detail preservation and contrast[123]. Complementarily, a content–noise complementary learning strategy enabled joint modeling of clean and noisy signal components, offering a more nuanced restoration[124]. In fetal MRI, a GAN-based motion correction method effectively reconstructed consistent anatomical structures from misaligned slices[125]. Recent developments have also explored diffusion-based models: regularized reverse diffusion improves denoising and super-resolution jointly[126], while alternate mask-guided diffusion in the pixel-frequency domain provided robust motion artifact removal without compromising structural fidelity[127]. Across imaging modalities, these models reflect a shift toward clinically grounded, structure-aware restoration that not only reduces noise but also preserves anatomical plausibility, which is essential for tasks such as lesion detection, segmentation, and treatment planning.

**Table S2**. Summary of publications on denoising and artifact removal.

| Publication (Year) | Model | Application | Loss Function | Link |
|---|---|---|---|---|
| PWGAN-WSHL (2021) [114] | GAN | Low-dose CT denoising | WGAN loss, L1 loss, MSE loss, structural loss | – |
| DenoMamba (2024) [117] | Mamba | Low-dose CT denoising | L1 loss | √ |
| m-WGAN (2019) [112] | GAN | CT image artifact removal | WGAN loss, MSE loss | – |
| TT U-Net (2023) [115] | Transformer | CT image artifact removal | L1 loss, adversarial loss | √ |
| CoreDiff (2024) [116] | Diffusion model | Low-dose CT denoising | Diffusion denoising loss | √ |
| PFGM++ (2024) [111] | Diffusion model | Photon-counting CT denoising | Diffusion denoising loss | √ |
| Yang et al. (2021) [119] | CNN | PET image denoising and artifact Removal | MSE loss | – |
| Hu et al. (2020) [118] | GAN | PET image denoising and artifact Removal | WGAN loss, MSE loss, gradient difference loss, content loss, ssim loss | – |
| PT-WGAN (2020) [120] | GAN | PET image denoising | Adversarial loss, MSE loss, ssim loss, perceptual loss | √ |
| Gong et al. (2024) [121] | Diffusion model | PET image denoising | Diffusion denoising loss | – |
| Yu et al. (2024) [122] | Diffusion model | PET image denoising | Diffusion denoising loss | – |
| RED-WGAN (2019) [123] | GAN | MRI image denoising | WGAN loss, MSE loss, perceptual loss, VGG loss | √ |
| Chung et al. (2022)[116] | Diffusion model | MRI image denoising | Diffusion denoising loss | – |
| CNCL (2022) [124] | GAN | MR, CT and PET image denoising | Content loss, noise loss, GAN loss | √ |
| Lim et al. (2023) [128] | GAN | MRI image artifact removal | WGAN loss, L1 loss, VGG loss | – |
| PFAD (2024) [129] | Diffusion model | MRI image artifact removal | Diffusion denoising loss | √ |

## S4.1.2. Accelerated image reconstruction

Generative models have significantly advanced medical image reconstruction, addressing limitations in image quality, sampling sparsity, and modality constraints. Across CT, PET, MRI, and emerging modalities, they enable high-fidelity recovery from low-dose or incomplete data, improving diagnostic accuracy and facilitating real-time clinical applications.

***CT Reconstruction**:* High-quality CT reconstruction is critical for accurate lesion detection and surgical planning, but dose reduction has introduced challenges such as projection sparsity and artifact amplification. Traditional algorithms often failed to meet clinical demands for detail preservation under low-dose protocols. To address these limitations, GAN-based sinogram restoration frameworks have been proposed, generating missing projection data with up to 95.88% SSIM under 60° scans for lung nodule screening[130]. Pix2pix GAN enabled cross-scanner kernel transfer, reducing quantification bias in emphysema studies and enhancing multicenter consistency[131]. In acute cases like intracranial hemorrhage DOLCE, a diffusion-based model, reconstructed high-fidelity images from limited-angle inputs while reducing metal artifacts[132]. Similarly, diffusion priors have been used for sparse-view CT to outperform traditional MBIR in subtle hemorrhage detection[133]. For orthopedic applications, the conditional GAN synthesized 3D CT from single-view X-rays, aiding acetabular cup planning in THA[134], while neural radiance fields enabled detailed 3D knee modeling from 2D data[135] . Material decomposition in dual-energy CT has also benefited from 3D generative networks for accurate calcification assessment[136]. In trauma imaging, Mamba-based Monte Carlo frameworks drastically reduced reconstruction time while maintaining quality[137]. Recent diffusion models have improved sparse-view consistency and accelerated convergence[138–140] , and SWORD enhanced textural details[77]. Despite these advances, challenges remain in balancing artifact suppression, low-dose sensitivity, device generalizability, and real-time feasibility for clinical deployment.

***PET Reconstruction**:* Positron emission tomography (PET) plays a unique role in assessing tumor metabolism, diagnosing neurodegenerative diseases, and monitoring treatment. However, its clinical application has been constrained by issues such as low signal-to-noise ratio, multimodal registration errors, and the complexity of dynamic imaging. Generative approaches helped address these barriers across reconstruction, attenuation correction, and personalization. For low-count PET, CycleGAN-based model maintained tumor metabolic volume consistency in pediatric oncology[141] with the normalized correlation coefficient improving from 0.970 to 0.996. Similarly, the method[142] developed a noise-aware adaptive loss function to balance noise distribution and anatomical fidelity in low-count data, enhancing the stability of the striatal dopamine transporter binding ratio in Parkinson's disease diagnosis. To reduce attenuation correction bias in PET/MRI, a joint reconstruction framework[143] simultaneously estimated activity and attenuation maps with $<1\%$ SUV error. Bimodal VAEs further reduced PET–MRI registration error by decoupling modality-specific features[144]. Dynamic PET benefited from deep generalized learning that restores fine details from sparse temporal frames[145]. In personalized therapy, a projection-domain CNN estimated individualized dose distributions for $^{177}$Lu treatment[146]. Lightweight architectures like Cycle-PET achieved sub-second reconstruction on embedded GPUs for emergency use[147]. And a projection generative network trained on Monte Carlo simulated data to reduce acquisition time without sacrificing tumor-to-background contrast[148]. Addition, studies[149,150] and more recent works[78,137,151,152] have further extended PET reconstruction techniques through task optimization, image synthesis, and diffusion modeling, providing stronger technical support for precision clinical diagnosis and treatment.

***MRI Reconstruction*:** The quality of MRI reconstruction critically affects early disease detection, diagnostic precision, and treatment planning. Yet, conventional methods remained limited by noise, motion artifacts, and data inconsistency, which hindered accurate visualization of small lesions and subtle anatomical details. In neuroimaging, where structural clarity is essential, recent generative approaches have markedly improved reconstruction quality[153,154]. In pediatric imaging, where reducing scan time is essential, GAN-based approaches integrated with compressed sensing have shown promise in generating high-resolution images with superior lesion contrast and improved edge definition[149,155,156]. Building on this, a transformer-based model using global self-attention has better captured multi-scale features, strengthening both qualitative and quantitative brain assessments[157]. For cardiac MRI, the dynamic nature of the heart demands real-time reconstruction and artifact suppression. Diffusion and score-based generative models have shown strong performance in restoring temporal details and reducing motion-induced distortions, even under sparse sampling[79,158–160]. Additionally, the Mamba framework, which incorporates uncertainty quantification, provided valuable insights into reconstruction reliability and thereby supported risk-aware clinical decision-making[161,162]. To improve consistency in low-field MRI and multicenter studies, where acquisition parameters and device variability are common, researchers have explored federated learning, regularization, and meta-learning hypernetworks[163–166]. These approaches not only ensured the recovery of image details under low-field conditions but also facilitated the integration of multicenter data, thus contributing to more consistent diagnostic outcomes in large-scale clinical studies[167]. Notably, the generative autoregressive transformer proposed in the work[168] enabled model-agnostic, privacy-preserving MRI reconstruction by capturing complex spatial–temporal patterns in distributed datasets. MambaRoll[98] combined autoregressive mechanisms with state space representations at the patch level, enabling efficient and high-quality reconstruction. AID[97] further explored sequential image generation for reconstructing MRI scans, offering improved spatial coherence and progressive refinement. These approaches have demonstrated the effectiveness of autoregressive models in capturing structured dependencies for medical image reconstruction. In summary, advanced MRI reconstruction techniques have demonstrated distinct advantages across neuroimaging, cardiac imaging, and low-field/multicenter applications. The future challenge lies in seamlessly integrating these advanced methods into clinical workflows to achieve real-time, efficient reconstruction and to validate their performance on large-scale, multicenter datasets.

***Others:*** Beyond conventional CT, PET, and MRI, generative models are increasingly been applied in ultrasound, photoacoustic, and EEG-based imaging, expanding the landscape of image reconstruction. In ultrasound, fast-sampling generative models enabled high-quality reconstruction with reduced data requirements, supporting real-time clinical applications[169] . Diffusion-based methods incorporated uncertainty quantification and variance modeling, improving structural fidelity and diagnostic confidence[170,171]. For single plane-wave data, generative models enhanced signal-to-noise ratio and contrast, boosting sensitivity in early lesion detection[172]. In photoacoustic imaging, reconstruction has been particularly challenging due to data sparsity and limited sampling angles. By combining diffusion models with iterative optimization[173], researchers have mitigated artifacts and loss of detail, while score-based generative models with rotational consistency constraints[174] ensured image consistency across different angles, providing reliable support for tumor and tissue function assessment. Moreover, the DM-RE2I framework[175] leveraged diffusion models to map EEG signals into image space, exploring the conversion from neural electrical activity to structural images and opening new avenues for early diagnosis and functional localization of neurological disorders. Taken together, generative models are reshaping image reconstruction across both established and emerging modalities. By enhancing image detail, reducing artifacts, and incorporating uncertainty modeling, these methods significantly improve diagnostic reliability. As diffusion and hybrid frameworks continue to evolve, generative reconstruction is poised to play an increasingly central role in precision diagnostics and personalized care.

**Table S3.** Summary of publications on medical image reconstruction.

| Publication (Year) | Model | Application | Loss Function | Link |
|---|---|---|---|---|
| **CT Reconstruction** | | | | |
| DL-recon (2022) [136] | GAN | CBCT-to-CT reconstruction | Adversarial loss, L1 loss | – |
| Pradhan et al. (2023) [134] | GAN | 2D-to-3D CT reconstruction | L1 loss, BCE loss, adversarial loss | – |
| HyperNeRFGAN (2024) [135] | GAN | X-ray-to-CT reconstruction | StyleGAN2Loss | √ |
| Krishnan et al. (2024) [131] | GAN | Low-dose CT reconstruction | L1 loss, adversarial loss, reconstruction loss | √ |
| MambaMIR (2025) [137] | GAN, Mamba | Low-dose CT/PET reconstruction | Adversarial loss, Charbonnier loss, image loss, frequency loss | – |
| SI-GAN (2019) [130] | GAN | Limited-angle CT reconstruction | Adversarialloss, sinogram loss, reconstruction loss | – |
| DOLCE (2023) [132] | Diffusion model | Limited-angle CT reconstruction | Diffusion denoising loss | √ |
| Lopez-Montes et al. (2024)[133] | Diffusion model | Limited-angle CT reconstruction | Diffusion denoising loss | – |
| TIFA (2024) [140] | Diffusion model | Limited-angle CT reconstruction | Diffusion denoising loss | √ |
| Xia et al. (2024) [138] | Diffusion model | Sparse-view CT reconstruction | Diffusion denoising loss | – |
| CDDM (2024) [139] | Diffusion model | Sparse-view CT reconstruction | Diffusion denoising loss | – |
| SWORD (2024) [77] | Diffusion model | Sparse-view CT reconstruction | Diffusion denoising loss | √ |
| **PET Reconstruction** | | | | |
| NADRU (2020) [142] | CNN | Low dose PET reconstruction | Dice loss, BCE loss, general and adaptive robust loss, ssim loss | – |
| Shi et al. (2023) [143] | CNN | Low dose PET reconstruction | L1 loss, image domain loss, gradient difference loss, LIP loss | √ |
| CPR-CNN (2024) [147] | CNN | Low dose PET reconstruction | Reconstruction loss, cycle consistency loss | – |
| DGLM (2024) [145] | CNN | Low count PET reconstruction | MSE loss, ssim loss | – |
| Lei et al. (2019) [141] | GAN | Low count PET reconstruction | Cycle-consistent adversarial loss, gradient descent loss, mean p-norm distance loss | – |
| Task-GAN (2019)[149] | GAN | Ultra-low dose PET reconstruction | L1 loss, adversarial loss, regression loss | – |
| AR-GAN (2022) [150] | GAN | Low dose PET reconstruction | L1 loss, adversarial loss, cross-entropy loss | – |
| DDPET-3D (2024) [152] | Diffusion model | Low dose PET reconstruction | Diffusion denoising loss | – |
| Wikberg et al. (2024) [146] | CNN | Sparsely acquired projections PET reconstruction | L1 loss, MSE loss | – |
| MMJSD (2024) [144] | VAE | Bimodal PET/MRI reconstruction | Negative log-likelihood loss, KL loss | – |
| Singh et al. (2024) [151] | Diffusion model | 2D/3D PET reconstruction | Poisson Log-Likelihood loss, Diffusion denoising loss | √ |
| MC-Diffusion (2024) [78] | Diffusion model | PET-MRI reconstruction | Diffusion denoising loss | √ |
| **PET Reconstruction** | | | | |
| Wang et al. (2019) [155] | GAN | MRI reconstruction | Content loss, perceptual loss, adversarial loss, dc loss | – |
| rsGAN (2020) [156] | GAN | Multi-contrast MRI reconstruction | L1 loss, perceptual loss, adversarial loss, dc loss | – |
| Kelkar et al. (2021) [164] | GAN | MRI reconstruction | MSE loss, log-likelihood loss, TV loss, dc loss | – |
| SwinMR (2022) [157] | Transformer | MRI reconstruction | Pixel-wise Charbonnier loss, frequency Charbonnier loss | √ |
| KM-MAML (2023) [166] | CNN | MRI reconstruction | L1 reconstruction loss, dc loss | √ |
| MambaMIR (2024) [161] | Mamba | MRI reconstruction | Adversarial loss, image loss, kspace loss, perceptual loss, dc loss | √ |
| DM-Mamba (2025)[162] | Mamba | MRI reconstruction | L1 loss, dc loss | √ |
| MambaRoll (2024) [98] | Mamba, AR | MRI reconstruction | Kspace loss, cascade loss, dc loss | √ |
| HFS-SDE (2024) [160] | Diffusion model | MRI reconstruction | Diffusion denoising loss, dc loss | √ |
| JSMoCo (2025) [159] | Diffusion model | MRI reconstruction | Diffusion denoising loss, dc loss | √ |
| AID (2025) [97] | Diffusion model, AR | MRI reconstruction | Diffusion denoising loss, dc loss | √ |
| Kofler et al. (2020) [163] | CNN | Cardiac cine MRI reconstruction | L2 loss, dc loss | – |
| Qiu et al. (2024) [79] | Diffusion model | Cardiac cine MRI reconstruction | Diffusion denoising loss, dc loss | – |
| DiffCMR (2024) [176] | Diffusion model | Cardiac cine MRI reconstruction | Diffusion denoising loss, dc loss | √ |
| FedGIMP (2023) [165] | GAN | Federated MRI reconstruction | Logistic adversarial loss, local reconstruction loss, dc loss | √ |
| FedGAT (2025) [168] | VAE, Transformer, AR | Federated MRI reconstruction | Perceptual loss, adversarial loss, cross-entropy loss, MSE loss, dc loss | √ |
| **Otherrs** | | | | |
| DDRM (2023) [170] | Diffusion model | US image reconstruction | Diffusion denoising loss | √ |
| DRUSvar (2024) [171] | Diffusion model | US image reconstruction | Diffusion denoising loss | √ |
| Lan et al. (2023) [169] | Diffusion model, GAN | US image reconstruction | Diffusion denoising loss | – |
| Merino et al. (2024) [172] | Diffusion model, GAN | US image reconstruction | Adversarial loss, Diffusion denoising loss | – |
| DM-RE2I (2023) [175] | Diffusion model | EEG to image reconstruction | Diffusion denoising loss | – |
| PAT-Diffusion (2023) [173] | Diffusion model | Photoacoustic tomography reconstruction | Diffusion denoising loss | √ |
| Tong et al. (2023) [174] | Diffusion model | Photoacoustic tomography reconstruction | Diffusion denoising loss | – |

## *S4.1.3. Super-resolution*

Medical image resolution is often constrained by acquisition time, radiation dose, and physical limitations of imaging systems. Super-resolution (SR) methods seek to reconstruct high-resolution images from low-quality inputs, and are commonly divided into temporal SR (for dynamic imaging) and spatial SR (for static structural enhancement) [177,178].

***Temporal super-resolution:*** In dynamic imaging, temporal super-resolution increases frame rates and enhances temporal consistency to capture continuous organ motion and suppress motion artifacts. High frame rate imaging enables more accurate functional assessments for rapidly moving organs such as the heart and lungs, aiding early disease diagnosis and treatment planning. Early efforts relied heavily on GAN-based frameworks with perceptual loss functions [179], which aimed to reconstruct intermediate frames by enhancing temporal fidelity and visual realism. These approaches were followed by optical flow–guided models[180,181], that introduced motion-aware learning mechanisms to better capture video inter-frame dependencies. Building upon these foundations, more advanced architectures have explored spatial–temporal interpolation. A deformation-based method enabled smooth motion transitions in 4D cardiac MRI by modeling local anatomical deformation[182], while the multi-pyramid voxel flow[183] improved interpolation performance under sparse temporal sampling. Subsequently, diffusion-based deformation models[184] offered more stable and noise-resilient frame generation, particularly useful in scenarios with irregular breathing or arrhythmias. Recent approaches such as the data-efficient interpolation network[185] and the dynamic dual-channel architecture[186] have further advanced temporal super-resolution by enhancing frame detail while minimizing artifacts. Meanwhile, hybrid models incorporating multi-level feedback loops and task-specific motion correction[180,187] have demonstrated strong generalization in clinical tasks such as myocardial ischemia evaluation, arrhythmia monitoring, and tumor motion tracking.

***Spatial super-resolution:*** Spatial super-resolution (SR) techniques aim to reconstruct high-resolution medical images from low-resolution inputs, improving the visibility of fine anatomical details crucial for lesion detection, tissue boundary delineation, and vascular assessment. Traditional interpolation methods are limited in restoring texture and structural accuracy. Recent deep learning approaches have addressed these limitations through multimodal fusion, frequency-domain modeling, and generative modeling. Among early methods, GAN-CIRCLE[188] integrated identical, residual, and cycle consistency constraints within a GAN framework to improve structural fidelity in CT images. In MRI, SOUP-GAN[189] enhanced perceptual quality and reduces aliasing by learning texture-aware representations.

Combining GANs with discrete wavelet transforms further improved texture realism and suppressed noise, enabling efficient SR in portable applications[190]. In X-ray imaging, frequency domain constraints have been shown to improve edge detail and suppress artifacts[191]. More recent methods leveraged hierarchical and diffusion-based models. Hierarchical amortized GAN allowed memory-efficient synthesis of high-resolution 3D medical images by capturing both global and local features[192], while local-to-global feature learning frameworks further enhanced anatomical consistency across scales[193]. Diffusion models have also been adopted: UHRCT_SR[194] employed a dual-stream structure-preserving network and an imaging enhancement operator for CT super-resolution, and partial diffusion model[195] accelerated the process by focusing on relevant components in brain MR images. In addition, the Deform-Mamba network [196] integrated deformable convolutions with state-space modeling to reconstruct high-quality MR images under limited resolution conditions. These methods reflect the ongoing shift from basic interpolation toward structurally informed reconstruction techniques. While notable improvements in image quality have been achieved, challenges such as modality variability, computational efficiency, and clinical integration remain areas of active research.

**Table S4.** Summary of publications on super-resolution.

| Publication (Year) | Model | Application | Loss Function | Link |
|---|---|---|---|---|
| **Temporal super-resolution** | | | | |
| Ren et al. (2021) [180] | CNN | Video super-resolution | L1 loss | – |
| Song (2022) [181] | CNN, Transformer | Video super-resolution | MSE loss | – |
| VSRResFeatGAN (2019) [179] | GAN | Video super-resolution | Adversarial loss, perceptual loss, charbonnier loss, | – |
| MFIN (2019)[187] | CNN | 4D MRI Temporal super-resolution | Cycle consistency loss, recon loss, ssim loss, | – |
| SVIN (2020) [182] | CNN | 4D MRI Temporal super-resolution | Similarity loss, smoothness regularization loss, regression loss | √ |
| DDoS-Unet (2024) [186] | CNN | 4D MRI Temporal super-resolution | Perceptual loss, L1 loss | √ |
| MPVF (2023) [183] | CNN, Transformer | 4D MRI Temporal super-resolution | Charbonnier loss | √ |
| UVI-Net (2024) [185] | CNN, Transformer | 4D MRI Temporal super-resolution | NCC loss, gradient loss | √ |
| DDM (2022) [184] | Diffusion model | 4D MRI Temporal super-resolution | Diffusion denoising loss, NCC loss, KL loss | √ |
| **Spatial super-resolution** | | | | |
| GAN-CIRCLE (2019) [188] | GAN | CT super-resolution | Adversarial loss, cycle-consistency loss, identity loss, joint sparsifying transform loss | √ |
| TTSR-FD (2021) [191] | GAN | X-ray super-resolution | Frequency domain loss, perpetual loss, adversarial loss, | – |
| SOUP-GAN (2022) [189] | GAN | MRI super-resolution | Adversarial loss, perceptual loss | √ |
| DWT-SRGAN (2022) [190] | GAN | MRI super-resolution | Perpetual loss, adversarial loss, wavelet loss | – |
| HA-GAN (2022) [192] | GAN | CT/MRI super-resolution | GAN loss, reconstruction loss | – |
| Huang et al. (2024) [193] | Transformer, CNN, | US/OCT/Endoscope/CT/MRI super-resolution | Charbonnier loss, L1 loss | – |
| UHRCT_SR (2023) [194] | Diffusion model | CT super-resolution | Diffusion denoising loss | √ |
| PartDiff (2023) [195] | Diffusion model | MRI super-resolution | Diffusion denoising loss | – |
| Deform-Mamba (2024) [196] | Mamba | MRI super-resolution | L1 loss, CE loss | – |

## S4.2. Diagnosis Phase: Enriching Diagnostic Imaging

### *S4.2.1. Unconditional synthesis*

Unconditional synthesis generates medical images from random noise or data distributions without specific constraints, enabling the creation of diverse and realistic samples without relying on annotated data. Early work in this area focused on generative adversarial networks (GANs), which produced synthetic images that resembled clinical data. Classical applications included the synthesis of blood vessel surfaces [197] and 3D brain MRI volumes[197], demonstrating GANs' capacity to model complex anatomical structures without supervision. However, classical GANs struggled with limited controllability, mode collapse, and low spatial resolution, which limited their use in anatomically precise tasks. To overcome these limitations, more advanced architectures such as StyleGAN[198,199] introduced structured latent spaces and style-based modulation, enabling fine-grained control over image attributes and significantly improving visual realism and resolution.

Driven by the need for higher fidelity and diversity, diffusion models have emerged as a robust alternative to GANs for unconditional medical image generation. Unlike GANs' single-shot generation, diffusion models employ a multi-step denoising process to iteratively transform random noise into structured images, offering greater training stability and output diversity. A denoising diffusion probabilistic model (DDPM) has shown strong performance in 3D medical imaging, particularly in enhancing clarity and resolution in volumetric CT and MRI data [200]. Model likes 3D MedDiffusion[76] further supported high-resolution, anatomically specific image synthesis for applications such as tumor detection. Moreover, diffusion models are well-suited for handling multimodal variability, improving robustness under challenging imaging conditions like noise and contrast shifts[201,202]. To further enhance visual fidelity and structural consistency, specialized variants have been developed. Some focused on CT-specific improvements[203], while others integrated frequency or attention-based priors to improve anatomical realism[91]. And the Deformation-Recovery Diffusion Model[204] introduced spatial controllability, facilitating downstream tasks such as segmentation and registration where anatomical precision is critical. These developments have highlighted the multifunctionality of diffusion models in generating data for a wide range of clinical applications[202].

In summary, unconditional image synthesis, especially using GANs and diffusion models, has shown considerable potential in producing diverse and high-quality medical images without the need for annotated datasets. These methods have proven instrumental in addressing data scarcity, improving the generalizability of diagnostic algorithms, and supporting downstream tasks such as early disease detection and treatment planning. However, their lack of explicit control over generated content can limit clinical utility in scenarios requiring precise anatomical, pathological, or modality-specific constraints. This has led to increasing interest in conditional synthesis approaches, which incorporate prior knowledge such as textual descriptions, anatomical maps, or clinical parameters to guide image generation toward specific diagnostic or therapeutic objectives.

**Table S5.** Summary of publications on medical image unconditional synthesis.

| Publication (Year) | Model | Application | Loss Function | Link |
|---|---|---|---|---|
| Danu et al. (2019) [205] | VAE，GAN | Blood vessel surfaces synthesis | MSE loss，adversarial loss | – |
| Syn-Net (2020) [199] | GAN | 2D brain MRI synthesis | L1 loss, perceptual loss, adversarial loss | – |
| Chong and Ho (2021) [197] | GAN | 3D brain MRI synthesis | Adversarial loss，GAN loss | – |
| 3D-StyleGAN (2021) [198] | GAN | 3D MRI synthesis | MSE loss, Logistic loss | √ |
| Txurio et al. (2023) [201] | Diffusion model | 2D CT synthesis | Diffusion denoising loss | – |
| MRGen (2024) [202] | Diffusion model， VAE | 2D MRI synthesis | Diffusion denoising loss | √ |
| VM-DDPM (2024) [91] | Diffusion model, Mamba | 2D X-ray/MRI synthesis | Diffusion denoising loss, GAN loss, BCE Loss | – |
| GH-DDM (2023) [203] | Diffusion model | 2D X-ray/CT/MRI/OCT synthesis | Diffusion denoising loss | – |
| Medicaldiffusion (2023) [200] | Diffusion model | 3D CT/MRI synthesis | Diffusion denoising loss | √ |
| DRDM (2024) [204] | Diffusion model | 3D CT/MRI synthesis | Distance error loss, angle error loss, regularization loss | √ |
| 3D MedDiffusion (2024) [76] | Diffusion model | 3D CT/MRI synthesis | Vector quantization loss, adversarial loss, tri-plane loss, Diffusion denoising loss | √ |

## *S4.2.2. Conditional synthesis*

In contrast to unconditional synthesis, which learns image distributions independently of external inputs, conditional synthesis incorporates domain-specific priors such as clinical text, imaging data, anatomical structures, or physiological parameters into the generative process. This improves the relevance, controllability, and diagnostic value of the synthesized outputs. Conditional methods can be broadly categorized into three types: text-to-image synthesis, image-to-image translation and completion, anatomically guided synthesis. Each reflects an evolving effort to bridge data-driven generation with clinical context, enhancing both interpretability and utility.

***Text-to-Image Synthesis***: In recent years, text-to-image synthesis has become an important direction in image generation, supported by the rapid development of latent diffusion models and their ability to integrate context. Initial efforts, such as AttnGAN[206], Mirrorgan[207], StackGAN[208], and Cogview[209], laid the groundwork by mapping simple textual inputs to image content. More recent models like DALL-E[210], Imagen[211], and Stable Diffusion[212] have significantly improved semantic consistency and resolution. In the medical domain, this modality facilitates the transformation of both free-text radiology reports and structured clinical variables such as age, sex, smoking history, blood pressure, and imaging modality into more diverse and representative medical images. Early efforts primarily focused on report-to-image synthesis, where structured or semi-structured radiology reports were used to condition chest x-ray generation. Models like Chest-diffusion[105] and Diff-CXR[66] translated radiology reports into synthetic chest X-rays while incorporating disease-specific priors, enhancing both data diversity and diagnostic interpretability. Building on this foundation, recent methods have expanded to include text-driven generation across multiple imaging modalities. For example, MediSyn[64] proposed a generalist framework for synthesizing a wide range of medical images, while TextoMorph[213] targeted tumor synthesis conditioned on lesion type and location. In the domain of CT generation, models such as GenerateCT[65], MAISI[214], MedSyn[215] enabled the generation of 3D chest CT volumes guided by textual descriptions, capturing both anatomical fidelity and modality-specific characteristics. Beyond free-text reports, some models adopted parametric conditioning using explicit clinical attributes. Cheart[216] generated cardiac anatomy from age, gender, body weight, and blood pressure, while EchoDiffusion[217] synthesized echocardiograms based on ejection fraction, a key cardiac biomarker. Similarly, the HeartBeat model[218] produced echocardiography videos from multimodal physiological signals, aiding cardiac function assessment. In neuroimaging, TaDiff [219] generated personalized and longitudinal brain MRIs conditioned on treatment variables to model glioma progression, while others[220,221] focused on synthesizing multimodal MRIs using acquisition sequences (e.g., T1w and FLAIR) along with pathological categories such as glioblastoma, sclerosis, and dementia. Meanwhile, conditioning on demographic and risk-related factors such as age, sex, and smoking status, as demonstrated by DGM-VLC[222], improved the pathological realism of chest imaging. Another model[70] generated follow-up CT scans for COVID-19 by integrating radiomic and clinical features. These developments reflect a shift from descriptive to clinically informed synthesis. Embedding structured clinical information into textual prompts enhances the control and realism of generated images while improving their alignment with diagnostic and prognostic workflows.

***Image-to-Image Synthesis**:* Image-to-image synthesis encompasses both modality translation and modality completion, offering practical solutions to missing or degraded imaging data. Unlike text-based methods, these approaches preserve spatial information and structural alignment, making them highly suitable for multimodal fusion. Modality translation transforms one imaging modality into another (e.g., MRI to CT), compensating for unavailable or low-quality scans[223]. Early models employed conditional GANs (cGANs)[224–226], achieving reasonable pixel-wise translations between CT, PET, and MRI. Later, unsupervised frameworks such as CycleGAN[227–229] addressed the lack of paired data by learning bidirectional mappings. Recent advances have incorporated sequential modeling (e.g., Mamba-enhanced transformers[230–233]) and diffusion probabilistic models (DPMs) [68,116,128,234] which offer improved structural preservation and generative flexibility. Conversely, modality completion seeks to recover missing image regions or entire modalities. AutoSyncoder[129] and ResViT[235] introduced deep architectures for automatic cross-modal inference, while the other method[236] utilized multi-scale transformer-based fusion. Hybrid approaches[237,238] combined pseudo-modalities (e.g., synthetic T2-weighted MRI) with coarse-to-fine refinement, enhancing robustness. More recent studies, including frequency-guided diffusion models[239] and unified multimodal synthesis frameworks[240], enhanced the fidelity and coherence of generated medical images. Additional studies[232,241,242] integrated conditional generation and cross-dimensional knowledge guidance, advancing the field from single-modality synthesis to multimodal unified modeling. Overall, the evolution from GAN-based methods to transformer-diffusion hybrids illustrates a growing emphasis on cross-modal alignment, data efficiency, and spatial consistency in clinical image synthesis.

***Anatomically-Guided Image Synthesis****:* Anatomically guided image synthesis integrates structural priors—such as segmentation masks, tissue boundaries, or lesion annotations—into generative models to improve anatomical consistency and clinical reliability in synthetic images. By embedding explicit spatial information, these methods aim to generate outputs that closely reflect real anatomical structures and pathological variations. Early work such as the CG-SAMR network [243] demonstrated the effectiveness of incorporating lesion and tissue confidence into the synthesis of multi-contrast MR images, producing anatomically faithful outputs across different imaging modalities. Building on this, a work[244] introduced attribute disentanglement to control nodule shape, size, and texture in chest X-rays, supporting fine-grained lesion-level augmentation. In the ophthalmic domain, synthesis methods have further expanded structural guidance. A two-stage model combining StyleGAN and GauGAN[245] generated diabetic fundus images from semantic lesion maps, while the vascular-guided GAN[246] leveraged vessel structure to preserve fine anatomical details in super-resolved retinal images. Beyond major organs, anatomical priors have also enabled synthesis in low-data or small-structure scenarios. LN-Gen[247] generated rectal lymph nodes by learning from anatomical features and shape representations, supporting data expansion in pelvic imaging. Similarly, SegGuidedDiff[248] conditioned the generation process on multi-class segmentation masks, enabling anatomically controllable synthesis across diverse regions and modalities. Together, these methods highlight the growing role of anatomical priors in enhancing structural accuracy, interpretability, and the practical value of synthetic data in medical imaging.

**Table S6.** Summary of publications on medical image conditional synthesis.

| Publication (Year) | Model | Application | Loss Function | Link |
|---|---|---|---|---|
| **Text-to-Image Synthesis** | | | | |
| Campello et al. (2022) [249] | GAN | Clinical information-to-MRI | Adversarial loss, cycle-consistency loss | √ |
| CHeart (2023) [216] | VAE | Clinical information-to-MRI | KL loss, log-likelihood loss | √ |
| TUMSyn (2024) [221] | Transformer, CNN | Clinical information-to-MRI | Contrastive loss, similarity loss | – |
| Del Castillo et al. (2025) [220] | Diffusion model, VAE | Clinical information-to-MRI | Diffusion denoising loss | – |
| TaDiff (2025) [219] | Diffusion model | Clinical information-to-MRI | Diffusion denoising loss，dice loss | – |
| MAISI (2024) [214] | Diffusion model | Clinical information-to-CT | Diffusion denoising loss | √ |
| EchoDiffusion (2023) [217] | Diffusion model | Clinical information-to-video | Diffusion denoising loss | √ |
| Kawata et al. (2024) [70] | Diffusion model, VAE | Clinical information-to-chest CT synthesis | Diffusion denoising loss, similarity loss | – |
| GenerateCT (2024) [65] | Diffusion model，Transformer | Report-to-chest CT synthesis | Diffusion denoising loss, perceptual loss, adversarial loss | √ |
| MedSyn (2024) [215] | Diffusion model, VAE | Report-to-chest CT synthesis | Diffusion denoising loss, KL loss | √ |
| DCM-VLC (2024) [222] | Diffusion model，GAN | Text-guided CT synthesis | Diffusion denoising loss，adversarial loss | √ |
| MediSyn (2025) [64] | Diffusion model, VAE | Text-guided diverse synthesis | Diffusion denoising loss | – |
| TextoMorph (2024) [213] | Diffusion model | Text-guided tumor synthesis | Diffusion denoising loss, contrastive loss | √ |
| Diff-CXR (2024) [66] | Diffusion model，Transformer | Report-to-CXR synthesis | Diffusion denoising loss，InfoNCE loss, BCE loss | √ |
| Chest-diffusion (2024) [105] | Diffusion model, VAE | Report-to-CXR synthesis | Diffusion denoising loss，contrast loss | – |
| **Image-to-Image Synthesis** | | | | |
| Ben-Cohen (2019) [224] | GAN | CT-to-PET translation | Adversarial loss, MSE loss, L1 loss | – |
| Jiao et al. (2020) [226] | GAN | US-to-MRI translation | Latent space loss, appearance loss, structural consistency loss, adversarial loss | – |
| sc-cycleGAN (2020) [228] | GAN | MR-to-CT translation | Adversarial loss, cycle-consistency loss, structure-consistency loss | – |
| Gong et al. (2020) [229] | GAN | MRI-to-PET translation | Adversarial loss, cycle-consistency loss | – |
| GLFC (2025) [233] | Mamba | CBCT-to-CT translation | Multiple contrast Loss | √ |
| EGDiff (2024) [128] | Diffusion model | CBCT-to-CT translation | Diffusion denoising loss, MSE loss | – |
| DiffMa (2024) [230] | Diffusion model, Mamba | CT-to-MRI translation | Diffusion denoising loss, infoNCE loss | √ |
| MIDiffusion (2024) [68] | Diffusion model | MRI cross-modality translation | Mutual information diffusion denoising loss | √ |
| Yan et al. (2022) [238] | GAN | Multimodal MRI completion | Adversarial loss, cycle-consistency loss | – |
| Raad et al. (2024) [241] | GAN | Multimodal MRI completion | Adversarial loss, MAE loss | – |
| CKG–GAN (2024) [242] | GAN | Multimodal MRI completion | Cross-dimensional knowledge loss + adversarial | √ |
| Zhang et al. (2024) [240] | GAN | Multimodal MRI completion | Synthesis loss, reconstruction loss, adversarial loss, | – |
| AutoSyncoder (2020) [129] | GAN，VAE | Multimodal MRI completion | Adversarial loss, negative log-likelihood loss | – |
| I2I-Mamba (2024) [232] | Mamba | Multimodal MRI completion | Adversarial loss, pixel-wise loss | √ |
| ResViT (2022) [235] | Transformer | Multimodal MRI completion | L1 loss, adversarial loss | √ |
| MMT (2023) [236] | Transformer | Multimodal MRI completion | Synthesis loss, reconstruction loss, adversarial loss, | – |
| FgC2F-UDiff (2024) [239] | Diffusion model | Multimodal MRI completion | Diffusion denoising loss | √ |

| Anatomically-Guided Image Synthesis | | | | |
|---|---|---|---|---|
| CG-SAMR (2021) [243] | GAN | Anatomy-guided CT synthesis | Adversarial loss, confidence map loss,feature matching loss, shape consistency loss | √ |
| Shen et al. (2023) [244] | GAN | Anatomy-guided CXR synthesis | Reconstruction loss, Perceptual loss, Adversarial loss | – |
| Hou et al. (2023) [245] | GAN | Anatomy-guided fundus image synthesis | Wasserstein GAN loss, feature matching loss, KL-loss | – |
| Real-ESRGAN (2024) [246] | GAN | Anatomy-guided fundus image synthesis | Adversarial loss, perceptual loss, L1 loss, L1_seg loss | – |
| LN-Gen (2024) [247] | Diffusion model | Anatomy-guided rectal lymph nodes synthesis | Diffusion denoising loss, adapter loss | √ |
| SegGuidedDiff (2024) [248] | Diffusion model | Anatomy-guided MRI synthesis | Diffusion denoising loss | √ |

## S4.3. Treatment Phase: Enabling Precision Interventions

In the treatment phase of clinical care, the integration of generative AI into radiotherapy and intraoperative navigation offers transformative potential for precision medicine. By modeling complex anatomical variations, capturing physiological motion, and supporting real-time clinical decision-making, generative models are increasingly bridging the gap between static preoperative imaging and dynamic, adaptive interventions. This section explores two key areas: dose prediction and planning in radiotherapy, and dynamic image synthesis for intraoperative navigation.

### *S4.3.1. Generation for treatment planning*

In the clinical phase of radiotherapy, achieving precise intervention not only enhances treatment efficacy but also minimizes damage to normal tissues. However, this process is challenged by several factors. Significant anatomical variations among patients, along with the motion of tumors and adjacent organs, introduce uncertainty in localization. Moreover, limitations in image quality and acquisition speed have hindered the real-time updating and optimization of treatment plans [250,251]. Generative AI models are increasingly employed to synthesize individualized dose maps and simulate anatomy in support of adaptive and personalized treatment.

Early models such as Cascade 3D U-Net introduced multi-scale CT feature fusion to model dose gradients in complex regions like the head and neck[252], while DoseNet applied a fully convolutional network to generate 3D dose distributions with high computational efficiency[253]. However, these models remained limited in their ability to capture long-range anatomical dependencies, such as the spatial relationships between pelvic tumors and neighboring organs. To address this, LSTM-based architectures[254] introduced sequence modeling for spatial continuity, and TransDose[255] combined transformer-based global feature extraction with super pixel graph convolution to enhance dose conformity around critical structures like the spinal cord. Further refinement came with SP-DiffDose[256], which fused Swin Transformer features with a projection network to improve local dose gradients in anatomically ambiguous regions, such as in pancreatic cancer. DiffDP[257] used conditional diffusion on CT and segmentation inputs to generate multiple plausible dose distributions, supporting flexible planning in high-risk cases (e.g., lung tumors near the heart). Leveraging the Mamba architecture, MD-dose[258] achieved faster sampling and more accurate predictions to support real-time adaptive radiotherapy. Concurrently, generative models have been extended to imaging tasks: self-improving foundation models[259], volumetric image generation[260], and CBCT synthesis for real-time tracking[261] have been used to augment input data and support decision-making. Applications including patient-specific simulation[250], heterogeneous tumor modeling[251], and conditional brain tumor generation[262] further highlighted the promise of multimodal fusion and deep generative learning in clinical radiotherapy workflows.

In summary, these dose prediction techniques offer significant advantages in terms of improving predictive accuracy, reducing trial-and-error costs, and optimizing treatment plans, thereby providing robust data support for personalized radiotherapy. Moreover, this body of research lays the foundation for subsequent advancements in real-time dynamic image synthesis and intraoperative navigation, which will be addressed in the following section.

### *S4.3.2. Intraoperative navigation: dynamic image synthesis*

Intraoperative navigation benefits from dynamic image synthesis, which aims to generate real-time, patient-specific visualizations that reflect both anatomical structure and physiological motion. These techniques support tasks such as cardiovascular function monitoring and radiotherapy adaptation by modeling temporal dynamics and suppressing motion artifacts.

In 2D+t synthesis, early research focused on generating dynamic sequences from sparse or single-frame data. For instance, a super-resolution GAN[263] accelerated cardiac MRI generation while preserving phase-specific morphology. DragNet[264] used deformable registration to reconstruct full cardiac cycles from static input, mitigating motion blur from sparse sampling. Video diffusion models[265] improved structural consistency in dynamic echocardiography, while HeartBeat[218] integrated ECG signals with hemodynamic features to produce personalized cardiac motion, illustrating the value of multimodal conditioning. Building on these efforts, a cascaded video diffusion model[217] introduced hierarchical refinement of motion and texture using semantic and anatomical cues, enhancing both temporal smoothness and spatial fidelity. Endora[266], a diffusion-based framework designed for dynamic medical procedures like endoscopy, exemplified the broader potential of generative models in medical video synthesis. By leveraging procedural priors and domain-specific context, it offered insights that may be applicable to advancing 2D+t generation techniques. Collectively, these models illustrate a shift toward semantically guided, feature-aware generation, enhancing the realism and controllability of dynamic synthesis. However, current 2D+t methods still face notable limitations in capturing complex out-of-plane motion, maintaining long-range temporal coherence, and modeling full volumetric dynamics from limited spatial views.

In 3D+t (4D) synthesis, methods focus on constructing evolving volumetric sequences from incomplete or low-resolution inputs. The temporally aware 3D GAN framework proposed in[267] combined respiratory motion compensation with dynamic cardiac MRI reconstruction, effectively capturing continuous cardiac motion and substantially reducing respiratory-induced artifacts. Regarding 4D cardiac MRI generation, most existing approaches[182–185] predicted deformation fields for the initial and final frames and interpolated these fields to generate intermediate frames; however, such methods were limited in handling large-scale motion. Newer approaches, such as 4D CT synthesis from sparse CBCT[268] and cross-modal CBCT-to-MRI translation[269] have respectively generated 4D synthetic CT from sparse-view CBCT and achieved cross-modal dynamic synthesis from 4D CBCT to 4D MRI, effectively addressing challenges in complex nonlinear motion mapping. Furthermore, the deep prior image-constrained motion compensation framework (DPI-MoCo)[270] incorporated a motion trajectory prediction module into 4D CBCT reconstruction, significantly reducing motion distortions. Similarly, another work[271] employed a feature disentanglement mechanism to extract differential features across multiple 3D/4D MRI sequences, achieving remarkable performance in quantitative tasks such as myocardial perfusion assessment.

Although these various strategies—including registration, GANs, and diffusion models have shown promising initial results in modeling spatiotemporal continuity and suppressing motion artifacts, they still face significant challenges. In particular, limitations in real-time performance, generalization to complex motion patterns, and cross-modal data consistency continue to restrict their clinical applicability and robustness. Notably, the text-driven 4D cardiac cine MRI synthesis method[272] introduces disease description texts as prior information to enable controlled synthesis of pathology-specific motion patterns. This innovative approach not only enhances the semantic accuracy and pathological specificity of the synthesized images, but also offers a promising pathway to overcome current limitations, potentially advancing intelligent diagnostics and precision therapy.

**Table S7**. Summary of publications on treatment phase: enabling precision interventions.

| Publication (Year) | Model | Application | Loss Function | Link |
|---|---|---|---|---|
| **Generation for Treatment Planning** | | | | |
| DoseNet (2018) [253] | CNN | Radiation dose prediction | L2 loss | √ |
| C3D (2021) [252] | CNN | Radiation dose prediction | L1 loss | √ |
| Radonic et al. (2024) [254] | CNN | Radiation dose prediction | MSE loss | – |
| TransDose (2023) [255] | Transformer | Radiation dose prediction | Cross entropy loss，Charbonnier Loss | – |
| VQGAN_TATrans (2024) [262] | GAN，VAE，Transformer | Brain tumor prediction | Pixel differences loss，perceptual loss，feature matching loss，gradient loss，codebook loss | √ |
| PC-DDPM (2024) [261] | Diffusion model | Real-time tumor tracking | Diffusion denoising loss，cycle-consistency loss | – |
| DiffDP (2023) [257] | Diffusion model | Radiation dose prediction | Diffusion denoising loss | √ |
| SP-DiffDose (2023) [256] | Diffusion model，Transformer | Radiation dose prediction | Diffusion denoising loss | – |
| MD-Dose (2024) [258] | Diffusion model，Mamba | Radiation dose prediction | Diffusion denoising loss | √ |
| **Intraoperative navigation: Dynamic image synthesis** | | | | |
| SVIN (2020) [182] | CNN | 4D dynamic MRI synthesis | Similarity loss, smoothness regularization loss, regression loss | √ |
| DragNet (2023) [264] | CNN | 2Dt cardiac MR synthesis | ELBO loss, KL loss, similarity loss | – |
| Quintero et al. (2024) [269] | CNN | 4D dynamic MRI synthesis | RMSE loss | – |
| MPVF (2023) [183] | CNN, Transformer | 4D dynamic MRI synthesis | Charbonnier loss | √ |
| UVI-Net (2024) [185] | CNN, Transformer | 4D dynamic MRI synthesis | NCC loss, gradient loss | √ |
| TAV-GAN (2021) [267] | GAN | 4D dynamic MRI synthesis | Temporally aware loss, SSIM loss, L1 loss | – |
| Thummerer et al. (2022) [268] | GAN | 4D CT synthesis | MSE loss | – |
| REGAIN (2023) [263] | GAN | 2Dt cardiac MRI enhancement | L1 fast-Fourier transform loss | – |
| Seq2Seq (2024) [271] | GAN | 3D/4D MRI synthesis | L1 loss, perceptual loss, adversarial loss, cycle-consistent loss | √ |
| DPI-MoCo (2024) [270] | GAN | 4D CBCT reconstruction | MSE loss, GAN loss, NCC loss, smooth loss | – |
| DDM (2022) [184] | Diffusion model | 4D dynamic MRI synthesis | Diffusion denoising loss, NCC loss, KL loss | √ |
| Reynaud et al. (2023) [217] | Diffusion model, Transformer | Echocardiography video synthesis | Diffusion denoising loss | √ |
| HeartBeat (2024) [218] | Diffusion model, VAE | Echocardiography video synthesis | Diffusion denoising loss | – |
| Endora (2024) [266] | Diffusion model, Transformer | Endoscopy video synthesis | Diffusion denoising loss | √ |

## S4.4. Prognosis Phase: Longitudinal & Personalized Medicine

Generative medical imaging techniques have demonstrated significant clinical potential in longitudinal prognostic analysis and personalized medicine. By leveraging deep modeling of patients' multi-temporal imaging data, these approaches can simulate dynamic disease progression, predict tissue degenerative changes, and quantify prognostic risk, thereby providing data-driven support for clinical decision-making.

### *S4.4.1. Tumor growth simulation and treatment response prediction*

In recent years, the integration of multimodal imaging data (e.g., MRI, CT) with patient-specific biomarkers (such as EGFR mutation status) has established a novel data foundation for predicting tumor growth and treatment response. For instance, a study[219] introduced a treatment-aware diffusion probabilistic model that simulated the 3D growth patterns and invasive behavior of gliomas using longitudinal MRI and molecular pathology data, boosting future tumor prediction accuracy (DSC from 0.556 to 0.719; +16.3%). Similarly, SADM[96] adopted a novel design that enabled learning of longitudinal dependencies even in the presence of missing data during training, and supported autoregressive generation of image sequences during inference. Another model[273] presented a universal tumor synthesis framework that fused cross-modal data to generate high-quality synthetic tumors, improving sensitivity to texture and morphological variations. In related work[274], radiomics features extracted from synthetic MRI enhanced glioblastoma survival prediction across multi-center datasets, supporting personalized radiotherapy planning. A patient-specific deep learning framework[275] presented a patient-specific deep learning framework for real-time, label-free tumor tracking, enabling non-invasive monitoring of treatment response. Complementing this, the cross-tumor CT foundation model[276] established a unified interpretive platform across cancer types, facilitating both treatment decision-making and prognostic evaluation. Together, these approaches demonstrate how generative models can effectively simulate tumor dynamics and predict individualized treatment outcomes, advancing the development of scalable, multimodal tools for personalized oncology.

### *S4.4.2. Spatiotemporal modeling of neurodegenerative disease progression*

For neurodegenerative diseases such as Alzheimer's disease (AD), tracking structural brain changes over time is critical for early diagnosis, disease staging, and therapeutic planning. Longitudinal MRI synthesis techniques have emerged as valuable tools for capturing subtle, progressive degeneration across brain regions, enabling precise spatiotemporal modeling of disease trajectories[277,278]. To address limitations in data availability and temporal resolution, a work[279] proposed a hybrid deep learning framework that integrated DCGAN and SRGAN to generate synthetic MRI sequences corresponding to different AD stages. This method achieved classification and prediction accuracies as high as 99.7%, demonstrating its potential in compensating for real-world data scarcity and improving progression staging. Building on this, TADM[280] incorporated a pre-trained Brain Age Estimator (BAE) to guide learning of intensity-based structural changes over time. By learning the distribution of inter-scan variations, TADM predicted future MRI volumes based on baseline scans. Compared to conventional approaches, it reduced mean brain volume error by 24% and improved similarity metrics by 4%, offering both anatomical accuracy and clinical reliability. These approaches not only provided a visual and quantitative representation of neurodegenerative progression, but also supported prognostic modeling for treatment response and optimal drug timing. As such, spatiotemporal generative modeling holds significant promise for advancing personalized, stage-aware intervention strategies in the management of AD and other neurodegenerative disorders.

### *S4.4.3. Translating multimodal generative prognostics into clinical practice*

A key challenge in prognostic modeling lies in the integration of diverse imaging modalities and clinical variables for robust risk stratification and treatment guidance. Early efforts leveraged generative models such as GANs to explore synthetic data-driven prognostic modeling. For example, a work[249] used a conditional GAN to generate synthetic cardiac aging images, aiding early diagnosis of diastolic dysfunction. Likewise, another work[281] employed a GAN-convolutional framework to predict long-term MRI changes, offering insights for aging-related prognostic assessment. In neurodegeneration, a new work[282] introduced a latent diffusion model that improved brain volume prediction in Alzheimer's patients by 22% and enhanced image similarity by 43%. In oncology, a self-evolving foundation model[259] was developed to improve HER2-positive breast cancer detection by 12–17% and increases EGFR mutation sensitivity, contributing to better patient stratification. For cerebrovascular prognosis, an end-to-end deep model[283] used synthetic CT to predict hematoma expansion, achieving an AUC of 0.91. In parallel, radiomics features derived from synthetic MRI[274] enhanced glioblastoma survival prediction across multi-center datasets, reinforcing the value of high-fidelity synthetic data in clinical radiotherapy planning. Collectively, these advances signal a paradigm shift toward generative, multimodal prognostics that move beyond static risk scores to dynamic, individualized disease forecasting. Bridging the gap to clinical practice will require not only technical improvements, such as domain adaptation and model interpretability, but also careful alignment with clinical workflows and decision-making requirements.

By capturing dynamic, multimodal disease trajectories, generative imaging models offer powerful tools for prognosis across tumor, neurological, and cardiovascular domains. Nonetheless, clinical translation at scale requires further work in domain adaptation, temporal modeling, and model interpretability. Future progress in these areas is expected to enhance robustness and generalizability across diverse clinical environments, reinforcing the role of generative models in precision medicine and personalized care.

**Table S8.** Summary of publications on prognosis phase: longitudinal & personalized medicine.

| Publication (Year) | Model | Application | Loss Function | Link |
|---|---|---|---|---|
| Moya-Sáez et al. (2022) [274] | CNN | Glioblastoma survival prediction | L1 loss | – |
| EfficientNet B0 (2024) [283] | CNN | Hematoma expansion prediction | Focal loss | √ |
| DaniNet (2019) [278] | GAN | Mimic disease progression | Biological constraints loss, Deformation loss | – |
| GP-GAN (2020) [277] | GAN | Brain tumor growth prediction | Adversarial loss, L1 loss, Dice loss | – |
| Song et al. (2023) [281] | GAN | Longitudinal MRI prediction | Adversarial loss, Binary cross-entropy loss, Gradient difference loss | – |
| DCGAN and SRGAN (2024) [279] | GAN | Alzheimer's disease progression | Adversarial loss, MSE loss, VGG Loss | – |
| TADM (2024) [280] | Diffusion model | Brain neurodegenerative prediction | Diffusion denoising loss | √ |
| BrLP (2024) [282] | Diffusion model | Disease progression prediction | Diffusion denoising loss | √ |
| DiffTumor (2024) [273] | Diffusion model，VAE | Generalizable tumor synthesis | Diffusion denoising loss | √ |
| PASTA (2025) [276] | Diffusion model，VAE | Tumor synthesis Foundation model | Diffusion denoising loss | √ |
| SADM (2023) [96] | Diffusion model, AR | Longitudinal MRI Generation | Diffusion denoising loss | √ |
| TaDiff (2025) [219] | Diffusion model | Longitudinal MRI Generation and Glioma Growth Prediction | Diffusion denoising loss | √ |

# S5. More Details on Overview of Public Datasets

The advancement of generative AI in medical imaging is closely linked to the availability of large-scale, high-quality, and multi-modal public datasets. These datasets not only provide essential resources for model training but also serve as standardized benchmarks for evaluating generalization, fidelity, and clinical applicability across diagnostic, therapeutic, and prognostic tasks. Comprehensive repositories such as the UK Biobank[284] offer diverse imaging modalities, including MRI, CT, ultrasound, and fundus photography, collected from over 500,000 participants. The Cancer Imaging Archive (TCIA)[285] provides expertly annotated CT, MRI, and PET scans across a wide range of tumor types. Although originally developed for tasks like classification, segmentation, and reconstruction, these datasets now underpin a wide range of generative applications, including image synthesis, modality translation, and anomaly simulation. In addition, open platforms such as Grand Challenge and Kaggle facilitate standardized access to datasets spanning X-ray, histopathology, cardiac MRI, and endoscopy. These platforms support reproducible benchmarking for generative models across tasks such as 2D/3D image generation, report synthesis, and multi-modal alignment. Overall, public datasets constitute the foundation of generative modeling in medical imaging. This section provides an overview of widely used datasets in the field, as summarized in Table S9.

***Organ- and region-specific CT and MRI dataset Datasets.*** Multi-organ and whole-body imaging datasets form the backbone of generative modeling across diverse anatomical regions. DeepLesion[286] and ULS[287] provide extensive lesion annotations and 3D whole-body CT volumes, supporting generative tasks such as lesion synthesis, multi-organ reconstruction, and cross-modality translation. Notably, the recently released PreCT-160K[288] dataset represents the largest known CT imaging corpus to date, with 160,000 CT scans spanning a wide range of anatomical sites and clinical conditions, offering unprecedented scale for training high-capacity generative models. The TotalSegmentator series[289,290] further provides fine-grained segmentation across over 10 anatomical structures, enabling anatomically faithful synthesis and training of anatomically guided generative models. Additionally, AutoPET and AutoPETIII[291] , based on paired PET-CT imaging, further support hybrid synthesis and modality fusion for tumor representation learning.

For head and neck imaging, datasets such as INSTANCE2022[292], SegRap2023[293], and HECKTOR2022[294] offer PET-CT and CT volumes with expert tumor and organ-at-risk (OAR) annotations, making them ideal for training cross-modality synthesis and radiotherapy planning models. In neuroimaging, collections like BraTS21[295], BraTS-MEN[296], fastMRI_Brain[297], IXI[298], and AOMIC[299] provide volumetric brain MRIs for brain tumor synthesis, progression modeling, and longitudinal prediction. Cardiac imaging is supported by datasets such as ACDC[300], M&Ms[301],[302], OCMR[303], and EchoNet-Dynamic[217], which contain both 2Dt and 3D MRI as well as ultrasound videos. These are crucial for developing dynamic generative models (e.g., 4D cine MRI synthesis, cardiac motion prediction) and dose adaptation frameworks in cardiovascular interventions. Abdominal datasets including FLARE[304,305] , AbdomenCT-1K[306], and PI-CAI[307] offer high-resolution CT/MRI volumes annotated for multi-organ segmentation, supporting synthetic image augmentation and simulation of disease-specific anatomical changes. These datasets provide foundational support for generative tasks like low-dose reconstruction, anatomical synthesis, and cross-organ correlation modeling.

***Pathology and High-resolution Microscopy.*** In the domain of computational pathology, datasets like CPIA[308], PatchCamelyon[309], BreakHis[310], NAFLD[311], and MIST-HER2[312] encompass millions of whole-slide images (WSIs), supporting self-supervised pretraining and high-resolution image generation. These datasets are crucial for exploring style-transfer synthesis (e.g., H&E to IHC), anomaly simulation, and histological progression modeling in cancer[313].

***Ultrasound, OCT, and Fundus Datasets.*** Ultrasound imaging, known for its portability and real-time capabilities yet susceptibility to noise, benefits from datasets such as TG3K[314], EchoNet-LVH[315], and EndoSLAM[316], which support generative tasks including denoising, quality enhancement, and anatomical segmentation. In ophthalmic imaging, OCT datasets like OCT2017[317] and Retinal OCT-C8[318] enable super-resolution, layer segmentation, and disease-specific synthesis. Additionally, fundus image datasets such as ODIR-5Kc, LAG[319], and AIROGS[320] provide large-scale, annotated images for generative modeling tasks like lesion synthesis, cross-disease domain translation, and image-to-report generation, further expanding the applicability of generative models in ocular disease analysis.

***Multimodal Image–Text Datasets.*** The integration of visual and textual modalities is becoming increasingly essential in generative medical imaging, particularly for tasks such as radiology report generation, text-conditioned synthesis, and multimodal retrieval. A variety of large-scale datasets have emerged to support these applications. In chest imaging, resources like CheXpertPlus[321], Medical-CXR-VQA[322], and PadChest[323] facilitate vision-language pretraining and VQA-style tasks. In pathology, datasets such as Quilt-1M[324] and OpenPath[325] provide over one million image–text pairs derived from clinical annotations and social media sources. More general-purpose datasets like Medtrinity-25M[326], MedICaT [327] and PMC-OA [328] offer paired images and captions from biomedical literature, widely used in training foundation models for captioning and contrastive learning. Additionally, multimodal datasets such as Duke Breast Cancer MRI[329] and I-SPY2[330] enable radiogenomic modeling by linking imaging features with genomic and clinical data.

As generative models in medical imaging continue to evolve, the importance of high-quality, well-curated training data has become increasingly evident. Public datasets play a central role in this progress by enabling model development, performance evaluation, and clinical validation across diverse modalities, anatomical regions, and disease types. While these datasets have contributed to improvements in robustness and generalizability, several challenges persist, such as domain discrepancies, annotation inconsistencies, and the limited availability of dynamic or longitudinal imaging data. Addressing these issues through improved standardization, collaborative curation, and the responsible integration of synthetic data will be key to supporting the reliable and clinically meaningful deployment of generative models in real-world settings.

**Table S9**. The publicly available datasets.

| Dataset | Modalities | Scale | Link |
|---|---|---|---|
| **Whole body** | | | |
| MedMNIST [331] | 2D & 3D | 708K 2D images & 10K 3D volumes | √ |
| DeepLesion[286] | 2D CT | 32.7K images | √ |
| ULS[287] | 3D CT | 38.8K volumes | √ |
| PreCT-160K[288] | 3D CT | 160K volumes | √ |
| FLARE24 Task1[332] | 3D CT | 10K volumes | √ |
| CT-ORG[333] | 3D CT | 140 volumes | √ |
| TotalSegmentator[289] | 3D CT | 1204 volumes | √ |
| TotalSegmentator v2[289] | 3D CT | 1228 volumes | √ |
| AutoPET[291] | 3D PET-CT | 1214 volumes | √ |
| AutoPETIII[291] | 3D PET-CT | 1614 volumes | √ |
| TotalSegmentator MRI[290] | 3D MRI | 298 volumes | √ |
| TotalSegmentator MRI v2[290] | 3D MRI | 616 volumes | √ |
| CPIA[308] | Pathology | 21.4M WSI | √ |
| NAFLD[311] | Pathology | 119.8K WSI | √ |
| **Head and Neck** | | | |
| INSTANCE2022[292] | 3D CT | 200 volumes | √ |
| SegRap2023[293] | 3D CT | 200 volumes | √ |
| HECKTOR2022[294] | 3D PET-CT | 882 volumes | √ |
| fastMRI_Brain[297] | 2D MRI | 6970 images | √ |
| IXI Dataset[298] | 3D MRI | 600 volumes | √ |
| AOMIC[299] | 3D MRI | 1370 volumes | √ |
| BraTS21[295] | 3D MRI | 2040 volumes | √ |
| BraTS2023-MEN[296] | 3D MRI | 1650 volumes | √ |
| CrossMoDA2021[334] | 3D MRI | 349 volumes | √ |
| CrossMoDA2023[334] | 3D MRI | 983 volumes | √ |
| TG3K[314] | 2D US | 3585 images | √ |
| TN3K[314] | 2D US | 3494 images | √ |
| TN-SCUI2020[335] | 2D US | 4554 images | √ |
| Ultrasound Nerve Segmentation[336] | 2D US | 11.1K images | √ |
| OCT2017[317] | OCT | 35.1K images | √ |
| Retinal OCT-C8[318] | OCT | 24K images | √ |
| ODIR-5K[337] | Fundus | 5000 images | √ |
| LAG[319] | Fundus | 11.7K images | √ |
| Diabetic Retinopathy Arranged[338] | Fundus | 35.1K images | √ |
| AIROGS[320] | Fundus | 101.4K images | √ |
| PatchCamelyon[309] | Pathology | 327.7K WSI | √ |
| OSCC[339] | Pathology | 1224 WSI | √ |
| **Chest** | | | |
| CheXchoNet[340] | 2D X-ray | 71.6K images | √ |
| BRAX[341] | 2D X-ray | 40.9K images | √ |
| SIIM-FISABIO-RSNA COVID-19[342] | 2D X-ray | 7597 images | √ |
| LIDC-IDRI[343] | 2D CT | 1010 images | √ |
| SARS-COV-2 Ct-Scan[344] | 2D CT | 2482 images | √ |
| ATM22[345] | 3D CT | 500 volumes | √ |
| LUNA16[346] | 3D CT | 888 volumes | √ |
| LNQ2023[347] | 3D CT | 513 volumes | √ |
| fastMRI_Breast[297] | 2D MRI | 300 images | √ |
| ISPY1-Tumor-SEG-Radiomics[348] | 3D MRI | 483 volumes | √ |
| ACRIN-Contralateral-Breast-MR[349] | 3D MRI | 984 volumes | √ |
| BUSI[350] | US | 780 images | √ |
| TDSC-ABUS2023[351] | US | 200 volumes | √ |
| Breakhis[310] | Pathology | 7909 WSI | √ |
| WSSS4LUAD[352] | Pathology | 10K WSI | √ |
| MIST-HER2[312] | Pathology | 22.7K WSI | √ |
| **Cardiac** | | | |
| ACDC[300] | 3D MRI | 150 volumes | √ |
| MICCAI 2024 CARE LAScarQS++[353] | 3D MRI | 194 volumes | √ |
| M&Ms Challenge[301] | 3D MRI | 375 volumes | √ |
| M&Ms-2 Challenge[302] | 3D MRI | 360 volumes | √ |
| OCMR[303] | 2Dt MRI | 165 volumes | √ |
| Harvard Cardiac MR Center Dataverse[354] | 2Dt MRI | 108 volumes | √ |
| CMRxRecon[355] | 2Dt MRI | 300 volumes | √ |
| Cardiac MRI Dataset[356] | 2Dt MRI | 7980 volumes | √ |
| Cardiac super-resolution label maps[357] | 2Dt MRI | 1331 volumes | √ |
| EchoNet-Dynamic[217] | US | 10K videos | √ |
| GANcMRI[358] | US | 45.5K videos | √ |
| EchoNet-LVH[315] | US | 12K videos | √ |
| EchoNet-Dynamic[359] | US | 10K videos | √ |
| **Abdomen** | | | |
| FLARE2022[304] | 3D CT | 2300 volumes | √ |
| FLARE2023[305] | 3D CT | 4500 volumes | √ |
| AbdomenCT-1K[306] | 3D CT | 1112 volumes | √ |
| AbdomenAtlas 1.0 Mini[360] | 3D CT | 5195 volumes | √ |
| UW-Madison GI Tract Image [361] | 2D MRI | 38.5K images | √ |
| PI-CAI[307] | 3D MRI | 1500 volumes | √ |
| FLARE 2024 Task3[362] | 3D MRI | 4817 volumes | √ |
| ISBI 2025 FUGC[363] | US | 890 images | √ |
| LIMUC[364] | US | 1043 videos | √ |
| SUN[365] | US | 1018 videos | √ |
| EndoSLAM[316] | US | 1020 videos | √ |
| RenalCell[366] | Pathology | 625.1K WSI | √ |
| GasHisSDB[367] | Pathology | 245.2K images | √ |
| SegPANDA200[368] | Pathology | 100.9K images | √ |
| NCT-CRC-HE[369] | Pathology | 100K WSI | √ |
| **Others** | | | |
| SPIDER (Spine)[370] | 3D MRI | 257 volumes | √ |
| Wrist Dataset[371] | 3Dt MRI | 55 volumes | √ |
| fastMRI_knee[297] | 2D MRI | 1398 images | √ |
| SKM-TEA (Knee)[372] | 3D MRI | 155 volumes | √ |
| MRNet (Knee)[200] | 3D MRI | 1370 volumes | √ |
| **Multimodal Dataset** | | | |
| Medical-CXR-VQA[322] | X-ray-Text | 377K images, 780K texts | √ |
| PadChest[323] | X-ray-Text | 160K images, 109K texts | √ |
| CheXpertPlus[321] | X-ray-Text | 223K images, 223K texts | √ |
| Quilt-1M[324] | Pathology-Text | 1M images, 1M texts | √ |
| OpenPath[325] | Pathology-Text | 208K images, 208K texts | √ |
| Medtrinity-25M[326] | Multimodal Images-Text | 25M images, 25M texts | √ |
| PMC-OA[328] | Multimodal Images-Text | 1.6M images, 1.6M texts | √ |
| MedICaT[327] | Multimodal Images-Text | 217K images, 217K texts | √ |
| Duke Breast Cancer MRI[329] | Genomic&MRI-Clinical data | 922 cases | √ |
| I-SPY2[330] | MRI-Clinical data | 719 cases | √ |

# S6. More Details on Evaluation Methods for Generative Models in Medical Imaging

Here we present more details on each of the evaluation methods that we introduced in the main paper.

## S6.1. Low-Level Evaluation: Pixel Fidelity

Low-level evaluation metrics focus on quantifying pixel-wise similarity between the generated image and the ground truth. They are most effective for tasks involving image reconstruction, denoising, or super-resolution, where spatial accuracy is essential.

***MSE / MAE / RMSE / PSNR***[373]***:*** Mean Squared Error (MSE) and Mean Absolute Error (MAE) are classical pixel-wise metrics that quantify the average intensity difference between generated and reference images. MSE applies a quadratic penalty, making it more sensitive to outliers, whereas MAE applies a uniform linear penalty, offering more robustness to extreme errors. Root Mean Squared Error (RMSE), the square root of MSE, restores the original intensity unit, improving interpretability. Peak Signal-to-Noise Ratio (PSNR), derived from MSE, expresses the logarithmic ratio between the maximum possible signal intensity and the power of corrupting noise. While these metrics are computationally simple and widely used in image restoration tasks such as denoising or super-resolution, they correlate poorly with human visual perception and often fail to reflect structural or semantic integrity.

***SSIM / MS-SSIM / FSIM / IW-SSIM:*** The Structural Similarity Index (SSIM)[374] addresses some of the shortcomings of pixel-based metrics by incorporating human visual perception principles. It evaluates image similarity by analyzing luminance, contrast, and structural components within local regions. Multi-Scale SSIM (MS-SSIM)[375] extends this concept by integrating information across multiple scales, improving sensitivity to global structural consistency and content variations. The Feature Similarity Index (FSIM)[376] enhances structural assessment by leveraging phase congruency and gradient magnitude, which are particularly effective for detecting edges and preserving texture, both of which are critical in medical imaging. Information Content Weighted SSIM (IW-SSIM)[377] incorporates a weighting scheme based on local information content, assigning greater importance to regions with higher perceptual or diagnostic value. This enhances its ability to assess clinically relevant areas such as lesions or organ boundaries, addressing the uniform sensitivity limitation of standard ssim. These metrics better approximate human perception compared to traditional error-based metrics, but their focus on local structure makes them less effective in detecting high-level semantic inconsistencies or anatomical implausibility.

***VIF / UQI / CACI****:* Visual Information Fidelity (VIF)[378] evaluates image quality using an information-theoretic model of the human visual system, estimating how much visual information is preserved in a distorted image relative to a reference. This offers a principled assessment of signal degradation, though at higher computational cost. Universal Quality Index (UQI)[379], in contrast, is a statistically-driven metric that integrates luminance, contrast, and structural similarity into a unified score. Both metrics provide a broader perspective on image degradation than pixel-based measures but are limited in capturing high-level semantic consistency. Complementing these, the Conservation and Correction Index (CACI)[380] jointly assesses structural preservation in healthy regions and effective correction in pathological areas by combining SSIM with lesion-based segmentation masks. These metric bridges low-level fidelity and task-specific evaluation, offering greater relevance in medical imaging scenarios that require both anatomical accuracy and pathology removal.

These metrics provide a useful assessment of structural integrity and visual fidelity, and are particularly effective in tasks such as image denoising, reconstruction, and compression. However, their reliance on low-level features limits their ability to detect high-level semantic inconsistencies, anatomical implausibility, or clinical irrelevance—factors that are critical in evaluating the diagnostic utility of generative models in medical imaging.

## S6.2. Mid-Level Evaluation: Feature and Distribution Consistency

Mid-level evaluation metrics assess the similarity between real and generated samples in a high-dimensional feature space, offering a more semantically informed and perceptually grounded perspective than pixel-based metrics. These methods typically rely on pretrained deep neural networks or kernel-based statistical comparisons to capture global distribution alignment, image realism, and sample diversity. They are especially valuable in tasks such as modality translation, unpaired image synthesis, and dynamic image generation.

***FID / KID / MMD / IS:*** Fréchet Inception Distance (FID)[381] compares real and generated image distributions by extracting deep features from a pretrained Inception-V3[382] network and modeling them as multivariate gaussians. It computes the wasserstein-2 distance[383] between these distributions, capturing both image quality and diversity. Kernel Inception Distance (KID)[384] improves robustness by employing polynomial-kernel-based maximum mean discrepancy, which avoids Gaussian assumptions and performs effectively even with limited data. Maximum Mean Discrepancy (MMD)[385] is a mid-level statistical metric that measures the distance between the feature distributions of real and generated images. It is commonly used as an alternative to FID when Gaussian assumptions are not desired, and forms the basis of kernel-based metrics such as KID. Inception Score (IS)[386] assesses image realism based on classification confidence and output diversity, calculated via Kullback–Leibler divergence. However, IS only evaluates individual images and depends heavily on the domain of the pretrained classifier. While these metrics are widely used, their reliance on natural-image-trained backbones (e.g., Inception V3[382] or VGG16[387]), limits their sensitivity to domain-specific structures and pathological variation, making them best suited for preliminary quality screening.

***LPIPS / CLIP Similarity / MedCLIP-score:*** To assess and mitigate hallucinations in generative medical imaging, where synthesized outputs may introduce or omit clinically important features, metrics such as Learned Perceptual Image Patch Similarity (LPIPS)[388], CLIP Similarity[389], and MedCLIP-score[106] have become essential tools. LPIPS evaluates perceptual similarity by comparing deep feature activations from pretrained convolutional networks, capturing structural coherence and visual realism across tasks such as anatomical inpainting and cross-modality synthesis. In comparison, CLIP Similarity and its domain-adapted variant MedCLIP-score embed both the generated image and its associated text prompt into a shared vision-language space using transformer-based encoders. The cosine similarity between the resulting embeddings reflects semantic alignment, which is particularly useful for detecting hallucinations when the generated content deviates from expected anatomical structures, pathological patterns, or spatial context. These metrics extend evaluation beyond pixel-level fidelity by capturing both visual and conceptual consistency. However, their effectiveness may depend on the domain specificity and robustness of the underlying pretrained models.

***RQI / AHI / BmU-I / BmU-V:*** To assess hallucination and semantic consistency in medical image generation, a set of complementary metrics has been proposed. The Restoration Quality Index (RQI)[380] evaluates low-level perceptual fidelity using LPIPS, measuring how closely restored images resemble healthy references in visual appearance. The Anomaly-to-Healthy Index (AHI)[380] assesses whether pathological images, once restored, align with the distribution of healthy data. Based on FID, AHI provides a statistical view of normalization effectiveness. At the semantic level, the Biomedical Understanding (BmU)[390] framework introduces a hallucination-aware evaluation strategy tailored for medical imaging. BmU-I applies a large language model to generate textual descriptions from image sequences, which are then compared with the original prompts using BERT-based[391] embedding similarity. This captures whether the generated content aligns with intended diagnostic meaning. BmU-V extends this approach to dynamic imaging by using video-language models such as Video-LLaMA[392] to evaluate consistency between temporal visual content and associated text. It is particularly suited to applications like cine MRI and surgical video generation. Collectively, these metrics provide a multi-level evaluation framework covering perceptual quality, statistical alignment, and semantic fidelity, enabling robust assessment of generative model reliability in clinical contexts.

***FVD / KVD / FVMD / VBench:*** Fréchet Video Distance (FVD)[393] extends FID to video by using 3D convolutional networks (e.g., I3D[394]) to extract spatiotemporal features, allowing the evaluation of both visual fidelity and motion consistency. Kernel Video Distance (KVD)[395] evaluates the similarity between real and generated videos by comparing their spatiotemporal features using a kernel-based approach. Fréchet Video Motion Distance (FVMD)[396] targets dynamic realism by tracking key points across frames, analyzing velocity and acceleration histograms, and comparing them using Fréchet distance. This is especially important in cine MRI, cardiac imaging, and moving organ simulations. VBench[397] offers a modular framework for video assessment, integrating pretrained models such as RAFT (motion)[398], MUSIQ (frame quality)[399], and ViCLIP (semantics)[400] with heuristic algorithms to evaluate spatial consistency, motion smoothness, dynamic degree, and temporal stylistic consistency. These metrics are crucial for dynamic imaging tasks but computationally intensive and often trained on non-medical data, potentially limiting clinical interpretability.

Mid-level evaluations bridge the gap between low-level pixel fidelity and high-level clinical relevance. While they offer more meaningful insight into perceptual and statistical realism, their effectiveness is still influenced by the domain alignment of the pretrained models used and the complexity of the evaluation task. In practice, these metrics are best used in combination with both low- and high-level assessments to ensure comprehensive validation of generative performance.

## S6.3. High-Level Evaluation: Expert and Clinical Assessment

High-level evaluation constitutes the final and most clinically significant stage in the assessment hierarchy of generative models for medical imaging. Unlike low- and mid-level metrics that emphasize pixel-level fidelity or feature-space alignment, this stage focuses on practical usability in real-world clinical workflows. It seeks to determine whether synthetic images can support essential medical tasks such as diagnosis, treatment planning, or disease monitoring. To achieve this, high-level evaluation combines two complementary approaches: expert assessment rooted in subjective diagnostic judgment and clinical validation based on downstream task performance, providing a comprehensive understanding of the model's clinical utility.

### S6.3.1. Expert Evaluation and Clinical Feedback

While quantitative metrics provide a baseline assessment of image fidelity and performance in downstream tasks, expert evaluation by radiologists and clinical specialists remains essential for determining the true clinical viability of generated medical images. These assessments offer nuanced insights into aspects such as diagnostic realism, anatomical consistency, and semantic plausibility, which are often beyond the reach of purely algorithmic evaluation. In these settings, clinicians are typically asked to blindly compare synthetic and real images, identify anatomical inaccuracies, assess lesion visibility, and judge whether an image meets diagnostic standards. This human-in-the-loop[401] approach helps uncover subtle issues such as unnatural tissue textures, inconsistent lesion morphology, or implausible anatomical deformations, which may not be penalized by numerical similarity metrics.

For example, in the GenerateCT study[65], a blinded evaluation was conducted in which two radiologists with 4 and 11 years of experience, respectively, independently reviewed a total of 200 chest CT volumes. The dataset included 100 real and 100 synthetic cases. They were tasked with identifying whether each image was real or synthetic, and assessing the semantic alignment between the CT volumes and their corresponding radiological prompts. Despite knowing that half the cases were synthetic, both radiologists misclassified over half of the synthetic volumes (59–64%) and a significant portion of real ones (26–29%), demonstrating the high realism of the generated data. Furthermore, 70% of synthetic volumes were judged to accurately match the prompts. This evaluation design highlights the ability of experts to uncover diagnostic realism and semantic integrity that are not captured by automated metrics. A more comprehensive expert-involved framework was proposed in MINIM, a generalist medical image–text generative model. In this study, clinicians were asked to evaluate synthetic images across four modalities (OCT, fundus, chest CT, and chest X-ray) by scoring them on a 1–3 scale: 1). from low quality, 2).to high quality but misaligned with the prompt, 3).to high quality and semantically aligned. Three rounds of evaluations were conducted. Initially, only 70.75% of images received a top score of 3. These clinician ratings were then used to train a reinforcement learning from human feedback (RLHF) reward model, which guided the model to self-improve in a closed-loop fashion. By the third round, 89.25% of synthetic images were rated 3, with significant improvements across all modalities (e.g., chest CT improved from 61% to 83%)[259]. This study exemplifies a dynamic and iterative human-in-the-loop framework where expert judgment not only evaluates but actively enhances the generative model's performance.

### S6.3.2. Clinical Validation via Downstream Tasks

Clinical evaluation focuses on verifying whether synthetic images retain essential anatomical and pathological features required for medical decision-making. A common strategy is to apply the generated data to downstream tasks such as segmentation, classification, or regression, and assess how well the models perform when trained or tested using the synthetic inputs. These tasks serve as indirect clinical proxies, reflecting whether the generated content contains sufficient and relevant clinical information.

***Segmentation tasks*** are used to assess whether synthetic images preserve anatomical fidelity that is essential for clinical interpretation. By training or evaluating segmentation models on synthetic data and comparing the results to real-image baselines, researchers can determine if the generated images maintain clear structural boundaries. Metrics such as the dice coefficient and Intersection over Union (IoU) quantify spatial alignment between predictions and expert annotations. For example, in the Med-DDPM[402] study, synthetic multi-modal MRIs were used to train a tumor segmentation model. The results indicate that the synthetic images preserved key anatomical features with sufficient accuracy to support downstream clinical tasks.

***Classification tasks*** evaluate whether synthetic images encode meaningful diagnostic signals that allow models to distinguish between disease subtypes or molecular phenotypes. Metrics such as accuracy, area under the curve (AUC), and sensitivity are typically used to quantify classification performance. A representative example is the MINIM framework[259], which showed that synthetic data can enhance downstream diagnostic accuracy. Incorporating synthetic breast cancer images into the training set increased HER2-positive tumor classification accuracy from 79.2% to 94.0%. For EGFR mutation prediction in lung cancer, adding synthetic images improved accuracy from 81.5% to 95.4%. These gains were most notable in underrepresented subgroups, suggesting that the generated data not only preserved diagnostic features but also improved model generalization in clinically relevant settings.

***Regression tasks*** help assess whether synthetic images retain continuous clinical attributes relevant to physiological function or disease progression. This type of evaluation is especially important in dynamic imaging, where visual plausibility must be supported by measurable clinical signals. In a study[217], synthetic echocardiogram sequences were used to augment training data for left ventricular ejection fraction prediction, a key indicator of cardiac performance. A regression model trained on 790 real samples achieved an $R^2$ score of 56%. When the training set was rebalanced by incorporating approximately 50% synthetic data to compensate for underrepresented LVEF ranges, the model's performance improved to an $R^2$ of 59% on a balanced validation set. These findings indicate that the generated images preserved physiologically meaningful variation and supported more accurate functional prediction, particularly in limited-data scenarios.

These task-driven evaluations demonstrate that synthetic medical images can effectively support a range of clinically relevant applications. By preserving structural, diagnostic, and physiological features, generated data enable reliable model performance across segmentation, classification, and regression tasks. Such results highlight the potential of generative models to enhance clinical workflows, particularly in scenarios with limited or imbalanced data availability.

# S7. More Details on The Emergence of Multimodal Foundation Models

Recently, foundation models (FMs) have been transforming medical image generation by enabling unified and transferable solutions across the clinical continuum. Pre-trained on large, diverse datasets, they exhibit strong generalization and zero-shot capabilities, making them effective across multi-modal and multi-stage imaging tasks. Unlike conventional models restricted to narrow objectives, FMs provide a flexible backbone for image synthesis, enhancement, and interpretation spanning diagnostic, therapeutic, and prognostic needs.

### *S7.1.1. Modality-specific foundation models*

Recent advances in modality-specific foundation models have significantly improved the quality and clinical utility of generative outputs across different medical imaging domains. For computed tomography (CT), MedDiff-FM[403] exemplified the potential of diffusion-based architectures, generating high-resolution CT volumes under varying acquisition conditions. This model supported diagnostic interpretation and radiotherapy planning through robust anatomical completion and denoising capabilities. Focusing on magnetic resonance imaging (MRI), BME-X[404] presented an MRI-specific foundation model designed to enhance low-field image quality. Expanding on this concept, Triad [405] introduced a task-driven foundation model trained on 3D MRI data, where segmentation, classification, and registration tasks were jointly optimized to learn a unified and anatomically structured representation space. By enforcing spatial and semantic consistency across tasks, the model enabled robust and interpretable organ-level analysis across diverse anatomical and pathological conditions. In the domain of ophthalmology, RETFound-DE[406] introduced a data-efficient foundation model specifically tailored for fundus imaging. Despite limited supervision during training, the model demonstrated robust generalization in diabetic retinopathy screening, underscoring the scalability and clinical relevance of specialized foundation models in resource-constrained settings. In computational pathology, BEPH[407] represented a foundation model trained on 11 million unlabeled whole-slide image (WSI) patches via self-supervised learning. The resulting representations generalized effectively to multiple tasks, including patch-level cancer detection, whole-slide classification, and survival prediction across various cancer types. Building upon this foundation, Prov-GigaPath[22] was trained on over 1.3 billion image tiles from 170,000 slides spanning 31 tissue types, achieving state-of-the-art performance across 26 pathology benchmarks, and setting a new standard for large-scale digital pathology modeling.

### *S7.1.2. Vision–language foundation models*

In parallel, vision–language foundation models have emerged as powerful tools for bridging clinical semantics and image generation. These models leverage large-scale multimodal pretraining to support interpretable, context-aware synthesis and diagnostic reasoning. For example, One study[408] introduced an image–text generative model grounded in medical literature, which enabled semantically meaningful synthesis even in rare or ambiguous diagnostic scenarios. Further advancing this paradigm, RoentGen[20] developed a vision–language model capable of generating chest X-rays conditioned on textual radiology reports, facilitating the development of standardized imaging protocols for early screening. In a complementary effort, another model[19] aligned visual and textual representations via masked contrastive learning with fine-grained supervision, enhancing radiographic interpretation and enabling automated report generation across diverse diseases. A notable advancement in this area is MINIM[259], a self-improving multimodal foundation model capable of synthesizing high-fidelity 3D CT, MRI, and OCT volumes based on partial inputs or clinical prompts. By integrating text-conditioned and multimodal pretraining strategies, MINIM has supported a wide range of downstream applications, including disease diagnosis, report generation, and self-supervised learning, demonstrating strong generalizability across imaging modalities and clinical contexts.

Foundation models mark a significant shift in medical image generation by offering a unified framework capable of synthesizing, enhancing, and interpreting multi-modal data across clinical stages. From text-conditioned image synthesis to volumetric modeling and outcome prediction, these models demonstrate strong potential to bridge data heterogeneity and support end-to-end clinical pipelines. However, this versatility comes at a cost, as current foundation models rely heavily on pretraining with vast and diverse datasets, often comprising millions or billions of image tiles or samples. This reliance introduces challenges related to data availability, annotation quality, and computational demands. Moreover, while their downstream performance is promising, issues such as hallucinations, interpretability, domain transferability, and integration into clinical workflows remain areas of active investigation. Future advances will require not only scaling models, but also refining strategies for efficient pretraining, federated learning, multi-level evaluation frameworks, and regulatory-aligned deployment to ensure foundation models can realize their full potential in real-world healthcare settings.

**Table S10.** Summary of publications on Foundation Models in clinical medical imaging.

| Publication (Year) | Model | Application | Loss Function | Link |
|---|---|---|---|---|
| MedDiff-FM (2024) [403] | Diffusion model | CT Foundation model | Denoising diffusion loss | – |
| RETFound-DE (2025) [406] | Diffusion model | Retinal Foundation model | Denoising diffusion loss | √ |
| RoentGen (2024) [20] | Diffusion model | Chest X-ray Foundation model | Denoising diffusion loss | √ |
| MINIM (2024) [259] | Diffusion model | OCT/CT/X-ray/MRI Foundation model | Denoising diffusion loss | √ |
| BME-X (2024) [404] | CNN | MRI Foundation model | Cross-entropy loss，MSE loss | √ |
| Triad (2025) [405] | Transformer, VAE | MRI Foundation Model | L1 loss, Log-ratio loss | √ |
| BEPH (2025) [407] | Transformer | Pathology Foundation model | MSE loss | √ |
| Prov-GigaPath (2024) [22] | Transformer | Pathology Foundation model | Contrastive loss，MSE loss | √ |
| MONET (2024) [408] | Transformer | Image-text Foundation model | Contrastive loss，Cross-entropy loss | √ |
| MaCo (2024) [19] | Transformer | Radiography-reports Foundation model | InfoNCE loss，MAE loss | √ |

# Reference

1. Akpinar, M. H. *et al.* Synthetic data generation via generative adversarial networks in healthcare: a systematic review of image- and signal-based studies. *IEEE Open J. Eng. Med. Biol.* (2024) doi:10.1109/OJEMB.2024.3508472.
2. Alamir, M. & Alghamdi, M. The role of generative adversarial network in medical image analysis: an In-depth survey. *ACM Comput. Surv.* **55**, 1–36 (2023).
3. Pan, Z. *et al.* Loss functions of generative adversarial networks (GANs): opportunities and challenges. *IEE Trans. Emerg. Topics Comput. Intell.* **4**, 500–522 (2020).
4. Shokraei Fard, A., Reutens, D. C. & Vegh, V. From CNNs to GANs for cross-modality medical image estimation. *Comput. Biol. Med.* **146**, 105556 (2022).
5. Apostolopoulos, I. D., Papathanasiou, N. D., Apostolopoulos, D. J. & Panayiotakis, G. S. Applications of generative adversarial networks (GANs) in positron emission tomography (PET) imaging: a review. *Eur. J. Nucl. Med. Mol. Imaging* **49**, 3717–3739 (2022).
6. Singh, N. K. & Raza, K. Medical image generation using generative adversarial networks. (2020) doi:10.1007/978-981-15-9735-0_5.
7. Rais, K., Amroune, M., Benmachiche, A. & Haouam, M. Y. Exploring variational autoencoders for medical image generation: A comprehensive study. Preprint at https://doi.org/10.48550/arXiv.2411.07348 (2024).
8. Ehrhardt, J. & Wilms, M. Autoencoders and variational autoencoders in medical image analysis. in *Biomedical Image Synthesis and Simulation* 129–162 (Elsevier, 2022).
9. Shi, Y. *et al.* Diffusion models for medical image computing: a survey. *Tsinghua Sci. Technol.* **30**, 357–383 (2024).
10. Alimisis, P., Mademlis, I., Radoglou-Grammatikis, P., Sarigiannidis, P. & Papadopoulos, G. T. Advances in diffusion models for image data augmentation: a review of methods, models, evaluation metrics and future research directions. Preprint at https://doi.org/10.48550/arXiv.2407.04103 (2024).
11. Fan, Y. *et al.* A survey of emerging applications of diffusion probabilistic models in MRI. Preprint at https://doi.org/10.48550/arXiv.2311.11383 (2024).
12. Hein, D., Bozorgpour, A., Merhof, D. & Wang, G. Physics-inspired generative models in medical imaging: a review. Preprint at https://doi.org/10.48550/arXiv.2407.10856 (2024).
13. Kazerouni, A. *et al.* Diffusion models in medical imaging: a comprehensive survey. Preprint at https://doi.org/10.48550/arXiv.2211.07804 (2023).
14. He, K. *et al.* Transformers in medical image analysis. *Intell. Med.* **3**, 59–78 (2023).
15. Heidari, M. *et al.* Computation-efficient era: a comprehensive survey of state space models in medical image analysis. Preprint at https://doi.org/10.48550/arXiv.2406.03430 (2024).
16. Xiong, J. *et al.* Autoregressive models in vision: a survey. Preprint at https://doi.org/10.48550/arXiv.2411.05902 (2024).
17. Huang, J. *et al.* Data and physics driven learning models for fast MRI -- fundamentals and methodologies from CNN, GAN to attention and transformers. Preprint at https://doi.org/10.48550/arXiv.2204.01706 (2022).

18. Christensen, M., Vukadinovic, M., Yuan, N. & Ouyang, D. Vision–language foundation model for echocardiogram interpretation. *Nat. Med.* **30**, 1481–1488 (2024).
19. Huang, W. *et al.* Enhancing representation in radiography-reports foundation model: a granular alignment algorithm using masked contrastive learning. *Nat. Commun.* **15**, 7620 (2024).
20. Bluethgen, C. *et al.* A vision–language foundation model for the generation of realistic chest x-ray images. *Nat. Biomed. Eng.* 1–13 (2024) doi:10.1038/s41551-024-01246-y.
21. Vorontsov, E. *et al.* A foundation model for clinical-grade computational pathology and rare cancers detection. *Nat. Med.* **30**, 2924–2935 (2024).
22. Xu, H. *et al.* A whole-slide foundation model for digital pathology from real-world data. *Nature* **630**, 181–188 (2024).
23. Kebaili, A., Lapuyade-Lahorgue, J. & Ruan, S. Deep learning approaches for data augmentation in medical imaging: A review. *J. Imaging* **9**, 81 (2023).
24. Cossio, M. Augmenting medical imaging: a comprehensive catalogue of 65 techniques for enhanced data analysis. Preprint at https://doi.org/10.48550/arXiv.2303.01178 (2023).
25. Pezoulas, V. C. *et al.* Synthetic data generation methods in healthcare: a review on open-source tools and methods. *Comput. Struct. Biotechnol. J.* (2024) doi:10.1016/j.csbj.2024.07.005.
26. Wang, T. *et al.* Medical imaging synthesis using deep learning and its clinical applications: a review. Preprint at https://doi.org/10.48550/arXiv.2004.10322 (2020).
27. Dayarathna, S. *et al.* Deep learning based synthesis of MRI, CT and PET: review and analysis. *Med. Image Anal.* **92**, 103046 (2024).
28. Liu, Y., Dwivedi, G., Boussaid, F. & Bennamoun, M. 3D brain and heart volume generative models: a survey. *ACM Comput. Surv.* **56**, 1–37 (2024).
29. Spadea, M. F., Maspero, M., Zaffino, P. & Seco, J. Deep learning based synthetic-CT generation in radiotherapy and PET: a review. *Med. Phys.* **48**, 6537–6566 (2021).
30. Lombardi, A. F. *et al.* Synthetic CT in musculoskeletal disorders: a systematic review. *Invest. Radiol.* **58**, 43–59 (2023).
31. Manjooran, G. P., Malakkaran, A. J., Joseph, A., Babu, H. M. & Meharban, M. S. A review on cross-modality synthesis from MRI to PET. in *ICSCCC - Int. Conf. Secur. Cyber Comput. Commun.* 126–131 (Institute of Electrical and Electronics Engineers Inc., 2021). doi:10.1109/ICSCCC51823.2021.9478170.
32. Boulanger, M. *et al.* Deep learning methods to generate synthetic CT from MRI in radiotherapy: a literature review. *Physica Med.* **89**, 265–281 (2021).
33. Wang, S. *et al.* Knowledge-driven deep learning for fast MR imaging: Undersampled MR image reconstruction from supervised to un-supervised learning. *Magn. Reson. Med.* **92**, 496–518 (2024).
34. Wang, S., Xiao, T., Liu, Q. & Zheng, H. Deep learning for fast MR imaging: A review for learning reconstruction from incomplete k-space data. *Biomed. Signal Process. Control* **68**, 102579 (2021).
35. Zeng, G. *et al.* A review on deep learning MRI reconstruction without fully sampled k-space. *BMC Med. Imaging* **21**, 195 (2021).

36. Chen, Y. *et al.* AI-based reconstruction for fast MRI—a systematic review and meta-analysis. *Proceedings of the IEEE* **110**, 224–245 (2022).
37. Deep learning techniques in PET/CT imaging: A comprehensive review from sinogram to image space. *Comput. Methods Programs Biomed.* **243**, 107880 (2024).
38. Samala, R. K. *et al.* AI and machine learning in medical imaging: key points from development to translation. *BJR| Artif. Intell.* **1**, ubae006 (2024).
39. Chaddad, A., Hu, Y., Wu, Y., Wen, B. & Kateb, R. Generalizable and explainable deep learning for medical image computing: an overview. *Curr. Opin. Biomed. Eng.* **33**, (2025).
40. Dimitriadis, A., Trivizakis, E., Papanikolaou, N., Tsiknakis, M. & Marias, K. Enhancing cancer differentiation with synthetic MRI examinations via generative models: a systematic review. *Insights into Imaging* **13**, (2022).
41. Koohi-Moghadam, M. & Bae, K. T. Generative AI in medical imaging: Applications, challenges, and ethics. *J. Med. Syst.* **47**, 94 (2023).
42. Goodfellow, I. *et al.* Generative adversarial nets.
43. Radford, A., Metz, L. & Chintala, S. Unsupervised representation learning with deep convolutional generative adversarial networks. Preprint at https://doi.org/10.48550/arXiv.1511.06434 (2016).
44. Kora Venu, S. & Ravula, S. Evaluation of deep convolutional generative adversarial networks for data augmentation of chest x-ray images. *Future Internet* **13**, 8 (2020).
45. Zhu, J.-Y., Park, T., Isola, P. & Efros, A. A. Unpaired image-to-image translation using cycle-consistent adversarial networks. in *Proceedings of the IEEE International Conference on Computer Vision* 2223–2232 (2017).
46. Kang, S. K. *et al.* Synthetic CT generation from weakly paired MR images using cycle-consistent GAN for MR-guided radiotherapy. *Biomed. Eng. Lett.* **11**, 263–271 (2021).
47. Karras, T., Laine, S. & Aila, T. A style-based generator architecture for generative adversarial networks.
48. Krishna, A. & Mueller, K. Medical (CT) image generation with style. in *Proc SPIE Int Soc Opt Eng* (eds. Matej S. & Metzler S.D.) vol. 11072 (SPIE, 2019).
49. Fetty, L. *et al.* Latent space manipulation for high-resolution medical image synthesis via the StyleGAN. *Z. Med. Phys.* **30**, 305–314 (2020).
50. Lai, M., Marzi, C., Mascalchi, M. & Diciotti, S. Brain MRI synthesis using Stylegan2-ADA. in *IEEE Comput. Soc. Conf. Comput. Vis. Pattern Recogn.* (IEEE Computer Society, 2024). doi:10.1109/ISBI56570.2024.10635279.
51. Kingma, D. P. & Welling, M. An introduction to variational autoencoders. *Found. Trends® Mach. Learn.* **12**, 307–392 (2019).
52. Van Erven, T. & Harremos, P. Rényi divergence and kullback-leibler divergence. *IEEE Trans. Inf. Theory* **60**, 3797–3820 (2014).
53. Sundar, V. K., Ramakrishna, S., Rahiminasab, Z., Easwaran, A. & Dubey, A. Out-of-distribution detection in multi-label datasets using latent space of β-VAE. in *2020 IEEE Security and Privacy Workshops (SPW)* 250–255 (2020). doi:10.1109/SPW50608.2020.00057.

54. Loizillon, S. *et al.* Detecting brain anomalies in clinical routine with the β-VAE: feasibility study on age-related white matter hyperintensities. in *Medical Imaging with Deep Learning - MIDL 2024* (Paris, France, 2024).
55. Harvey, W., Naderiparizi, S. & Wood, F. Conditional image generation by conditioning variational auto-encoders. Preprint at https://doi.org/10.48550/arXiv.2102.12037 (2022).
56. Pesteie, M., Abolmaesumi, P. & Rohling, R. N. Adaptive augmentation of medical data using independently conditional variational auto-encoders. *IEEE Trans. Med. Imag.* **38**, 2807–2820 (2019).
57. Van Den Oord, A. & Vinyals, O. Neural discrete representation learning. *Adv. Neural Inf. Process. Syst.* **30**, (2017).
58. Ramanathan, S. & Ramasundaram, M. Vector quantized convolutional autoencoder network for LDCT image reconstruction with hybrid loss. *SN Comput. Sci.* **5**, 2 (2023).
59. Zhao, A. *et al.* 4D VQ-GAN: synthesising medical scans at any time point for personalised disease progression modelling of idiopathic pulmonary fibrosis. Preprint at https://doi.org/10.48550/arXiv.2502.05713 (2025).
60. Ibrahim, B. I., Nicolae, D. C., Khan, A., Ali, S. I. & Khattak, A. VAE-GAN based zero-shot outlier detection. in *Proceedings of the 2020 4th International Symposium on Computer Science and Intelligent Control* 1–5 (ACM, Newcastle upon Tyne United Kingdom, 2020). doi:10.1145/3440084.3441180.
61. Foroozandeh, M. & Eklund, A. Synthesizing brain tumor images and annotations by combining progressive growing GAN and SPADE. Preprint at https://doi.org/10.48550/arXiv.2009.05946 (2020).
62. Volokitin, A. *et al.* Modelling the distribution of 3D brain MRI using a 2D slice VAE. in *Medical Image Computing and Computer Assisted Intervention – MICCAI 2020* (eds. Martel, A. L. et al.) vol. 12267 657–666 (Springer International Publishing, Cham, 2020).
63. Ho, J., Jain, A. & Abbeel, P. Denoising diffusion probabilistic models. *Adv. Neural Inf. Process. Syst.* **33**, 6840–6851 (2020).
64. Cho, J. *et al.* MediSyn: A generalist text-guided latent diffusion model for diverse medical image synthesis. Preprint at https://doi.org/10.48550/arXiv.2405.09806 (2025).
65. Hamamci, I. E. *et al.* GenerateCT: Text-conditional generation of 3D chest CT volumes. Preprint at https://doi.org/10.48550/arXiv.2305.16037 (2024).
66. Huang, P. *et al.* Diff-CXR: Report-to-CXR generation through a disease-knowledge enhanced diffusion model. Preprint at https://doi.org/10.48550/arXiv.2410.20165 (2024).
67. Arslan, F., Kabas, B., Dalmaz, O., Ozbey, M. & Çukur, T. Self-consistent recursive diffusion bridge for medical image translation. Preprint at https://doi.org/10.48550/arXiv.2405.06789 (2024).
68. Wang, Z. *et al.* Mutual information guided diffusion for zero-shot cross-modality medical image translation. *IEEE Trans. Med. Imag.* **43**, 2825–2838 (2024).
69. Daum, D. *et al.* On differentially private 3D medical image synthesis with controllable latent diffusion models. in *Deep Generative Models* (eds. Mukhopadhyay, A., Oksuz, I.,

Engelhardt, S., Mehrof, D. & Yuan, Y.) vol. 15224 139–149 (Springer Nature Switzerland, Cham, 2025).

70. Kawata, N. *et al.* Generation of short-term follow-up chest CT images using a latent diffusion model in COVID-19. *Jpn. J. Radiol.* (2024) doi:10.1007/s11604-024-01699-w.
71. Sanderson, D., Olmos, P. M., Del Cerro, C. F., Desco, M. & Abella, M. Diffusion X-ray image denoising. in *Medical imaging with deep learning* (2024).
72. Xiang, T., Yurt, M., Syed, A. B., Setsompop, K. & Chaudhari, A. DDM$^2$: self-supervised diffusion MRI denoising with generative diffusion models. Preprint at https://doi.org/10.48550/arXiv.2302.03018 (2023).
73. Saharia, C. *et al.* Image super-resolution via iterative refinement. *IEEE Trans. Pattern Anal. Mach. Intell.* **45**, 4713–4726 (2022).
74. Xia, B. *et al.* Diffir: efficient diffusion model for image restoration. in *Proceedings of the IEEE/CVF International Conference on Computer Vision* 13095–13105 (2023).
75. Dorjsembe, Z., Odonchimed, S. & Xiao, F. Three-dimensional medical image synthesis with denoising diffusion probabilistic models. in *Medical Imaging with Deep Learning* (2022).
76. Wang, H. *et al.* 3D MedDiffusion: A 3D medical diffusion model for controllable and high-quality medical image generation. Preprint at https://doi.org/10.48550/arXiv.2412.13059 (2024).
77. Xu, K., Lu, S., Huang, B., Wu, W. & Liu, Q. Stage-by-stage wavelet optimization refinement diffusion model for sparse-view CT reconstruction. *IEEE Trans. Med. Imag.* **43**, 3412–3424 (2024).
78. Xie, T. *et al.* Joint diffusion: mutual consistency-driven diffusion model for PET-MRI co-reconstruction. *Phys. Med. Biol.* **69**, 155019 (2024).
79. Qiu, S. *et al.* Spatiotemporal diffusion model with paired sampling for accelerated cardiac cine MRI. Preprint at https://doi.org/10.48550/arXiv.2403.08758 (2024).
80. Guo, Z. *et al.* Diffusion models in bioinformatics and computational biology. *Nat. Rev. Bioeng.* **2**, 136–154 (2024).
81. Dosovitskiy, A. *et al.* An image is worth 16x16 words: transformers for image recognition at scale. *Arxiv Prepr. Arxiv:2010,11929* (2020).
82. Gu, A. & Dao, T. Mamba: linear-time sequence modeling with selective state spaces. *Arxiv Prepr. Arxiv:2312,00752* (2023).
83. Van Den Oord, A., Kalchbrenner, N. & Kavukcuoglu, K. Pixel recurrent neural networks. in *International conference on machine learning* 1747–1756 (PMLR, 2016).
84. Gregor, K., Danihelka, I., Mnih, A., Blundell, C. & Wierstra, D. Deep autoregressive networks. in *International Conference on Machine Learning* 1242–1250 (PMLR, 2014).
85. Gao, Y. *et al.* A data-scalable transformer for medical image segmentation: architecture, model efficiency, and benchmark. Preprint at https://doi.org/10.48550/arXiv.2203.00131 (2023).
86. Zhao, X., Yang, T., Li, B. & Zhang, X. SwinGAN: a dual-domain swin transformer-based generative adversarial network for MRI reconstruction. *Comput. Biol. Med.* **153**, (2023).

87. Kottu, J. A vision transformer-driven method for generating medical reports based on x-ray radiology. (California State University, Sacramento, 2024).
88. Adams, L. C. *et al.* What does DALL-E 2 know about radiology? *J. Med. Internet Res.* **25**, e43110 (2023).
89. Liu, J. *et al.* Swin-UMamba†: adapting mamba-based vision foundation models for medical image segmentation. *IEEE Trans. Med. Imag.* 1–1 (2024) doi:10.1109/TMI.2024.3508698.
90. Yue, Y. & Li, Z. MedMamba: vision mamba for medical image classification. Preprint at https://doi.org/10.48550/arXiv.2403.03849 (2024).
91. Ju, Z. & Zhou, W. VM-DDPM: vision mamba diffusion for medical image synthesis. Preprint at https://doi.org/10.48550/arXiv.2405.05667 (2024).
92. Fu, L. *et al.* MD-dose: a diffusion model based on the mamba for radiation dose prediction. in *2024 IEEE International Conference on Bioinformatics and Biomedicine (BIBM)* 911–918 (IEEE, 2024). doi:10.1109/BIBM62325.2024.10822581.
93. Li, K. *et al.* VideoMamba: state space model for efficient video understanding. in *Computer Vision – ECCV 2024* (eds. Leonardis, A. et al.) vol. 15084 237–255 (Springer Nature Switzerland, Cham, 2025).
94. Chen, Z., Wang, S., Yan, D. & Li, Y. A spatio-temporl deepfake video detection method based on TimeSformer-CNN. in *2024 Third International Conference on Distributed Computing and Electrical Circuits and Electronics (ICDCECE)* 1–6 (IEEE, 2024).
95. Behrouz, A., Santacatterina, M. & Zabih, R. MambaMixer: efficient selective state space models with dual token and channel selection. Preprint at https://doi.org/10.48550/arXiv.2403.19888 (2024).
96. Yoon, J. S., Zhang, C., Suk, H.-I., Guo, J. & Li, X. SADM: sequence-aware diffusion model for longitudinal medical image generation. in *Information Processing in Medical Imaging* (eds. Frangi, A., de Bruijne, M., Wassermann, D. & Navab, N.) 388–400 (Springer Nature Switzerland, Cham, 2023). doi:10.1007/978-3-031-34048-2_30.
97. Luo, G., Huang, S. & Uecker, M. Autoregressive image diffusion: generation of image sequence and application in MRI. Preprint at https://doi.org/10.48550/arXiv.2405.14327 (2025).
98. Kabas, B. *et al.* Physics-driven autoregressive state space models for medical image reconstruction. Preprint at https://doi.org/10.48550/arXiv.2412.09331 (2024).
99. Park, J. *et al.* Can mamba learn how to learn? A comparative study on In-context learning tasks. Preprint at https://doi.org/10.48550/arXiv.2402.04248 (2024).
100. Cardoso, M. J. *et al.* MONAI: an open-source framework for deep learning in healthcare. Preprint at https://doi.org/10.48550/arXiv.2211.02701 (2022).
101. He, K. *et al.* Masked autoencoders are scalable vision learners. in *Proceedings of the IEEE/CVF Conference on Computer Vision and Pattern Recognition* 16000–16009 (2022).
102. Reichenpfader, D., Müller, H. & Denecke, K. A scoping review of large language model based approaches for information extraction from radiology reports. *npj Digital Med.* **7**, 222 (2024).

103. Thirunavukarasu, A. J. *et al.* Large language models in medicine. *Nat. Med.* **29**, 1930–1940 (2023).
104. Liu, F. *et al.* A medical multimodal large language model for future pandemics. *npj Digital Med.* **6**, 226 (2023).
105. Huang, P. *et al.* Chest-diffusion: A light-weight text-to-image model for report-to-CXR generation. in *2024 IEEE International Symposium on Biomedical Imaging (isbi)* 1–5 (2024). doi:10.1109/ISBI56570.2024.10635417.
106. Wang, Z., Wu, Z., Agarwal, D. & Sun, J. Medclip: contrastive learning from unpaired medical images and text. in *Proceedings of the Conference on Empirical Methods in Natural Language Processing. Conference on Empirical Methods in Natural Language Processing* vol. 2022 3876 (2022).
107. Shiri, M., Beyan, C. & Murino, V. MadCLIP: few-shot medical anomaly detection with CLIP. Preprint at https://doi.org/10.48550/arXiv.2506.23810 (2025).
108. Khattak, M. U., Kunhimon, S., Naseer, M., Khan, S. & Khan, F. S. UniMed-CLIP: towards a unified image-text pretraining paradigm for diverse medical imaging modalities. Preprint at https://doi.org/10.48550/arXiv.2412.10372 (2024).
109. Tiu, E. *et al.* Expert-level detection of pathologies from unannotated chest X-ray images via self-supervised learning. *Nat. Biomed. Eng.* **6**, 1399–1406 (2022).
110. Moor, M. *et al.* Foundation models for generalist medical artificial intelligence. *Nature* **616**, 259–265 (2023).
111. Hein, D. *et al.* Noise suppression in photon-counting computed tomography using unsupervised poisson flow generative models. *Visual Comput. Ind., Biomed., Art* **7**, (2024).
112. Hu, Z. *et al.* Artifact correction in low-dose dental CT imaging using wasserstein generative adversarial networks. *Med. Phys.* **46**, 1686–1696 (2019).
113. Jiang, C. *et al.* Wasserstein generative adversarial networks for motion artifact removal in dental CT imaging. in *Progr. Biomed. Opt. Imaging Proc. SPIE* (eds. Schmidt T.G., Chen G.-H., & Bosmans H.) vol. 10948 (SPIE, 2019).
114. Wang, G. & Hu, X. Low-dose CT denoising using a progressive wasserstein generative adversarial network. *Comput. Biol. Med.* **135**, (2021).
115. Deng, Z. *et al.* TT U-net: temporal transformer U-net for motion artifact reduction using PAD (pseudo all-phase clinical-dataset) in cardiac CT. *IEEE Trans. Med. Imag.* **42**, 3805–3816 (2023).
116. Gao, Q., Li, Z., Zhang, J., Zhang, Y. & Shan, H. CoreDiff: contextual error-modulated generalized diffusion model for low-dose CT denoising and generalization. *IEEE Trans. Med. Imag.* **43**, 745–759 (2024).
117. Öztürk, Ş., Duran, O. C. & Çukur, T. DenoMamba: a fused state-space model for low-dose CT denoising. Preprint at https://doi.org/10.48550/arXiv.2409.13094 (2024).
118. Hu, Z. *et al.* Obtaining PET/CT images from non-attenuation corrected PET images in a single PET system using wasserstein generative adversarial networks. *Phys. Med. Biol.* **65**, 215010 (2020).
119. Yang, J., Sohn, J. H., Behr, S. C., Gullberg, G. T. & Seo, Y. CT-less direct correction of attenuation and scatter in the image space using deep learning for whole-body FDG PET: potential benefits and pitfalls. *Radiol.: Artif. Intell.* **3**, e200137 (2021).

120. Gong, Y. *et al.* Parameter-transferred wasserstein generative adversarial network (PT-WGAN) for low-dose PET image denoising. *IEEE Trans. Radiat. Plasma Med. Sci.* **5**, 213–223 (2021).
121. Gong, K., Johnson, K., El Fakhri, G., Li, Q. & Pan, T. PET image denoising based on denoising diffusion probabilistic model. *Eur. J. Nucl. Med. Mol. Imaging* **51**, 358–368 (2024).
122. Yu, B. & Gong, K. Adaptive whole-body PET image denoising using 3D diffusion models with ControlNet. Preprint at https://doi.org/10.48550/arXiv.2411.05302 (2024).
123. Ran, M. *et al.* Denoising of 3D magnetic resonance images using a residual encoder–decoder wasserstein generative adversarial network. *Med. Image Anal.* **55**, 165–180 (2019).
124. Geng, M. *et al.* Content-noise complementary learning for medical image denoising. *IEEE Trans. Med. Imag.* **41**, 407–419 (2022).
125. Lim, A., Lo, J., Wagner, M. W., Ertl-Wagner, B. & Sussman, D. Motion artifact correction in fetal MRI based on a generative adversarial network method. *Biomed. Signal Process. Control* **81**, 104484 (2023).
126. Chung, H., Lee, E. S. & Ye, J. C. MR image denoising and super-resolution using regularized reverse diffusion. *IEEE Trans. Med. Imag.* **42**, 922–934 (2022).
127. Xu, J. *et al.* Motion artifact removal in pixel-frequency domain via alternate masks and diffusion model. Preprint at https://doi.org/10.48550/arXiv.2412.07590 (2024).
128. Fu, L. *et al.* Energy-guided diffusion model for CBCT-to-CT synthesis. *Comput. Med. Imaging Graphics* **113**, (2024).
129. Raju, J. C., Murugesan, B., Ram, K. & Sivaprakasam, M. AutoSyncoder: an adversarial AutoEncoder framework for multimodal MRI synthesis. in *Lect. Notes Comput. Sci.* (eds. Deeba F., Johnson P., Würfl T., & Ye J.C.) vol. 12450 LNCS 102–110 (Springer Science and Business Media Deutschland GmbH, 2020).
130. Li, Z. *et al.* Promising generative adversarial network based sinogram inpainting method for ultra-limited-angle computed tomography imaging. *Sens. Switz.* **19**, (2019).
131. Krishnan, A. R. *et al.* Lung CT harmonization of paired reconstruction kernel images using generative adversarial networks. *Med. Phys.* **51**, 5510–5523 (2024).
132. Liu, J. *et al.* Dolce: a model-based probabilistic diffusion framework for limited-angle ct reconstruction. in *Proceedings of the IEEE/CVF International Conference on Computer Vision* 10498–10508 (2023).
133. Lopez-Montes, A. *et al.* Stationary CT imaging of intracranial hemorrhage with diffusion posterior sampling reconstruction. Preprint at https://doi.org/10.48550/arXiv.2407.11196 (2024).
134. Pradhan, N. *et al.* Conditional generative adversarial network model for conversion of 2 dimensional radiographs into 3 dimensional views. *IEEE Access* **11**, 96283–96296 (2023).
135. Kania, A. *et al.* HyperNeRFGAN: hypernetwork approach to 3D NeRF GAN. Preprint at https://doi.org/10.48550/arXiv.2301.11631 (2024).
136. Zhang, X. *et al.* DL-recon: combining 3D deep learning image synthesis and model uncertainty with physics-based image reconstruction. in *Proc SPIE Int Soc Opt Eng* (ed. Stayman J.W.) vol. 12304 (SPIE, 2022).

137. Huang, J. *et al.* Enhancing global sensitivity and uncertainty quantification in medical image reconstruction with Monte Carlo arbitrary-masked mamba. *Medical Image Analysis* **99**, 103334 (2025).
138. Xia, W. *et al.* Parallel diffusion model-based sparse-view cone-beam breast CT. Preprint at https://doi.org/10.48550/arXiv.2303.12861 (2024).
139. Chen, H., Hao, Z., Guo, L. & Xiao, L. Mitigating data consistency induced discrepancy in cascaded diffusion models for sparse-view CT reconstruction. Preprint at https://doi.org/10.48550/arXiv.2403.09355 (2024).
140. Wang, Y., Li, Z. & Wu, W. Time-reversion fast-sampling score-based model for limited-angle CT reconstruction. *IEEE Trans. Med. Imag.* **43**, 3449–3460 (2024).
141. Lei, Y. *et al.* Whole-body PET estimation from low count statistics using cycle-consistent generative adversarial networks. *Phys. Med. Biol.* **64**, (2019).
142. Xiang, L., Wang, L., Gong, E., Zaharchuk, G. & Zhang, T. Noise-aware standard-dose PET reconstruction using general and adaptive robust loss. in *Lect. Notes Comput. Sci.* (eds. Liu M., Lian C., Yan P., & Cao X.) vol. 12436 LNCS 654–662 (Springer Science and Business Media Deutschland GmbH, 2020).
143. Shi, L. *et al.* Deep learning-based attenuation map generation with simultaneously reconstructed PET activity and attenuation and low-dose application. *Phys. Med. Biol.* **68**, 35014 (2023).
144. Gautier, V. *et al.* Bimodal PET/MRI generative reconstruction based on VAE architectures. *Phys. Med. Biol.* **69**, (2024).
145. Zhang, Q. *et al.* Deep generalized learning model for PET image reconstruction. *IEEE Trans. Med. Imag.* **43**, 122–134 (2024).
146. Wikberg, E. *et al.* Improvements of 177Lu SPECT images from sparsely acquired projections by reconstruction with deep-learning-generated synthetic projections. *EJNMMI Phys.* **11**, (2024).
147. Zhai, M., Wang, H., Han, J., Wu, T. & Ye, H. Generating PET images from low-dose data using a cycle PET reconstruction convolutional neural network. in *IEEE Int. Conf. Mechatronics Autom., ICMA* 369–374 (Institute of Electrical and Electronics Engineers Inc., 2024). doi:10.1109/ICMA61710.2024.10632916.
148. Leube, J., Gustafsson, J., Lassmann, M., Salas-Ramirez, M. & Tran-Gia, J. Analysis of a deep learning-based method for generation of SPECT projections based on a large monte carlo simulated dataset. *EJNMMI Phys.* **9**, (2022).
149. Ouyang, J. *et al.* Task-GAN: improving generative adversarial network for image reconstruction. in *Lect. Notes Comput. Sci.* (eds. Knoll F., Maier A., Rueckert D., & Ye J.C.) vol. 11905 LNCS 193–204 (Springer, 2019).
150. Luo, Y. *et al.* Adaptive rectification based adversarial network with spectrum constraint for high-quality PET image synthesis. *Med. Image Anal.* **77**, (2022).
151. Singh, I. R. *et al.* Score-based generative models for PET image reconstruction. *Mach. Learn. Biomed. Imaging* **2**, 547–585 (2024).
152. Xie, H. *et al.* Dose-aware diffusion model for 3D low-dose PET: multi-institutional validation with reader study and real low-dose data. Preprint at https://doi.org/10.48550/arXiv.2405.12996 (2024).

153. Zhao, Y. *et al.* Whole-body magnetic resonance imaging at 0.05 tesla. *Science* **384**, eadm7168 (2024).
154. Man, C. *et al.* Deep learning enabled fast 3D brain MRI at 0.055 tesla. *Sci. Adv.* **9**, eadi9327 (2023).
155. Wang, G., Gong, E., Banerjee, S., Pauly, J. & Zaharchuk, G. Accelerated MRI reconstruction with dual-domain generative adversarial network. in *Lect. Notes Comput. Sci.* (eds. Knoll F., Maier A., Rueckert D., & Ye J.C.) vol. 11905 LNCS 47–57 (Springer, Cham, 2019).
156. Dar, S. U. H. *et al.* Prior-guided image reconstruction for accelerated multi-contrast mri via generative adversarial networks. *IEEE J. Sel. Top. Sign. Proces.* **14**, 1072–1087 (2020).
157. Huang, J. *et al.* Swin transformer for fast MRI. *Neurocomputing* **493**, 281–304 (2022).
158. Hou, R., Li, F. & Zeng, T. Fast and reliable score-based generative model for parallel MRI. *IEEE Trans. Neural Netw. Learn. Syst.* 1–14 (2023) doi:10.1109/TNNLS.2023.3333538.
159. Chen, L. *et al.* Joint coil sensitivity and motion correction in parallel MRI with a self-calibrating score-based diffusion model. *Med. Image Anal.* **102**, 103502 (2025).
160. Cao, C. *et al.* High-frequency space diffusion model for accelerated MRI. *IEEE Trans. Med. Imag.* **43**, 1853–1865 (2024).
161. Huang, J. *et al.* MambaMIR: an arbitrary-masked mamba for joint medical image reconstruction and uncertainty estimation. Preprint at https://doi.org/10.48550/arXiv.2402.18451 (2024).
162. Meng, Y., Yang, Z., Song, Z. & Shi, Y. DM-mamba: dual-domain multi-scale mamba for MRI reconstruction. Preprint at https://doi.org/10.48550/arXiv.2501.08163 (2025).
163. Kofler, A. *et al.* Neural networks-based regularization for large-scale medical image reconstruction. *Phys. Med. Biol.* **65**, 135003 (2020).
164. Kelkar, V. A., Bhadra, S. & Anastasio, M. A. Compressible latent-space invertible networks for generative model-constrained image reconstruction. *IEEE Trans. Comput. Imaging* **7**, 209–223 (2021).
165. Elmas, G. *et al.* Federated learning of generative image priors for MRI reconstruction. *IEEE Trans. Med. Imag.* **42**, 1996–2009 (2023).
166. Ramanarayanan, S., Palla, A., Ram, K. & Sivaprakasam, M. Generalizing supervised deep learning MRI reconstruction to multiple and unseen contrasts using meta-learning hypernetworks[formula presented]. *Appl. Soft Comput.* **146**, (2023).
167. Donners, R. *et al.* Deep learning reconstructed new-generation 0.55 T MRI of the knee-a prospective comparison with conventional 3 T MRI. *Invest. Radiol.* **59**, 823–830 (2024).
168. Nezhad, V. A., Elmas, G., Kabas, B., Arslan, F. & Çukur, T. Generative autoregressive transformers for model-agnostic federated MRI reconstruction. Preprint at https://doi.org/10.48550/arXiv.2502.04521 (2025).
169. Lan, H., Li, Z., He, Q. & Luo, J. Fast sampling generative model for ultrasound image reconstruction. Preprint at https://doi.org/10.48550/arXiv.2312.09510 (2023).

170. Zhang, Y., Huneau, C., Idier, J. & Mateus, D. Diffusion reconstruction of ultrasound images with informative uncertainty. Preprint at https://doi.org/10.48550/arXiv.2310.20618 (2023).
171. Zhang, Y., Huneau, C., Idier, J. & Mateus, D. Ultrasound imaging based on the variance of a diffusion restoration model. in *2024 32nd European Signal Processing Conference (EUSIPCO)* 760–764 (IEEE, 2024).
172. Merino, S., Salazar, I. & Lavarello, R. Generative models for ultrasound image reconstruction from single plane-wave simulated data. in *2024 IEEE UFFC Latin America Ultrasonics Symposium (LAUS)* 1–4 (IEEE, 2024). doi:10.1109/LAUS60931.2024.10553012.
173. Song, X. *et al.* Sparse-view reconstruction for photoacoustic tomography combining diffusion model with model-based iteration. *Photoacoustics* **33**, 100558 (2023).
174. Tong, S., Lan, H., Nie, L., Luo, J. & Gao, F. Score-based generative models for photoacoustic image reconstruction with rotation consistency constraints. Preprint at https://doi.org/10.48550/arXiv.2306.13843 (2023).
175. Zeng, H. *et al.* DM-RE2I: a framework based on diffusion model for the reconstruction from EEG to image. *Biomed. Signal Process. Control* **86**, 105125 (2023).
176. Xiang, T. *et al.* DiffCMR: fast cardiac MRI reconstruction with diffusion probabilistic models. in *Statistical Atlases and Computational Models of the Heart. Regular and Cmrxrecon Challenge Papers* (eds. Camara, O. et al.) vol. 14507 380–389 (Springer Nature Switzerland, Cham, 2024).
177. Yang, H. *et al.* Deep learning in medical image super resolution: a review. *Appl. Intell.* **53**, 20891–20916 (2023).
178. Xiao, H. *et al.* Deep learning for medical imaging super-resolution: a comprehensive review. *Neurocomputing* 129667 (2025).
179. Lucas, A., Lopez-Tapia, S., Molina, R. & Katsaggelos, A. K. Generative adversarial networks and perceptual losses for video super-resolution. *IEEE Trans. Image Process.* **28**, 3312–3327 (2019).
180. Ren, S., Li, J., Guo, K. & Li, F. Medical video super-resolution based on asymmetric back-projection network with multilevel error feedback. *IEEE Access* **9**, 17909–17920 (2021).
181. Song, X. *et al.* Deformable transformer for endoscopic video super-resolution. *Biomed. Signal Process. Control* **77**, 103827 (2022).
182. Guo, Y. *et al.* A spatiotemporal volumetric interpolation network for 4d dynamic medical image. in *Proceedings of the IEEE/CVF Conference on Computer Vision and Pattern Recognition* 4726–4735 (2020).
183. Wei, T.-T., Kuo, C., Tseng, Y.-C. & Chen, J.-J. MPVF: 4D medical image inpainting by multi-pyramid voxel flows. *IEEE J. Biomed. Health Inform.* **27**, 5872–5882 (2023).
184. Kim, B. & Ye, J. C. Diffusion deformable model for 4D temporal medical image generation. in *Lect. Notes Comput. Sci.* (eds. Wang L., Dou Q., Fletcher P.T., Speidel S., & Li S.) vol. 13431 LNCS 539–548 (Springer Science and Business Media Deutschland GmbH, 2022).
185. Kim, J., Yoon, H., Park, G., Kim, K. & Yang, E. Data-efficient unsupervised interpolation without any intermediate frame for 4D medical images. in *2024 IEEE/CVF*

*Conference on Computer Vision and Pattern Recognition (CVPR)* 11353–11364 (IEEE, Seattle, WA, USA, 2024). doi:10.1109/CVPR52733.2024.01079.

186. Chatterjee, S., Sarasaen, C., Rose, G., Nürnberger, A. & Speck, O. Ddos-unet: incorporating temporal information using dynamic dual-channel unet for enhancing super-resolution of dynamic mri. *IEEE Access* (2024) doi:10.1109/ACCESS.2024.3427674.
187. Karani, N., Zhang, L., Tanner, C. & Konukoglu, E. An image interpolation approach for acquisition time reduction in navigator-based 4D MRI. *Med. Image Anal.* **54**, 20–29 (2019).
188. You, C. *et al.* CT Super-Resolution GAN Constrained by the Identical, Residual, and Cycle Learning Ensemble (GAN-CIRCLE). *IEEE Trans. Med. Imag.* **39**, 188–203 (2020).
189. Zhang, K. *et al.* SOUP-GAN: super-resolution MRI using generative adversarial networks. *Tomography* **8**, 905–919 (2022).
190. Balasubramanian, A., Dhanasekaran, H., Raghu, B. & Kumarasamy, K. MRI super-resolution using generative adversarial network and discrete wavelet transform. in *Proc. - Int. Conf. Augment. Intell. Sustain. Syst., ICAISS* 1314–1318 (Institute of Electrical and Electronics Engineers Inc., 2022). doi:10.1109/ICAISS55157.2022.10010995.
191. Ma, Q., Koh, J. C. & Lee, W. S. A frequency domain constraint for synthetic and real X-ray image super resolution. in *Lect. Notes Comput. Sci.* (eds. Haq N., Johnson P., Maier A., Würfl T., & Yoo J.) vol. 12964 LNCS 120–129 (Springer Science and Business Media Deutschland GmbH, 2021).
192. Sun, L. *et al.* Hierarchical amortized GAN for 3D high resolution medical image synthesis. *IEEE J. Biomed. Health Inform.* **26**, 3966–3975 (2022).
193. Huang, W. *et al.* Deep local-to-global feature learning for medical image super-resolution. *Comput. med. imaging graph. : off. j. Comput. Med. Imaging Soc.* **115**, 102374 (2024).
194. Chu, Y., Zhou, L., Luo, G., Qiu, Z. & Gao, X. Topology-preserving computed tomography super-resolution based on dual-stream diffusion model. in *Medical Image Computing and Computer Assisted Intervention – MICCAI 2023* (eds. Greenspan, H. et al.) vol. 14229 260–270 (Springer Nature Switzerland, Cham, 2023).
195. Zhao, K. *et al.* Mri super-resolution with partial diffusion models. *IEEE Trans. Med. Imag.* (2024).
196. Ji, Z., Zou, B., Kui, X., Vera, P. & Ruan, S. Deform-mamba network for MRI super-resolution. in *Medical Image Computing and Computer Assisted Intervention – MICCAI 2024* (eds. Linguraru, M. G. et al.) vol. 15007 242–252 (Springer Nature Switzerland, Cham, 2024).
197. Chong, C. K. & Ho, E. T. W. Synthesis of 3D MRI brain images with shape and texture generative adversarial deep neural networks. *IEEE Access* **9**, 64747–64760 (2021).
198. Hong, S. *et al.* 3D-StyleGAN: a style-based generative adversarial network for generative modeling of three-dimensional medical images. in *Deep Generative Models, and Data Augmentation, Labelling, and Imperfections: First Workshop, DGM4MICCAI 2021, and First Workshop, DALI 2021, Held in Conjunction with MICCAI 2021, Strasbourg, France, October 1, 2021, Proceedings 1* 24–34 (Springer, 2021).

199. Zuo, L. *et al.* Synthesizing realistic brain mr images with noise control. in *Lect. Notes Comput. Sci.* (eds. Burgos N., Svoboda D., Wolterink J.M., & Zhao C.) vol. 12417 LNCS 21–31 (Springer Science and Business Media Deutschland GmbH, 2020).
200. Khader, F. *et al.* Medical diffusion: Denoising diffusion probabilistic models for 3D medical image generation. Preprint at https://doi.org/10.48550/arXiv.2211.03364 (2023).
201. Txurio, M. S. *et al.* Diffusion models for realistic CT image generation. in *Smart Innov. Syst. Technol.* (eds. Chen Y.-W., Tanaka S., Howlett R.J., & Jain L.C.) vol. 357 SIST 335–344 (Springer Science and Business Media Deutschland GmbH, 2023).
202. Wu, H., Zhao, Z., Zhang, Y., Xie, W. & Wang, Y. MRGen: diffusion-based controllable data engine for MRI segmentation towards unannotated modalities. Preprint at https://doi.org/10.48550/arXiv.2412.04106 (2024).
203. Zhang, S., Liu, J., Hu, B. & Mao, Z. GH-DDM: the generalized hybrid denoising diffusion model for medical image generation. *Multimedia Syst.* **29**, 1335–1345 (2023).
204. Zheng, J.-Q. *et al.* Deformation-recovery diffusion model (DRDM): instance deformation for image manipulation and synthesis. Preprint at https://doi.org/10.48550/arXiv.2407.07295 (2024).
205. Danu, M., Nita, C.-I., Vizitiu, A., Suciu, C. & Itu, L. M. Deep learning based generation of synthetic blood vessel surfaces. in *Int. Conf. Syst. Theory, Control Comput., ICSTCC - Proc.* (ed. Precup R.-E.) 662–667 (Institute of Electrical and Electronics Engineers Inc., 2019). doi:10.1109/ICSTCC.2019.8885576.
206. Xu, T. *et al.* AttnGAN: fine-grained text to image generation with attentional generative adversarial networks. Preprint at https://doi.org/10.48550/arXiv.1711.10485 (2017).
207. Qiao, T., Zhang, J., Xu, D. & Tao, D. MirrorGAN: Learning text-to-image generation by redescription. Preprint at https://doi.org/10.48550/arXiv.1903.05854 (2019).
208. Sahithi, Y. L., Sunny, N., Deepak, M. M. L. & Amrutha, S. Text-to-image synthesis using stackGAN. in *Glob. Conf. Inf. Technol. Commun., GCITC* (Institute of Electrical and Electronics Engineers Inc., 2023). doi:10.1109/GCITC60406.2023.10426184.
209. Ding, M. *et al.* CogView: Mastering text-to-image generation via transformers.
210. Reddy, M. D. M., Basha, M. S. M., Hari, M. M. C. & Penchalaiah, M. N. Dall-e: Creating images from text. *Ugc Care Group J.* **8**, 71–75 (2021).
211. Saharia, C. *et al.* Photorealistic text-to-image diffusion models with deep language understanding. *Adv. Neural Inf. Process. Syst.* **35**, 36479–36494 (2022).
212. Rombach, R., Blattmann, A., Lorenz, D., Esser, P. & Ommer, B. High-resolution image synthesis with latent diffusion models. in *Proceedings of the Ieee/cvf Conference on Computer Vision and Pattern Recognition* 10684–10695 (2022).
213. Li, X. *et al.* Text-driven tumor synthesis. Preprint at https://doi.org/10.48550/arXiv.2412.18589 (2024).
214. Guo, P. *et al.* MAISI: Medical AI for synthetic imaging. Preprint at https://doi.org/10.48550/arXiv.2409.11169 (2024).
215. Xu, Y. *et al.* MedSyn: Text-guided anatomy-aware synthesis of high-fidelity 3D CT images. Preprint at https://doi.org/10.48550/arXiv.2310.03559 (2024).
216. Qiao, M. *et al.* Cheart: a conditional spatio-temporal generative model for cardiac anatomy. *IEEE Trans. Med. Imag.* **43**, 1259–1269 (2023).

217. Reynaud, H. *et al.* Feature-conditioned cascaded video diffusion models for precise echocardiogram synthesis. in *Medical Image Computing and Computer Assisted Intervention – MICCAI 2023* vol. 14229 142–152 (Springer Nature Switzerland, Cham, 2023).
218. Zhou, X. *et al.* HeartBeat: towards controllable echocardiography video synthesis with multimodal conditions-guided diffusion models. in *Lect. Notes Comput. Sci.* (eds. Linguraru M.G. et al.) vol. 15007 LNCS 361–371 (Springer Science and Business Media Deutschland GmbH, 2024).
219. Liu, Q. *et al.* Treatment-aware diffusion probabilistic model for longitudinal MRI generation and diffuse glioma growth prediction. *IEEE Trans. Med. Imag.* 1–1 (2025) doi:10.1109/TMI.2025.3533038.
220. Castillo, M. H. G. del, Garcia, R. M., Mazón, M. J. C., Garcia, E. A. & Fernández-Miranda, P. M. Diffusion models for conditional MRI generation. Preprint at https://doi.org/10.48550/arXiv.2502.18620 (2025).
221. Wang, Y. *et al.* Towards general text-guided image synthesis for customized multimodal brain MRI generation. Preprint at https://doi.org/10.48550/arXiv.2409.16818 (2024).
222. Xing, X., Ning, J., Nan, Y. & Yang, G. Deep generative models unveil patterns in medical images through vision-language conditioning. Preprint at https://doi.org/10.48550/arXiv.2410.13823 (2024).
223. Zhou, X. *et al.* Multimodality MRI synchronous construction based deep learning framework for MRI-guided radiotherapy synthetic CT generation. *Comput. Biol. Med.* **162**, 107054 (2023).
224. Ben-Cohen, A. *et al.* Cross-modality synthesis from CT to PET using FCN and GAN networks for improved automated lesion detection. *Eng. Appl. Artif. Intell.* **78**, 186–194 (2019).
225. Haubold, J. *et al.* Contrast agent dose reduction in computed tomography with deep learning using a conditional generative adversarial network. *Eur. Radio.* **31**, 6087–6095 (2021).
226. Jiao, J., Namburete, A. I., Papageorghiou, A. T. & Noble, J. A. Self-supervised ultrasound to MRI fetal brain image synthesis. *IEEE Trans. Med. Imag.* **39**, 4413–4424 (2020).
227. Dong, X. *et al.* Synthetic CT generation from non-attenuation corrected PET images for whole-body PET imaging. *Phys. Med. Biol.* **64**, (2019).
228. Yang, H. *et al.* Unsupervised MR-to-CT synthesis using structure-constrained CycleGAN. *IEEE Trans. Med. Imag.* **39**, 4249–4261 (2020).
229. Gong, K. *et al.* MR-based attenuation correction for brain PET using 3-D cycle-consistent adversarial network. *IEEE Trans. Radiat. Plasma Med. Sci.* **5**, 185–192 (2021).
230. Wang, Z., Zhang, L., Wang, L. & Zhang, Z. Soft masked mamba diffusion model for CT to MRI conversion. Preprint at https://doi.org/10.48550/arXiv.2406.15910 (2024).
231. Liu, C. *et al.* Frequency space mamba enhanced bidirectional generative network for dual-source CBCT. in *Proc SPIE Int Soc Opt Eng* (eds. Luo Q., Li X., Gu Y., & Zhu D.) vol. 13242 61–68 (SPIE, 2024).

232. Atli, O. F. *et al.* I2I-mamba: multi-modal medical image synthesis via selective state space modeling. Preprint at https://doi.org/10.48550/arXiv.2405.14022 (2024).
233. Zhou, X. *et al.* GLFC: unified global-local feature and contrast learning with mamba-enhanced UNet for synthetic CT generation from CBCT. Preprint at https://doi.org/10.48550/arXiv.2501.02992 (2025).
234. Ferreira, V. R. S. *et al.* Diffusion model for generating synthetic contrast enhanced CT from non-enhanced heart axial CT images. in *International Conference on Enterprise Information Systems, ICEIS - Proceedings* (eds. Filipe J., Smialek M., Brodsky A., & Hammoudi S.) vol. 1 857–864 (Science and Technology Publications, Lda, 2024).
235. Dalmaz, O., Yurt, M. & Cukur, T. ResViT: residual vision transformers for multimodal medical image synthesis. *IEEE Trans. Med. Imag.* **41**, 2598–2614 (2022).
236. Liu, J. *et al.* One model to synthesize them all: multi-contrast multi-scale transformer for missing data imputation. in *IEEE Trans. Med. Imaging* vol. 42 2577–2591 (Institute of Electrical and Electronics Engineers Inc., 2023).
237. Kaplan, S. *et al.* Synthesizing pseudo-T2w images to recapture missing data in neonatal neuroimaging with applications in rs-fMRI. *Neuroimage* **253**, 119091 (2022).
238. Yan, K. *et al.* Coarse-to-fine learning framework for semi-supervised multimodal MRI synthesis. in *Lect. Notes Comput. Sci.* (eds. Wallraven C., Liu Q., & Nagahara H.) vol. 13189 LNCS 370–384 (Springer Science and Business Media Deutschland GmbH, 2022).
239. Xiao, X., Hu, Q. V. & Wang, G. FgC2F-UDiff: frequency-guided and coarse-to-fine unified diffusion model for multi-modality missing MRI synthesis. *IEEE Trans. Comput. Imaging* (2024) doi:10.1109/TCI.2024.3516574.
240. Zhang, Y. *et al.* Unified multi-modal image synthesis for missing modality imputation. *IEEE Trans. Med. Imag.* (2024).
241. Raad, R. *et al.* Conditional generative learning for medical image imputation. *Sci. Rep.* **14**, 171 (2024).
242. Zhou, B., Zhou, Q., Miao, C., Liu, Y. & Guo, Y. Cross-dimensional knowledge-guided synthesizer trained with unpaired multimodality MRIs. *Soft Comput.* **28**, 8393–8408 (2024).
243. Guo, P. *et al.* Anatomic and molecular MR image synthesis using confidence guided CNNs. *IEEE Trans. Med. Imag.* **40**, 2832–2844 (2020).
244. Shen, Z. *et al.* Image synthesis with disentangled attributes for chest x-ray nodule augmentation and detection. *Med. Image Anal.* **84**, 102708 (2023).
245. Hou, B. High-fidelity diabetic retina fundus image synthesis from freestyle lesion maps. *Biomed. Opt. Express* **14**, 533 (2023).
246. Jia, Y., Chen, G. & Chi, H. Retinal fundus image super-resolution based on generative adversarial network guided with vascular structure prior. *Sci. Rep.* **14**, 22786 (2024).
247. Guo, W. *et al.* LN-gen: rectal lymph nodes generation via anatomical features. Preprint at https://doi.org/10.48550/arXiv.2408.14977 (2024).
248. Konz, N., Chen, Y., Dong, H. & Mazurowski, M. A. Anatomically-controllable medical image generation with segmentation-guided diffusion models. Preprint at https://doi.org/10.48550/arXiv.2402.05210 (2024).

249. Campello, V. M. *et al.* Cardiac aging synthesis from cross-sectional data with conditional generative adversarial networks. *Front. Cardiovasc. Med.* **9**, (2022).
250. Lai, X. Modelling, inference and simulation of personalised breast cancer treatment. (2019).
251. Savić, M., Kurbalija, V., Balaz, I. & Ivanović, M. Heterogeneous tumour modeling using PhysiCell and its implications in precision medicine. in *Cancer, Complexity, Computation* (eds. Balaz, I. & Adamatzky, A.) vol. 46 157–189 (Springer International Publishing, Cham, 2022).
252. Liu, S., Zhang, J., Li, T., Yan, H. & Liu, J. Technical note: a cascade 3D U-net for dose prediction in radiotherapy. *Med. Phys.* **48**, 5574–5582 (2021).
253. Kearney, V., Chan, J. W., Haaf, S., Descovich, M. & Solberg, T. D. DoseNet: a volumetric dose prediction algorithm using 3D fully-convolutional neural networks. *Phys. Med. Biol.* **63**, 235022 (2018).
254. Radonic, D. *et al.* Proton dose calculation with LSTM networks in presence of a magnetic field. *Phys. Med. Biol.* **69**, 215019 (2024).
255. Jiao, Z. *et al.* TransDose: transformer-based radiotherapy dose prediction from CT images guided by super-pixel-level GCN classification. *Med. Image Anal.* **89**, 102902 (2023).
256. Fu, L. *et al.* SP-DiffDose: a conditional diffusion model for radiation dose prediction based on multi-scale fusion of anatomical structures, guided by SwinTransformer and projector. Preprint at https://doi.org/10.48550/arXiv.2312.06187 (2023).
257. Feng, Z. *et al.* DiffDP: radiotherapy dose prediction via a diffusion model. in *Medical Image Computing and Computer Assisted Intervention – MICCAI 2023* (eds. Greenspan, H. et al.) vol. 14225 191–201 (Springer Nature Switzerland, Cham, 2023).
258. Fu, L. *et al.* MD-dose: a diffusion model based on the mamba for radiotherapy dose prediction. *Arxiv E-prints* arXiv-2403 (2024).
259. Wang, J. *et al.* Self-improving generative foundation model for synthetic medical image generation and clinical applications. *Nat. Med.* **31**, 1–9 (2024).
260. Pan, S. *et al.* Data-driven volumetric image generation from surface structures using a patient-specific deep leaning model. *Arxiv* arXiv-2304 (2023).
261. Pan, S. *et al.* Patient-specific CBCT synthesis for real-time tumor tracking in surface-guided radiotherapy. Preprint at https://doi.org/10.48550/arXiv.2410.23582 (2024).
262. Zhou, M. & Khalvati, F. Conditional generation of 3d brain tumor regions via VQGAN and temporal-agnostic masked transformer. in *Medical Imaging with Deep Learning* (2024).
263. Yoon, S. *et al.* Accelerated cardiac MRI cine with use of resolution enhancement generative adversarial inline neural network. *Radiology* **307**, e222878 (2023).
264. Zakeri, A. *et al.* DragNet: learning-based deformable registration for realistic cardiac MR sequence generation from a single frame. *Med. Image Anal.* **83**, (2023).
265. Pellicer, A. O. *et al.* Generation of synthetic echocardiograms using video diffusion models. in *Proc IEEE Southwest Symp Image Anal Interpret* 33–36 (Institute of Electrical and Electronics Engineers Inc., 2024). doi:10.1109/SSIAI59505.2024.10508643.

266. Li, C. *et al.* Endora: Video generation models as endoscopy simulators. Preprint at https://doi.org/10.48550/arXiv.2403.11050 (2024).
267. Ghodrati, V. *et al.* Temporally aware volumetric generative adversarial network-based MR image reconstruction with simultaneous respiratory motion compensation: initial feasibility in 3D dynamic cine cardiac MRI. *Magn. Reson. Med.* **86**, 2666–2683 (2021).
268. Thummerer, A. *et al.* Deep learning–based 4D-synthetic CTs from sparse-view CBCTs for dose calculations in adaptive proton therapy. *Med. Phys.* **49**, 6824–6839 (2022).
269. Quintero, P., Wu, C., Otazo, R., Cervino, L. & Harris, W. On-board synthetic 4D MRI generation from 4D CBCT for radiotherapy of abdominal tumors: a feasibility study. *Med. Phys.* **51**, 9194–9206 (2024).
270. Hu, D. *et al.* DPI-MoCo: deep prior image constrained motion compensation reconstruction for 4D CBCT. *IEEE Trans. Med. Imag.* (2024) doi:10.1109/TMI.2024.3483451.
271. Han, L. *et al.* Synthesis-based imaging-differentiation representation learning for multi-sequence 3D/4D MRI. *Med. Image Anal.* **92**, (2024).
272. Liu, C., Yuan, X., Yu, Z. & Wang, Y. Texdc: text-driven disease-aware 4d cardiac cine mri images generation. in *Proceedings of the Asian Conference on Computer Vision* 3005–3021 (2024).
273. Chen, Q. *et al.* Towards generalizable tumor synthesis. in *Proceedings of the IEEE/CVF Conference on Computer Vision and Pattern Recognition* 11147–11158 (2024).
274. Moya-Sáez, E. *et al.* Synthetic MRI improves radiomics-based glioblastoma survival prediction. *NMR Biomed.* **35**, (2022).
275. Takahashi, W., Oshikawa, S. & Mori, S. Real-time markerless tumour tracking with patient-specific deep learning using a personalised data generation strategy: proof of concept by phantom study. *Br. J. Radiol.* **93**, (2020).
276. Lei, W. *et al.* A data-efficient pan-tumor foundation model for oncology CT interpretation. Preprint at https://doi.org/10.48550/arXiv.2502.06171 (2025).
277. Elazab, A. *et al.* GP-GAN: brain tumor growth prediction using stacked 3D generative adversarial networks from longitudinal MR images. *Neural Networks* **132**, 321–332 (2020).
278. Ravi, D., Alexander, D. C. & Oxtoby, N. P. Degenerative adversarial NeuroImage nets: generating images that mimic disease progression. in *Lect. Notes Comput. Sci.* (eds. Shen D. et al.) vol. 11766 LNCS 164–172 (Springer Science and Business Media Deutschland GmbH, 2019).
279. SinhaRoy, R. & Sen, A. A hybrid deep learning framework to predict alzheimer's disease progression using generative adversarial networks and deep convolutional neural networks. *Arabian J. Sci. Eng.* **49**, 3267–3284 (2024).
280. Litrico, M., Guarnera, F., Giuffrida, M. V., Ravì, D. & Battiato, S. TADM: temporally-aware diffusion model for neurodegenerative progression on brain MRI. in *Medical Image Computing and Computer Assisted Intervention – MICCAI 2024* (eds. Linguraru, M. G. et al.) vol. 15002 444–453 (Springer Nature Switzerland, Cham, 2024).
281. Song, L., Wang, Q., Li, H., Fan, J. & Hu, B. Longitudinal structural MRI data prediction in nondemented and demented older adults via generative adversarial convolutional network. *Neural Process. Lett.* **55**, 989–999 (2023).

282. Puglisi, L., Alexander, D. C. & Ravì, D. Enhancing spatiotemporal disease progression models via latent diffusion and prior knowledge. in *Medical Image Computing and Computer Assisted Intervention – MICCAI 2024* (eds. Linguraru, M. G. et al.) vol. 15002 173–183 (Springer Nature Switzerland, Cham, 2024).
283. Yalcin, C. *et al.* Hematoma expansion prediction in intracerebral hemorrhage patients by using synthesized CT images in an end-to-end deep learning framework. *Comput. Med. Imaging Graphics* **117**, (2024).
284. Bycroft, C. *et al.* The UK biobank resource with deep phenotyping and genomic data. *Nature* **562**, 203–209 (2018).
285. Clark, K. *et al.* The cancer imaging archive (TCIA): maintaining and operating a public information repository. *J. Digital Imaging* **26**, 1045–1057 (2013).
286. Yan, K., Wang, X., Lu, L. & Summers, R. M. DeepLesion: automated mining of large-scale lesion annotations and universal lesion detection with deep learning. *J. Med. Imaging* **5**, 36501–36501 (2018).
287. Shi, K., Li, Y., Ho, B., Wang, J. & Guo, K. Universal lesion segmentation challenge 2023: a comparative research of different algorithms. Preprint at https://doi.org/10.48550/arXiv.2502.10608 (2025).
288. Wu, L., Zhuang, J. & Chen, H. Large-scale 3D medical image pre-training with geometric context priors. Preprint at https://doi.org/10.48550/arXiv.2410.09890 (2024).
289. Wasserthal, J. *et al.* TotalSegmentator: robust segmentation of 104 anatomic structures in CT images. *Radiol.: Artif. Intell.* **5**, e230024 (2023).
290. D'Antonoli, T. A. *et al.* TotalSegmentator MRI: robust sequence-independent segmentation of multiple anatomic structures in MRI. *Radiology* **314**, e241613 (2025).
291. Gatidis, S. *et al.* A whole-body FDG-PET/CT dataset with manually annotated tumor lesions. *Sci. Data* **9**, 601 (2022).
292. Li, X. *et al.* The state-of-the-art 3D anisotropic intracranial hemorrhage segmentation on non-contrast head CT: the INSTANCE challenge. Preprint at https://doi.org/10.48550/arXiv.2301.03281 (2023).
293. Luo, X. *et al.* Segrap2023: a benchmark of organs-at-risk and gross tumor volume segmentation for radiotherapy planning of nasopharyngeal carcinoma. *Med. Image Anal.* **101**, 103447 (2025).
294. Andrearczyk, V. *et al.* Overview of the HECKTOR challenge at MICCAI 2022: automatic head and neck tumor segmentation and outcome prediction in PET/CT. in *Head and Neck Tumor Segmentation and Outcome Prediction* (eds. Andrearczyk, V., Oreiller, V., Hatt, M. & Depeursinge, A.) vol. 13626 1–30 (Springer Nature Switzerland, Cham, 2023).
295. Baid, U. *et al.* The RSNA-ASNR-MICCAI BraTS 2021 benchmark on brain tumor segmentation and radiogenomic classification. Preprint at https://doi.org/10.48550/arXiv.2107.02314 (2021).
296. LaBella, D. *et al.* The ASNR-MICCAI brain tumor segmentation (BraTS) challenge 2023: intracranial meningioma. Preprint at https://doi.org/10.48550/arXiv.2305.07642 (2023).
297. Zbontar, J. *et al.* fastMRI: An Open Dataset and Benchmarks for Accelerated MRI. Preprint at https://doi.org/10.48550/arXiv.1811.08839 (2019).

298. Jiang, D. *et al.* Denoising of 3D magnetic resonance images with multi-channel residual learning of convolutional neural network. *Jpn. J. Radiol.* **36**, 566–574 (2018).
299. Snoek, L. *et al.* The Amsterdam open MRI collection, a set of multimodal MRI datasets for individual difference analyses. *Sci. Data* **8**, 85 (2021).
300. Bernard, O. *et al.* Deep learning techniques for automatic MRI cardiac multi-structures segmentation and diagnosis: is the problem solved? *IEEE Trans. Med. Imag.* **37**, 2514–2525 (2018).
301. Campello, V. M. *et al.* Multi-centre, multi-vendor and multi-disease cardiac segmentation: the M&ms challenge. *IEEE Trans. Med. Imag.* **40**, 3543–3554 (2021).
302. Galazis, C. *et al.* Tempera: spatial transformer feature pyramid network for cardiac MRI segmentation. in *Statistical Atlases and Computational Models of the Heart. Multi-disease, Multi-view, and Multi-center Right Ventricular Segmentation in Cardiac MRI Challenge* (eds. Puyol Antón, E. et al.) vol. 13131 268–276 (Springer International Publishing, Cham, 2022).
303. Chen, C. *et al.* OCMR (v1.0)--open-access multi-coil k-space dataset for cardiovascular magnetic resonance imaging. Preprint at https://doi.org/10.48550/arXiv.2008.03410 (2020).
304. Ma, J. *et al.* Unleashing the strengths of unlabeled data in pan-cancer abdominal organ quantification: the FLARE22 challenge. Preprint at https://doi.org/10.48550/arXiv.2308.05862 (2023).
305. Ma, J. *et al.* Automatic organ and pan-cancer segmentation in abdomen CT: the FLARE 2023 challenge. Preprint at https://doi.org/10.48550/arXiv.2408.12534 (2024).
306. Ma, J. *et al.* Abdomenct-1k: is abdominal organ segmentation a solved problem? *IEEE Trans. Pattern Anal. Mach. Intell.* **44**, 6695–6714 (2021).
307. Li, X. *et al.* The prostate imaging: cancer AI (PI-CAI) 2022 grand challenge (PIMed team). *Dep. Radiol. Stanf. Univ. Stanf. CA 94305 USA; Dep. Urol. Stanf. Univ. Stanf. CA 94305 USA; Inst. Comput. Math. Eng. Stanf. CA 94305 USA*.
308. Ying, N. *et al.* CPIA dataset: a comprehensive pathological image analysis dataset for self-supervised learning pre-training. Preprint at https://doi.org/10.48550/arXiv.2310.17902 (2023).
309. Veeling, B. S., Linmans, J., Winkens, J., Cohen, T. & Welling, M. Rotation equivariant CNNs for digital pathology. in *Medical Image Computing and Computer Assisted Intervention – MICCAI 2018* (eds. Frangi, A. F., Schnabel, J. A., Davatzikos, C., Alberola-López, C. & Fichtinger, G.) vol. 11071 210–218 (Springer International Publishing, Cham, 2018).
310. Spanhol, F. A., Oliveira, L. S., Petitjean, C. & Heutte, L. A dataset for breast cancer histopathological image classification. *IEEE Trans. Bio-Med. Eng.* **63**, 1455–1462 (2015).
311. Zingman, I., Stierstorfer, B., Lempp, C. & Heinemann, F. Learning image representations for anomaly detection: application to discovery of histological alterations in drug development. *Med. Image Anal.* **92**, 103067 (2024).
312. Li, F., Hu, Z., Chen, W. & Kak, A. Adaptive supervised PatchNCE loss for learning H&E-to-IHC stain translation with inconsistent groundtruth image pairs. in *Medical*

*Image Computing and Computer Assisted Intervention – MICCAI 2023* (eds. Greenspan, H. et al.) vol. 14225 632–641 (Springer Nature Switzerland, Cham, 2023).

313. Song, A. H. *et al.* Artificial intelligence for digital and computational pathology. *Nat. Rev. Bioeng.* **1**, 930–949 (2023).
314. Gong, H. *et al.* Multi-task learning for thyroid nodule segmentation with thyroid region prior. in *2021 IEEE 18th International Symposium on Biomedical Imaging (ISBI)* 257–261 (IEEE, 2021). doi:10.1109/ISBI48211.2021.9434087.
315. Duffy, G. *et al.* High-throughput precision phenotyping of left ventricular hypertrophy with cardiovascular deep learning. *JAMA Cardiol.* **7**, 386–395 (2022).
316. Ozyoruk, K. B. *et al.* EndoSLAM dataset and an unsupervised monocular visual odometry and depth estimation approach for endoscopic videos. *Med. Image Anal.* **71**, 102058 (2021).
317. Kermany, D. Labeled optical coherence tomography (oct) and chest x-ray images for classification. *Mendeley Data* (2018).
318. Subramanian, M., Shanmugavadivel, K., Naren, O. S., Premkumar, K. & Rankish, K. Classification of retinal oct images using deep learning. in *2022 International Conference on Computer Communication and Informatics (ICCCI)* 1–7 (IEEE, 2022). doi:10.1109/ICCCI54379.2022.9740985.
319. Li, L., Xu, M., Wang, X., Jiang, L. & Liu, H. Attention based glaucoma detection: a large-scale database and CNN model. in *Proceedings of the IEEE/CVF Conference on Computer Vision and Pattern Recognition* 10571–10580 (2019).
320. De Vente, C. *et al.* Airogs: artificial intelligence for robust glaucoma screening challenge. *IEEE Trans. Med. Imag.* **43**, 542–557 (2023).
321. Chambon, P. *et al.* CheXpert plus: augmenting a large chest X-ray dataset with text radiology reports, patient demographics and additional image formats. Preprint at https://doi.org/10.48550/arXiv.2405.19538 (2024).
322. Hu, X. *et al.* Interpretable medical image visual question answering via multi-modal relationship graph learning. *Med. Image Anal.* **97**, 103279 (2024).
323. Bustos, A., Pertusa, A., Salinas, J.-M. & De La Iglesia-Vaya, M. Padchest: a large chest x-ray image dataset with multi-label annotated reports. *Med. Image Anal.* **66**, 101797 (2020).
324. Ikezogwo, W. *et al.* Quilt-1m: one million image-text pairs for histopathology. *Adv. Neural Inf. Process. Syst.* **36**, 37995–38017 (2023).
325. Huang, Z., Bianchi, F., Yuksekgonul, M., Montine, T. J. & Zou, J. A visual–language foundation model for pathology image analysis using medical twitter. *Nat. Med.* **29**, 2307–2316 (2023).
326. Xie, Y. *et al.* Medtrinity-25m: a large-scale multimodal dataset with multigranular annotations for medicine. *Arxiv Prepr. Arxiv:2408,02900* (2024).
327. Subramanian, S. *et al.* MedICaT: a dataset of medical images, captions, and textual references. Preprint at https://doi.org/10.48550/arXiv.2010.06000 (2020).
328. Lin, W. *et al.* PMC-CLIP: contrastive language-image pre-training using biomedical documents. in *Medical Image Computing and Computer Assisted Intervention – MICCAI 2023* (eds. Greenspan, H. et al.) vol. 14227 525–536 (Springer Nature Switzerland, Cham, 2023).

329. Saha, A. *et al.* A machine learning approach to radiogenomics of breast cancer: a study of 922 subjects and 529 DCE-MRI features. *Br. J. Cancer* **119**, 508–516 (2018).
330. Li, W., Newitt, D. C. & Gibbs, J. I-SPY 2 breast dynamic contrast enhanced MRI trial (ISPY2). *Cancer Imaging Arch.* (2023).
331. Yang, J. *et al.* MedMNIST v2 - a large-scale lightweight benchmark for 2D and 3D biomedical image classification. *Sci. Data* **10**, 41 (2023).
332. Zhu, J. Method for MICCAI FLARE24 challenge. in *MICCAI 2024 FLARE Challenge*.
333. Rister, B., Yi, D., Shivakumar, K., Nobashi, T. & Rubin, D. L. CT-ORG, a new dataset for multiple organ segmentation in computed tomography. *Sci. Data* **7**, 381 (2020).
334. Shapey, J. *et al.* Segmentation of vestibular schwannoma from MRI, an open annotated dataset and baseline algorithm. *Sci. Data* **8**, 286 (2021).
335. Gireesha, H. M. & Nanda, S. Thyroid nodule segmentation and classification in ultrasound images. *Int. J. Eng. Res. Technol.* (2014).
336. Anna Montoya, Hasnin, kaggle446, shirzad, Will Cukierski, and yffud. Ultrasound nerve segmentation. https://kaggle.com/ultrasound-nerve-segmentation.
337. Herrera-Chavez, A. I. *et al.* Multi-label image classification for ocular disease diagnosis using K-fold cross-validation on the ODIR-5K dataset. in *2024 IEEE 33rd International Symposium on Industrial Electronics (ISIE)* 1–6 (IEEE, 2024). doi:10.1109/ISIE54533.2024.10595740.
338. Pires, R. *et al.* A data-driven approach to referable diabetic retinopathy detection. *Artif. Intell. Med.* **96**, 93–106 (2019).
339. Rahman, T. Y., Mahanta, L. B., Chakraborty, C., Das, A. K. & Sarma, J. D. Textural pattern classification for oral squamous cell carcinoma. *J. Microsc.* **269**, 85–93 (2018).
340. Elias, P. & Bhave, S. CheXchoNet: a chest radiograph dataset with gold standard echocardiography labels.
341. Reis, E. P. *et al.* BRAX, brazilian labeled chest x-ray dataset. *Sci. Data* **9**, 487 (2022).
342. Lakhani, P. *et al.* The 2021 SIIM-FISABIO-RSNA machine learning COVID-19 challenge: annotation and standard exam classification of COVID-19 chest radiographs. *J. Digital Imaging* **36**, 365–372 (2022).
343. Pehrson, L. M., Nielsen, M. B. & Ammitzbøl Lauridsen, C. Automatic pulmonary nodule detection applying deep learning or machine learning algorithms to the LIDC-IDRI database: a systematic review. *Diagnostics* **9**, 29 (2019).
344. Soares, E., Angelov, P., Biaso, S., Froes, M. H. & Abe, D. K. SARS-CoV-2 CT-scan dataset: a large dataset of real patients CT scans for SARS-CoV-2 identification. *Medrxiv* 2020–4 (2020).
345. Zhang, M. *et al.* Multi-site, multi-domain airway tree modeling. *Med. Image Anal.* **90**, 102957 (2023).
346. Setio, A. A. A. *et al.* Validation, comparison, and combination of algorithms for automatic detection of pulmonary nodules in computed tomography images: the LUNA16 challenge. *Med. Image Anal.* **42**, 1–13 (2017).
347. Dorent, R. *et al.* LNQ 2023 challenge: benchmark of weakly-supervised techniques for mediastinal lymph node quantification. *Mach. Learn. Biomed. Imaging* **3**, 1–15 (2025).
348. Chitalia, R. *et al.* Expert tumor annotations and radiomic features for the ispy1/acrin 6657 trial data collection. *Cancer Imaging Arch.* (2022).

349. You, C. *et al.* Artificial intelligence in breast imaging: current situation and clinical challenges. *Exploration* **3**, 20230007 (2023).
350. Al-Dhabyani, W., Gomaa, M., Khaled, H. & Fahmy, A. Dataset of breast ultrasound images. *Data Brief* **28**, 104863 (2020).
351. Luo, G. *et al.* Tumor detection, segmentation and classification challenge on automated 3D breast ultrasound: the TDSC-ABUS challenge. Preprint at https://doi.org/10.48550/arXiv.2501.15588 (2025).
352. Han, C. *et al.* WSSS4LUAD: grand challenge on weakly-supervised tissue semantic segmentation for lung adenocarcinoma. Preprint at https://doi.org/10.48550/arXiv.2204.06455 (2022).
353. Li, L., Zimmer, V. A., Schnabel, J. A. & Zhuang, X. AtrialJSQnet: a new framework for joint segmentation and quantification of left atrium and scars incorporating spatial and shape information. *Med. Image Anal.* **76**, 102303 (2022).
354. El-Rewaidy, H. *et al.* Multi-domain convolutional neural network (MD-CNN) for radial reconstruction of dynamic cardiac MRI. *Magn. Reson. Med.* **85**, 1195–1208 (2021).
355. Wang, C. *et al.* CMRxRecon: a publicly available k-space dataset and benchmark to advance deep learning for cardiac MRI. *Sci. Data* **11**, 687 (2024).
356. Andreopoulos, A. & Tsotsos, J. K. Efficient and generalizable statistical models of shape and appearance for analysis of cardiac MRI. *Med. Image Anal.* **12**, 335–357 (2008).
357. Chen, Z., Ren, H., Li, Q. & Li, X. Motion correction and super-resolution for multi-slice cardiac magnetic resonance imaging via an end-to-end deep learning approach. *Comput. Med. Imaging Graphics* **115**, 102389 (2024).
358. Vukadinovic, M., Kwan, A. C., Li, D. & Ouyang, D. GANcMRI: cardiac magnetic resonance video generation and physiologic guidance using latent space prompting. in *Machine Learning for Health (ML4H)* 594–606 (PMLR, 2023).
359. Ouyang, D. *et al.* Video-based AI for beat-to-beat assessment of cardiac function. *Nature* **580**, 252–256 (2020).
360. Qu, C. *et al.* Abdomenatlas-8k: annotating 8,000 ct volumes for multi-organ segmentation in three weeks. *Adv. Neural Inf. Process. Syst.* **36**, 36620–36636 (2023).
361. Zhou, H., Lou, Y., Xiong, J., Wang, Y. & Liu, Y. Improvement of deep learning model for gastrointestinal tract segmentation surgery. *Front. Comput. Intell. Syst.* **6**, 103–106 (2023).
362. Pardo, J., Liu, J., Ramón-Ferrer, V., Amador-Domínguez, E. & Calleja, P. K-flares: a K-adapter based approach for the FLARES challenge. in *In Proceedings of the Iberian Languages Evaluation Forum (iberlef 2024), Co-located with the 40th Conference of the Spanish Society for Natural Language Processing (SEPLN 2024), CEURWS. Org* (2024).
363. Bai, J. *et al.* PSFHS challenge report: pubic symphysis and fetal head segmentation from intrapartum ultrasound images. *Med. Image Anal.* **99**, 103353 (2025).
364. Polat, G. *et al.* Improving the computer-aided estimation of ulcerative colitis severity according to mayo endoscopic score by using regression-based deep learning. *Inflamm. Bowel Dis.* **29**, 1431–1439 (2023).

365. Misawa, M. *et al.* Development of a computer-aided detection system for colonoscopy and a publicly accessible large colonoscopy video database (with video). *Gastrointest. Endosc.* **93**, 960–967 (2021).
366. Brummer, O., Pölönen, P., Mustjoki, S. & Brück, O. Integrative analysis of histological textures and lymphocyte infiltration in renal cell carcinoma using deep learning. *Biorxiv* 2022–8 (2022).
367. Hu, W. *et al.* GasHisSDB: a new gastric histopathology image dataset for computer aided diagnosis of gastric cancer. *Comput. Biol. Med.* **142**, 105207 (2022).
368. Kawai, M., Ota, N. & Yamaoka, S. Large-scale pretraining on pathological images for fine-tuning of small pathological benchmarks. in *Medical Image Learning with Limited and Noisy Data* (eds. Xue, Z. et al.) vol. 14307 257–267 (Springer Nature Switzerland, Cham, 2023).
369. Tsai, M.-J. & Tao, Y.-H. Deep learning techniques for the classification of colorectal cancer tissue. *Electronics* **10**, 1662 (2021).
370. van der Graaf, J. W. *et al.* Lumbar spine segmentation in MR images: a dataset and a public benchmark. *Sci. Data* **11**, 264 (2024).
371. Sharafi, A., Arpinar, V. E., Nencka, A. S. & Koch, K. M. Development and stability analysis of carpal kinematic metrics from 4D magnetic resonance imaging. *Skeletal Radiol.* **54**, 57–65 (2025).
372. Desai, A. D. *et al.* SKM-TEA: a dataset for accelerated MRI reconstruction with dense image labels for quantitative clinical evaluation. Preprint at https://doi.org/10.48550/arXiv.2203.06823 (2022).
373. Eskicioglu, A. M. & Fisher, P. S. Image quality measures and their performance. *IEEE Trans. Commun.* **43**, 2959–2965 (1995).
374. Wang, Z., Bovik, A. C., Sheikh, H. R. & Simoncelli, E. P. Image quality assessment: from error visibility to structural similarity. *IEEE Trans. Image Process.* **13**, 600–612 (2004).
375. Wang, Z., Simoncelli, E. P. & Bovik, A. C. Multiscale structural similarity for image quality assessment. in *The Thrity-seventh Asilomar Conference on Signals, Systems & Computers, 2003* vol. 2 1398–1402 (Ieee, 2003).
376. Zhang, L., Zhang, L., Mou, X. & Zhang, D. FSIM: a feature similarity index for image quality assessment. *IEEE Trans. Image Process.* **20**, 2378–2386 (2011).
377. Wang, Z. & Li, Q. Information content weighting for perceptual image quality assessment. *IEEE Trans. Image Process.* **20**, 1185–1198 (2010).
378. Sheikh, H. R. & Bovik, A. C. A visual information fidelity approach to video quality assessment. in *The First International Workshop on Video Processing and Quality Metrics for Consumer Electronics* vol. 7 2117–2128 (sn, 2005).
379. Wang, Z. & Bovik, A. C. A universal image quality index. *IEEE Signal Process. Lett.* **9**, 81–84 (2002).
380. Bercea, C. I., Wiestler, B., Rueckert, D. & Schnabel, J. A. Evaluating normative representation learning in generative AI for robust anomaly detection in brain imaging. *Nat. Commun.* **16**, 1624 (2025).
381. Yu, Y., Zhang, W. & Deng, Y. Frechet inception distance (fid) for evaluating gans. *China Univ. Min. Technol. Beijing Grad. Sch.* **3**, (2021).

382. Wang, C. *et al.* Pulmonary image classification based on inception-v3 transfer learning model. *IEEE Access* **7**, 146533–146541 (2019).
383. Korotin, A., Egiazarian, V., Asadulaev, A., Safin, A. & Burnaev, E. Wasserstein-2 generative networks. Preprint at https://doi.org/10.48550/arXiv.1909.13082 (2020).
384. Wenliang, L., Moskovitz, T., Kanagawa, H. & Sahani, M. Amortised learning by wake-sleep. in *International Conference on Machine Learning* 10236–10247 (PMLR, 2020).
385. Gretton, A., Borgwardt, K. M., Rasch, M. J., Schölkopf, B. & Smola, A. A kernel two-sample test. *J. Mach. Learn. Res.* **13**, 723–773 (2012).
386. Barratt, S. & Sharma, R. A note on the inception score. Preprint at https://doi.org/10.48550/arXiv.1801.01973 (2018).
387. Simonyan, K. & Zisserman, A. Very deep convolutional networks for large-scale image recognition. Preprint at https://doi.org/10.48550/arXiv.1409.1556 (2015).
388. Dosovitskiy, A. & Brox, T. Generating images with perceptual similarity metrics based on deep networks. *Adv. Neural Inf. Process. Syst.* **29**, (2016).
389. Radford, A. *et al.* Learning transferable visual models from natural language supervision. in *International Conference on Machine Learning* 8748–8763 (PmLR, 2021).
390. Sun, W. *et al.* Bora: Biomedical generalist video generation model. Preprint at https://doi.org/10.48550/arXiv.2407.08944 (2024).
391. Devlin, J., Chang, M.-W., Lee, K. & Toutanova, K. Bert: pre-training of deep bidirectional transformers for language understanding. in *Proceedings of the 2019 Conference of the North American Chapter of the Association for Computational Linguistics: Human Language Technologies, Volume 1 (long and Short Papers)* 4171–4186 (2019).
392. Zhang, H., Li, X. & Bing, L. Video-LLaMA: an instruction-tuned audio-visual language model for video understanding. Preprint at https://doi.org/10.48550/arXiv.2306.02858 (2023).
393. Unterthiner, T. *et al.* FVD: a new metric for video generation. (2019).
394. Carreira, J. & Zisserman, A. Quo vadis, action recognition? a new model and the kinetics dataset. in *Proceedings of the IEEE Conference on Computer Vision and Pattern Recognition* 6299–6308 (2017).
395. Unterthiner, T. *et al.* Towards accurate generative models of video: a new metric & challenges. Preprint at https://doi.org/10.48550/arXiv.1812.01717 (2019).
396. Liu, J. *et al.* Fréchet video motion distance: a metric for evaluating motion consistency in videos. Preprint at https://doi.org/10.48550/arXiv.2407.16124 (2024).
397. Huang, Z. *et al.* VBench: comprehensive benchmark suite for video generative models. in 21807–21818 (2024).
398. Teed, Z. & Deng, J. RAFT-3D: scene flow using rigid-motion embeddings. in 8375–8384 (2021).
399. Griffin, A. *et al.* Revalidation and quality assurance: the application of the MUSIQ framework in independent verification visits to healthcare organisations. *BMJ Open* **7**, e014121 (2017).
400. Xu, H. *et al.* VideoCLIP: contrastive pre-training for zero-shot video-text understanding. Preprint at https://doi.org/10.48550/arXiv.2109.14084 (2021).

401. Budd, S., Robinson, E. C. & Kainz, B. A survey on active learning and human-in-the-loop deep learning for medical image analysis. *Med. Image Anal.* **71**, 102062 (2021).
402. Dorjsembe, Z., Pao, H.-K., Odonchimed, S. & Xiao, F. Conditional diffusion models for semantic 3D medical image synthesis. Preprint at https://doi.org/10.36227/techrxiv.23723787 (2023).
403. Yu, Y., Gu, Y., Zhang, S. & Zhang, X. MedDiff-FM: A diffusion-based foundation model for versatile medical image applications. Preprint at https://doi.org/10.48550/arXiv.2410.15432 (2024).
404. Sun, Y., Wang, L., Li, G., Lin, W. & Wang, L. A foundation model for enhancing magnetic resonance images and downstream segmentation, registration and diagnostic tasks. *Nat. Biomed. Eng.* 1–18 (2024) doi:10.1038/s41551-024-01283-7.
405. Wang, S. *et al.* Triad: vision foundation model for 3D magnetic resonance imaging. Preprint at https://doi.org/10.48550/arXiv.2502.14064 (2025).
406. Sun, Y. *et al.* A data-efficient strategy for building high-performing medical foundation models. *Nat. Biomed. Eng.* 1–13 (2025) doi:10.1038/s41551-025-01365-0.
407. Yang, Z. *et al.* A foundation model for generalizable cancer diagnosis and survival prediction from histopathological images. *Nat. Commun.* **16**, 2366 (2025).
408. Kim, C. *et al.* Transparent medical image AI via an image–text foundation model grounded in medical literature. *Nat. Med.* **30**, 1154–1165 (2024).